\documentclass[a4paper]{article}
\usepackage{epsfig}
\usepackage{graphicx}  % standard LaTeX graphics tool
\usepackage{color}     % for color figures
\usepackage{comment}
\usepackage{color}
\usepackage{lscape}
\usepackage{amssymb}
\date{\today}
 \normalsize

\newcommand{\bmat}{\left(\begin{array}}
\newcommand{\emat}{\end{array}\right)}
\newcommand{\be}{\begin{equation}}
\newcommand{\ee}{\end{equation}}
\newcommand{\bea}{\begin{eqnarray}}
\newcommand{\eea}{\end{eqnarray}}

\def\gtwid{\mathrel{\raise.3ex\hbox{$>$\kern-.75em\lower1ex\hbox{$\sim$}}}}
\def\ltwid{\mathrel{\raise.3ex\hbox{$<$\kern-.75em\lower1ex\hbox{$\sim$}}}}
\def\gev{{\rm \, Ge\kern-0.125em V}}
\def\tev{{\rm \, Te\kern-0.125em V}}

\def    \be            {\begin{equation}}
\def    \ee            {\end{equation}}
\def    \bea           {\begin{eqnarray}}
\def    \eea           {\end{eqnarray}}
\def\eps{\epsilon}
\def\a{\alpha}
\def\b{\beta}
\def\g{\gamma}
\def\d{\delta}
\def\n{\nu}

\def\sig{\sigma}

\def\th{\theta}

\def\m{\mu}
\def\nn{\nonumber}
\def\d{\delta}
\def\D{\Delta}
\def\s{\sigma}
\def\r{\rho}
\def\t{\theta}

\def\la{\lambda}
\def\me{\langle m\rangle_e}
\def\mee{\langle m\rangle_{ee}}
\begin{document}
\renewcommand{\thefootnote}{\fnsymbol{footnote}}

\vspace{.3cm}

\title{\Large\bf Neutrino Mass Textures and Partial $\mu$\,--\,$\tau$ Symmetry
}

\author
{ \it \bf  E. I. Lashin$^{1,2,3}$\thanks{slashin@zewailcity.edu.eg, elashin@ictp.it},   N.
Chamoun$^{4,5}$\thanks{nchamoun@th.physik.uni-bonn.de}, C. Hamzaoui$^{6}$\thanks{hamzaoui.cherif@uqam.ca}, and
S. Nasri$^{7,8}$\thanks{snasri@uaeu.ac.ae}
\\
\small$^1$ Ain Shams University, Faculty of Science, Cairo 11566,
Egypt.\\
\small$^2$ Centre for Theoretical Physics, Zewail City of Science and
Technology,\\
\small Sheikh Zayed, 6 October City, 12588, Giza, Egypt.\\
\small$^3$ The Abdus Salam ICTP, P.O. Box 586, 34100 Trieste, Italy. \\
\small$^4$  Physics Department, HIAST, P.O.Box 31983, Damascus,
Syria.\\
 \small$^5$  Physikalisches Institut der Universit$\ddot{a}$t Bonn, Nu$\ss$alle 12, D-53115 Bonn, Germany.\\
\small$^6$ Groupe de Physique Th\'eorique des Particules,\\
\small D\'epartement des Sciences de la Terre et de L'Atmosph\`ere,\\
\small Universit\'e du Qu\'ebec \`a Montr\'eal, Case Postale 8888, Succ. Centre-Ville,\\
\small  Montr\'eal, Qu\'ebec, Canada, H3C 3P8. \\
\small$^7$ Department of Physics, UAE University, P.O.Box 17551, Al-Ain, United Arab Emirates.\\
\small$^8$ Laboratoire de Physique Th\'eorique, ES-SENIA University, DZ-31000 Oran, Algeria.\\
}
\maketitle

\begin{center}
\small{\bf Abstract}\\[3mm]
\end{center}
We discuss the viability of the $\mu$--$\tau$ interchange symmetry imposed on the neutrino mass matrix in the flavor space. Whereas the
 exact symmetry is shown to lead to textures of completely degenerate spectrum which is incompatible with the  neutrino
oscillation data,  introducing small perturbations into the preceding textures, inserted in a minimal way, lead however to four deformed textures representing an approximate
$\mu$--$\tau$  symmetry. We motivate the form of these `minimal' textures, which disentangle the effects of the perturbations, and present some
 concrete realizations assuming exact $\mu$--$\tau$ at the Lagrangian level but at the expense of adding new symmetries and matter fields.
 We find that all these deformed textures are capable to
 accommodate the experimental data, and in all types of neutrino mass hierarchies,  in particular the non-vanishing value for the smallest mixing angle.\\
\vspace{1.1cm}{\bf Keywords}: Neutrino masses,
\vspace{1.1cm}{\bf PACS numbers}: 14.60.Pq; 11.30.Hv; 14.60.St
\begin{minipage}[h]{14.0cm}
\end{minipage}
\vskip 0.3cm \hrule \vskip 0.5cm
%%%%%%%%%%%%%%%%%%%%%%%%%%%%%%%%%%
\section{Introduction}
The elusive neutrino particles proved, so far, to be the only feasible window for the physics beyond the Standard Model (SM) of
particle physics. The observed  solar and atmospheric neutrino oscillations in the Super-Kamiokande \cite{SK} experiment
constitute a compelling evidence for the massive nature of neutrinos which is a clear departure from the SM particle physics.
In the flavor basis where  the charged lepton mass matrix is diagonal, the mixing can be solely attributed
to the effective neutrino  mass matrix $M_\nu$. In such a case the neutrino mass matrix $M_\nu$ can be parameterized by nine free
parameters: three masses ($m_1$, $m_2$ and $m_3$), three mixing angles
($\t_x$, $\t_y$ and $\t_z$) and three phases (two Majorana-type $\r$, $\s$ and one Dirac-type $\d$).
The culmination of experimental data \cite{SNO,KM,K2K,CHOOZ} amounts to constraining the masses and the mixing angles, while for
the phases there is no, so far, a feasible experimental set for their determination. The recent results from the T2K\cite{t2k}, MINOS\cite{mino},
and Double Chooz\cite{CHOOZ2} experiments reveal a nonzero value of $\t_z$. The more recent Daya
Bay \cite{daya} and RENO\cite{reno} experiments confirm a sizable value with relatively high precision.
The discovery of relatively large mixing angle $\t_z$ has a tremondous impact on searching for a sizable
CP-viloation effect  in neutrino oscillations that  enables measuring the Dirac phase $\d$. The impact could
also extend to our understanding of matter-antimatter asymmetry that shaped our universe.

In order to cope with a relatively large mixing angle $\t_z$, one might be compelled to introduce  new ideas in
model building that may enrich our theoretical understanding of the neutrino flavor problem or the flavor problem in general
in case we are fortunate enough. One of the common ideas, often discussed in the literature\cite{proc2012}, is using flavor symmetries, and
 one of the most  attractive ideas in this regard is the $\mu$--$\tau$ symmetry \cite{MUTAU1,MUTAU2}. This symmetry  is enjoyed by many popular mixing patterns
such as tri-bimaximal mixing (TBM) \cite{tbm}, bimaximal mixing (BM) \cite{bm}, hexagonal mixing (HM) \cite{hm} and scenarios of $A_5$ mixing \cite{a5}, and it was largely studied in the
literature \cite{literature}. Actually, many sorts of these symmetries happen to be `accidental' - just a numerical coincidence of parameters without
underlying symmetry, but rather a symmetry resulting from a mutual influence of different and independent factors. The authors of \cite{smirnov13} showed
that the TBM symmetry falls under this category in that large deviations from its predictions are allowed experimentally. Nonetheless, one can adopt
a more `fundamental' approach and construct models incorporating the symmetry in question at the Lagrangian level. In this context, recent,
particularly simple, choices for discrete and continuous flavor symmetry addressing the non-vanishing $\t_z$ question were respectively worked out in
\cite{z2lash} and \cite{u1lash}.

For the $\mu$--$\tau$ symmetry, it is well known that the exact form often requires vanishing $\t_z$ and, thus, the recent results on non-vanishing $\t_z$ force us
to abandon the idea of exact $\mu$--$\tau$ symmetry and to invoke small perturbation violating it. The idea of introducing perturbations
over a $\mu$ -- $\tau$ symmetric mass matrix was recently introduced in \cite{marf,gup,adhikary}, where the authors analyzed the effect of perturbations and the correlation
of their sizes with those  corresponding to the deviation of $\t_z$  and $\t_y-{\pi\over 4}$ from zero. In \cite{marf}, the perturbations are introduced into the
$\mu$-$\tau$ symmetric neutrino mass matrix at all entries, while in \cite{gup} the perturbations are introduced only at the mass matrix entries which are related
through $\mu$--$\tau$ symmetry. The perturbations in \cite{adhikary} were imposed on four and three zero neutrino Yukawa textures. In fact, approximate interchange symmetry
between second and third generation fields goes back to \cite{joshipura} where $\mu$--$\tau$ symmetry was extended to all fermions with a
concrete realization in a two-doublets Higgs model.

In this present work, we follow a similar procedure as in \cite{gup}, and insert the perturbations only at mass matrix entries
related by $\mu$--$\tau$ symmetry. In our approach, however, the deformed relations are thought of as defining textures, and this way of thinking
 provides deep insight about the $\mu$--$\tau$ symmetry itself and its breaking. The two relations defining the approximately $\mu$--$\tau$ symmetric texture
 contain two parameters, generally complex,
controlling the strength of the symmetry breaking. For the sake of simplicity and clarity, we disentangle each parameter  to be kept alone  in the relations
defining the texture. The `minimal' textures obtained in this way (minimal in the sense of containing just one symmetry breaking parameter) may be considered
as a `basis' for all perturbations. Moreover, the numerical study of \cite{gup} with normal hierarchy spectrum required one of the two symmetry breaking parameters to be small
with respect to the other, and this motivated us to consider the extreme case where one of the two symmetry breaking parameters is absent.

As we shall see, the exact $\mu$ ---$\tau$ symmetry can be realized in two different ways as equating to zero two linear combinations of the mass matrix entries.
  Thus, upon deforming these two defining linear combinations, in each of the possible two ways of realizing $\mu$ ---$\tau$ symmetry, by two parameters (each parameter affecting one linear combination) and separating the two parameters effects, we end up with
 four possible textures.  The two equations defining each textures
provide us with four real equations, which are used to reduce the independent parameters of the neutrino mass matrix in this specific texture from nine to five. We choose the five input parameters to be the mixing angles ($\theta_x, \theta_y,  \theta_z$), the Dirac phase $\d$ and the solar mass square difference $\d m^2$, and we vary them within their experimentally acceptable regions. Moreover, we vary also the complex parameter defining the deformation. Therefore, in this way we can reconstruct the neutrino mass matrix out of 7-dimensional parameter
space, and compute the unknown mass spectrum $(m_1, m_2,  m_3)$ and the two Majorana phases
$\r$ and $\s$. We perform consistency check with the other experimental results, and find that all possible four textures could accommodate the data.
However,  no singular models, where one of the masses equals zero, could be viable.

In contrast to the analysis of \cite{gup} which stated that normal type hierarchy is not compatible with small perturbations ($\eps < 20\%$), we found all
the patterns viable in all types of mass hierarchies (normal, inverted and quasi-degenerate) for even smaller perturbations ($\chi = 2\eps  < 20\%$). The
different conclusions are due to two factors. First, in \cite{gup} the phase angles are varied whereas the mixing angles and the other observables are fixed
to their central values, which corresponds to narrow slices in the parameter space we adopted in our work. Second, the definition of normal hierarchy in our work
($m_1/m_3 < m_2/m_3 < 0.7$) is less restricted than the definition adopted in \cite{gup} ($m_1 \ll m_2 \ll m_3$). Thus we believe our analysis is more thorough and our
conclusions are more solid.

As to the origin of the perturbations, there are few strategies to follow. First, one can add terms violating explicitly
 the $\mu$ ---$\tau$ symmetry in the Lagrangian, as was done in \cite{explicit_breaking}. Second, one may assume exact symmetry, leading to $\t_z =0$, at high scale. Then renormalization group (RG)
 running of the neutrino mass matrix elements creates a term which breaks the $\mu$ ---$\tau$ symmetry at the electroweak scale. However, many studies showed that the RG effects are negligible.
  In \cite{RG_doublet}, this process of symmetry breaking via RG running within multiple Higgs doublets model was only valid, for a sizable $\t_z$, in a quasi-degenrate spectrum. In \cite{RG_MSSM}, the
  same conclusion, about the inability of radiative breaking to generate relatively large $\t_z$, was reached in minimal supersymmetric standard model (MSSM) schemes.
  Thus, we shall not consider RG effects, but impose approximate  $\mu$ ---$\tau$ symmetry at high scale (seesaw scale, say) which would remain valid at measurable
  electroweak scale. Third, as was done in \cite{grimus12}, the  $\mu$ --$\tau$ symmetry is replaced by another symmetry including the former as a subgroup. In this spirit and in
  line with
\cite{gup,joshipura}, we address in detail the question of the perturbations root and present some concrete examples at the Lagrangian level for the `minimal' texture form having only one breaking parameter
by means of adding extra Higgs fields and symmetries, in both types I and II of seesaw mechanisms. In type II seesaw, we achieve the desired perturbed form by adding a new $Z_2$-symmetry to
the one characterizing the $\mu$-$\tau$ symmetry (which we denote henceforth by $S$) and three Higgs triplets responsible for giving masses to the left-handed (LH) neutrinos and by substituting
three Higgs doublets for the SM Higgs field for the charged lepton masses. On the other hand, we achieve the desired form in type I seesaw by considering a flavor symmetry of the form $S \times Z_8$ and by having three SM-like Higgs doublets for the charged leptons masses, four other Higgs doublets for the Dirac neutrino mass matrix and additional
two Higgs singlets for the Majorana right-handed (RH) neutrino mass matrix.

The plan of the paper is as follows: in section $2$, we review
the standard notation for the neutrino mass matrix
and its relation to the experimental constraints. In section $3$,
we present the $\mu$--$\tau$ symmetry and its implications. The realization of $\mu$--$\tau$ symmetry as textures and its consequences for non-singluar and singular
cases are respectively worked out in section $4$ and $5$. In section $6$, we present the minimal possible ways for breaking the $\mu$--$\tau$ symmetry leading to
four cases being interpreted as four possible textures, and we classify all the hierarchy patterns regarding the mass spectra.
The detailed relevant formulae and  the results of the phenomenological analysis of each texture are presented in Section $7$ (for nonsingular cases) and
Section $8$ (for singular ones). In section ~$9$, we present a possible Lagrangian for the approximate $\mu$--$\tau$ leading
to the `minimal' textures we adopted.  The last section~$10$ is devoted for discussions and conclusions.

\section{Standard notation}

In the flavor basis, where the charged lepton mass matrix is diagonal, we
diagonalize the symmetric
neutrino mass matrix $M_\nu$ by a unitary transformation,
\be
V^{\dagger} M_\nu\; V^{*} \; = \; \left (\matrix{ m_1 & 0 & 0 \cr 0 & m_2 & 0
\cr 0 & 0 & m_3 \cr} \right ), \;
\label{diagM}
\ee
with $m_i$ (for $i=1,2,3$) real and positive. We introduce the mixing
angles $(\theta_x, \theta_y, \theta_z)$ and the phases ($\delta,\rho,\sigma$)
such that \cite{Xing}:
\bea
V &=& UP \nn\\
P &=& \mbox{diag}(e^{i\rho},e^{i\sigma},1) \nn\\
U \; &=& \;
\left ( \matrix{ c_x c_z & s_x c_z & s_z \cr - c_x s_y
s_z - s_x c_y e^{-i\delta} & - s_x s_y s_z + c_x c_y e^{-i\delta}
& s_y c_z \cr - c_x c_y s_z + s_x s_y e^{-i\delta} & - s_x c_y s_z
- c_x s_y e^{-i\delta} & c_y c_z \cr } \right ) \; ,
\label{defv}
\eea
(with $s_x \equiv \sin\theta_x \ldots$)
to have
\begin{equation}
M_\nu \; =\; U \left ( \matrix{ \lambda_1 & 0 & 0 \cr 0 &
\lambda_2 & 0 \cr 0 & 0 & \lambda_3 \cr} \right ) U^T. \;
\label{massdef}
\end{equation}
with
\begin{equation}
\lambda_1 \; =\; m_1 e^{2i\rho} \; , ~~~ \lambda_2 \; =\; m_2
e^{2i\sigma} \; , ~~~ \lambda_3 = m_3. \;
\label{deflam}
\end{equation}

In this parametrization, the mass matrix elements
are given by:
\bea
M_{\n\,11}&=& m_1 c_x^2 c_z^2 e^{2\,i\,\r} + m_2 s_x^2 c_z^2 e^{2\,i\,\s}
+ m_3\,s_z^2,\nn\\
%%%%%%%%%%%%%%%%%%%%%%%%%%%%%%%%%%%%%
M_{\n\,12}&=& m_1\left( - c_z s_z c_x^2 s_y e^{2\,i\,\r}
- c_z c_x s_x c_y e^{i\,(2\,\r-\d)}\right)
+ m_2\left( - c_z s_z s_x^2 s_y e^{2\,i\,\s}
+ c_z c_x s_x c_y e^{i\,(2\,\s-\d)}\right) + m_3 c_z s_z s_y,\nn\\
%%%%%%%%%%%%%%%%%%%%%%%%%%%%%%%%%%
M_{\n\,13}&=& m_1\left( - c_z s_z c_x^2 c_y e^{2\,i\,\r}
+ c_z c_x s_x s_y e^{i\,(2\,\r-\d)}\right)
+ m_2\left( - c_z s_z s_x^2 c_y e^{2\,i\,\s}
- c_z c_x s_x s_y e^{i\,(2\,\s-\d)}\right) + m_3 c_z s_z c_y,\nn\\
%%%%%%%%%%%%%%%%%%%%%%%%%%%%%%%%%
M_{\n\,22}&=& m_1 \left( c_x s_z s_y  e^{i\,\r}
+ c_y s_x e^{i\,(\r-\d)}\right)^2 + m_2 \left( s_x s_z s_y  e^{i\,\s}
- c_y c_x e^{i\,(\s-\d)}\right)^2 + m_3 c_z^2 s_y^2, \nn\\
%%%%%%%%%%%%%%%%%%%%%%%%%%%%%%%%%%%%%%%%%%%%
M_{\n\,33}&=& m_1 \left( c_x s_z c_y  e^{i\,\r}
- c_y s_x e^{i\,(\r-\d)}\right)^2 + m_2 \left( s_x s_z c_y  e^{i\,\s}
+ s_y c_x e^{i\,(\s-\d)}\right)^2 + m_3 c_z^2 c_y^2, \nn\\
%%%%%%%%%%%%%%%%%%%%%%%%%%
M_{\n\, 23} &=& m_1\left( c_x^2 c_y s_y s_z^2  e^{2\,i\,\r}
 + s_z c_x s_x (c_y^2-s_y^2) e^{i\,(2\,\r-\d)} - c_y s_y s_x^2
e^{2\,i\,(\r-\d)}\right)
\nn\\
&& +  m_2\left( s_x^2 c_y s_y s_z^2  e^{2\,i\,\s}
 + s_z c_x s_x (s_y^2-c_y^2) e^{i\,(2\,\s-\d)} - c_y s_y c_x^2
e^{2\,i\,(\s-\d)}\right)
+ m_3 s_y c_y c_z^2.
\label{melements}
\eea
Note  that under the
transformation given by
\bea
T_1: &&\t_y\rightarrow {\pi \over 2} - \t_y\;\; \mbox{and}\;\; \d\rightarrow \d
\pm \pi,
\label{sy1}
\eea
the mass matrix elements are transformed amongst themselves by swapping the indices
$2$ and $3$ and keeping the index $1$ intact:
\bea
M_{\n 11} \leftrightarrow M_{\n 11}, && M_{\n 12} \leftrightarrow M_{\n 13}\nn\\
M_{\n 22} \leftrightarrow M_{\n 33}, && M_{\n 23} \leftrightarrow M_{\n23}.
\label{tr1}
\eea
On the other hand, the mass matrix is transformed into its complex conjugate i.e
\bea
M{_\n}_{ij}\left(T_2(\d,\r,\s)\right) = M^*_{\n ij} \left((\d,\r,\s)\right)
\eea
under the mapping given by:
\bea
T_2: \r \rightarrow \pi - \r, & \s \rightarrow \pi - \s, & \d \rightarrow
2\,\pi-\d,
\label{sy2}
\eea

The above two symmetries $T_{1,2}$ are
quite useful in classifying the models and in connecting the phenomenological
analysis of patterns related
by them.

 It is straightforward  to relate our parametrization convention Eq.~(\ref{defv}) to the more familiar  one
 used in the recent data analysis of \cite{fog1}. In fact, the mixing angles in the two parameterizations are equal
 \begin{equation}
\theta_x \; \equiv \; \theta_{12} \; , ~~~~~ \theta_y \;
\equiv \; \theta_{23} \; , ~~~~~ \theta_z \; \equiv \;
\theta_{13}.
\end{equation}
whereas there is a simple linear relation, discussed in \cite{z2lash,tex0lash}, between the phases defined in our parametrization and those corresponding to the
standard one.

The solar and
atmospheric neutrino mass-squared differences are characterized by two independent neutrino mass-squared differences\cite{fog1}:
\begin{equation}
\delta m^2 \; \equiv \; m_2^2-m_1^2 \; , \; \left|\Delta m^2\right|  \;\equiv \;
\left|m_3^2-{1\over 2}\left(m_1^2+m_2^2\right)\right|\;\; ,
\label{sqm}
\end{equation}
whereas the parameter
\begin{equation}
R_\nu \; \equiv \;  \frac{\delta m^2} {\left|\Delta
m^2\right|}.
\label{rnu}
\end{equation}
characterizes the hierarchy of these two quantities.

The neutrino mass scales are constrained in the reactor
nuclear experiments on beta-decay kinematics and
neutrinoless double-beta decay by two parameters which are the
effective electron-neutrino mass:
\begin{equation}
\langle
m\rangle_e \; = \; \sqrt{\sum_{i=1}^{3} \displaystyle \left (
|V_{e i}|^2 m^2_i \right )} \;\; ,
%       (2.23)
\end{equation}
and the effective Majorana mass term
$\langle m \rangle_{ee} $:
\begin{equation} \label{mee}
\langle m \rangle_{ee} \; = \; \left | m_1
V^2_{e1} + m_2 V^2_{e2} + m_3 V^2_{e3} \right | \; = \; \left | M_{\n 11} \right
|.
\end{equation}

Another parameter with an upper bound coming from cosmological observations is
the `sum'
parameter $\Sigma$:
\be
\Sigma = \sum_{i=1}^{3} m_i.
\ee

Moreover, the  Jarlskog rephasing invariant quantity
is given by\cite{jarlskog}:
\begin{equation}\label{jg}
J = s_x\,c_x\,s_y\, c_y\, s_z\,c_z^2 \sin{\delta}
\end{equation}

There are no experimental bounds on the phase angles, and we take the principal
value range for $\d, 2\r$ and $2 \s$ to be $[0,2\pi]$.
As to the other oscillation parameters, the experimental constraints give the
 values stated in Table~(\ref{tab1}) with 1, 2, and 3-$\sigma$ errors \cite{fog1,fog2}. Actually, the fits of oscillation data found in \cite{fog1} and \cite{fog2} are consistent
 with each other except that the latter fits are stricter for $\t_z$. In our numerical analysis, we prefer to use the former fit having a  wider range for $\t_z$ in order
 to easily catch the pattern of variation depending on $\t_z$. Other groups \cite{schwetz,garcia} have also carried out global fits for the oscillation data and their findings are in line with those of the group of \cite{fog1}.
\begin{table}[h]
 \begin{center}
{\small
 \begin{tabular}{c c c c c}
\hline
\hline
\mbox{Parameter} &$\mbox{Best fit}$ & $1\s$ \mbox{range} & $2\s$ \mbox{range} &
$3\s$ \mbox{range}\\
\hline
 $\d m^2 (10^{-5}\mbox{eV}^2)$ &$7.58$ & $\left[7.32, 7.80\right]$ &$\left[7.16,
7.99\right]$ &
 $\left[6.99, 8.18\right]$ \\
 \hline
 $\left|\D m^2\right|(10^{-3}\mbox{eV}^2)$& $2.35$ &$\left[2.26, 2.47\right]$
&$\left[2.17, 2.57\right]$ &
 $\left[2.06, 2.67\right]$ \\
 \hline
$\th_x$ & $33.58^o$ &$\left[32.96^o, 35.00^o\right]$ &$\left[31.95^o,
36.09^o\right]$ & $\left[30.98^o, 37.11^o\right]$ \\
\hline
$\th_y$ &$40.40^o$ &$\left[38.65^o, 45.00^o\right]$ &$\left[36.87^o,
50.77^o\right]$ & $\left[35.67^o,53.13^o\right]$\\
\hline
$\th_z$ & $8.33^o$ &$\left[7.71^o, 10.30^o\right]$ &$\left[6.29^o,
11.68^o\right]$ &$\left[4.05^o, 12.92^o\right]$ \\
& $8.99^o$ & $(8.45^o,9.39^o)$ & $(7.99^o,9.82^o)$ & $(7.47^o,10.80^o)$ \\
\hline
$R_{\nu}$ & $0.0323$ & $\left[0.0296 , 0.0345\right]$&$\left[0.0279 ,
0.0368\right]$ & $\left[0.0262 , 0.0397\right]$\\
\hline
 \end{tabular}
 }
 \end{center}
\caption{\small The global-fit results of three neutrino
mixing angles $(\th_x, \th_y, \th_z)$ and two neutrino mass-squared differences
$\d m^2$ and $\D m^2$ as defined in Eq.~(\ref{sqm}). The results $[\cdots]$ and $(\cdots)$ as respectively extracted
from \cite{fog1} and \cite{fog2}. In \cite{fog1}, it is assumed that $\cos{\d}= \pm 1$ and that new reactor fluxes have been
used, while in \cite{fog2} $\d$ is not restricted and the old reactor flux is used.}
 \label{tab1}
\end{table}

 We adopt the
less conservative 2-$\sigma$ range as reported in \cite{fog3}
for the
non oscillation parameters $\me$, $\Sigma$, whereas for the other non-oscillation parameter $\mee$ we use values found in \cite{Cuoricino}:
\bea
\langle m\rangle_e &<& 1.8\; \mbox{eV}, \nonumber \\
\Sigma &<& 1.19 \;\mbox{eV}, \nonumber \\
\langle m\rangle_{ee} & < & 0.34-0.78\; \mbox{eV}.
\label{nosdata}
\eea

\section{The $\mu$--$\tau$ symmetry and neutrino mass matrix}
The $\mu$--$\tau$ symmetry can be described by the following general set of
conditions\cite{marf},
\be
\left|V_{\mu\,i}\right| =  \left|V_{\tau\,i}\right|,\;\;\; \mbox{for}\, i
=1,2,3.
\label{parcond1}
\ee
According to our adopted parameterizations for $V$ in Eq.(\ref{defv}) these
conditions imply two classes of solutions. The first class, hereafter labeled by
class I,  is characterized by,
\be
\t_y = {\pi\over 4}, \;\;\;\; 2\, s_x\,c_x\,s_z\,c_\d=0,
\label{class1c}
\ee
while the second class, hereafter labeled by class II, is determined by,
\be
\t_z = {\pi\over 2}, \;\;\;\; s_{2\,x}\,s_{2\,y}\,c_\d = c_{2\,y}\,c_{2\,x},
\label{class2c}
\ee
The two classes, I and II, are distinguished by the possible allowed values for
mixing angles $\t_y$ and $\t_z$.  In class I, the mixing angle $\t_y$ is fixed
to be ${\pi\over 4}$, while for class II the mixing angle $\t_z$ is fixed to be
${\pi\over 2}$. These restrictions are the only nontrivial consequence of the
$\mu$--$\tau$ symmetry. Regarding the other mixing angles and phases for each
class, the restriction imposed through the symmetry is rather loose.
However, according to the allowed values for mixing angles and phases, the class
II cannot be divided into a finite number of sub-classes in contrast to the class
I which can be divided into four sub-classes as follows,
\begin{description}
\item{\textbf{(a)}} $\t_y= {\pi\over 4}$ and $\t_x=0$ while $\t_z, \d, \r$ and
$\s$ are free,
\item{\textbf{(b)}} $\t_y= {\pi\over 4}$ and $\t_x= {\pi\over 2}$ while $\t_z,
\d, \r$ and $\s$ are free,
\item{\textbf{(c)}} $\t_y= {\pi\over 4}$ and $\t_z=0$ while $\t_x, \d, \r$ and
$\s$ are free,
\item{\textbf{(d)}} $\t_y= {\pi\over 4}$ and $\d=\pm {\pi\over 2}$ while $\t_x,
\t_z, \r$ and $\s$ are free.
\end{description}
The sub-classes (a) and (b) seem unsatisfactory because the predicted  $\t_x$ is
far from the experimentally preferred value. The remedy for this defect is to
introduce a small perturbation having a large effect on $\t_x$ as was done in \cite{marf}.
As to the sub-class (c),
it seems to be the most interesting class, from a phenomenological point of view, when joint by
fixing $\t_x$ near the experimentally preferred value. In a sense, it can contain
models with tri-bimaximal, bimaximal, hexagonal, and $A_5$ symmetries.
The last remaining sub-class (d), predicting maximal $CP$ violation, can include
the tetramaximal symmetry \cite{Xingtetra}. The class II is phenomenologically
disfavored since $\t_z={\pi\over 2}$ is far from the experimentally preferred
value, which might justify dropping this whole class in the analysis
carried out in\cite{marf}.

We can get more insight into the $\mu\,$--$\,\tau$ symmetry by writting its
implications on the neutrino mass matrix entries.
The class I and its sub-classes are found to imply
\begin{description}
\item{\textbf{(a)}} $M_{\n\,12}=M_{\n\,13}$ and $M_{\n\,22}=M_{\n\,33}$,
\item{\textbf{(b)}} $M_{\n\,12}=M_{\n\,13}$ and $M_{\n\,22}=M_{\n\,33}$,
\item{\textbf{(c)}} $M_{\n\,12}=-M_{\n\,13}$ and $M_{\n\,22}=M_{\n\,33}$,
\item{\textbf{(d)}} $M_{\n\,12}=M^*_{\n\,13}$ and $M_{\n\,22}=M^*_{\n\,33}$ for
vanishing Majorana phases, otherwise no simple algebraic relation between the mass
entries is found.
\end{description}
In the second class II, the implied mass relations are,
\be
M_{\n\,12}=M_{\n\,13}=0,\; \mbox{and}\;\;
\left|M_{\n\,22}\right|=\left|M_{\n\,33}\right|.
\ee
The above mentioned considerations motivate us to take as a starting point  one of the
following mass relations as defining the $\mu$\,--\,$\tau$ symmetry.
The first relation is taken to be
\be
M_{\n\,12}=M_{\n\,13},\; \mbox{and}\;\; M_{\n\,22} = M_{\n\,33}.
\label{s+}
\ee
while the second one is
 \be
M_{\n\,12}= - M_{\n\,13},\; \mbox{and}\;\; M_{\n\,22} = M_{\n\,33}.
\label{s-}
\ee
These two alternative ways for imposing $\mu$\,--\,$\tau$ symmetry in
Eq.(\ref{s+}) and Eq.(\ref{s-}) are respectively designated by $S_+$ and $S_-$
in order to ease the corresponding referral.
The other possible relations like ($M_{\n\,12}=M^*_{\n\,13}$ and
$M_{\n\,22}=M^*_{\n\,33}$) or ($M_{\n\,12}=M_{\n\,13}=0$ and
$\left|M_{\n\,22}\right|=\left|M_{\n\,33}\right|$)
are disfavored because they involve non analytical algebraic relation between
mass entries that cannot be generated by usual discrete flavor symmetries.  There
is still a further motivation
for imposing $\mu$\,--\,$\tau$ symmetry via $S_+$ or $S_-$ which can be easily
inferred from the symmetry properties enjoyed by the neutrino mass matrix as
explained in
section~2. In fact, the transformation rule in Eq.(\ref{sy1}) singles out $\t={\pi\over
4}$ as a fixed point for the transformation and the mass relations in
Eq.(\ref{tr1}) already links the
mass matrix entries relevant for the $\mu$\,--\,$\tau$ symmetry. The difference in
sign between the two alternative realizations, $ M_{\n\,12}= \pm M_{\n\,13}$,
can be attributed to the different phases assigned to the third neutrino filed
$\n_\tau$.

\section{The exact $\mu$\,--\,$\tau$ symmetry as a texture for
non singular neutrino mass matrix}
The exact exact $\mu$\,--\,$\tau$ symmetry can be treated as a texture defined
by,
\bea
M_{\n\,12} \mp M_{\n\,13}&=&0,\\\nn
M_{\n\,22}-M_{\n\,33}&=&0,
\label{smpt1}
\eea
where the minus and plus sign correspond respectively to the cases of Eq.(\ref{s+}) and Eq.(\ref{s-}).

Using Eqs.~(\ref{defv}-\ref{deflam}), the relation defining the texture can be
expressed as
\bea
M_{\n\,12} \mp M_{\n\,13}=0,&\Rightarrow& \sum_{j=1}^{3}
\left(U_{1j}\,U_{2j} \mp U_{1j}\,U_{3j}\right)\; \lambda_j=0 \nn\\
&\Rightarrow& A_1^{\mp}\,\la_1 +  A_2^{\mp}\,\la_2 +  A_3^{\mp}\,\la_3 =0 \nn\\
M_{\n\,22}-M_{\n\,33} = 0,&\Rightarrow& \sum_{j=1}^{3}
\left(U_{2j}\,U_{2j} - U_{3j}\,U_{3j}\right)\; \lambda_j=0,\nn\\
&\Rightarrow& B_1\,\la_1 +  B_2\,\la_2 +  B_3\,\la_3 =0
\label{smpt2}
\eea
where
\bea
\label{ABc}
A_j^{\mp} = U_{1j}\;\left(U_{2j} \mp U_{3j}\right), &\mbox{and}&
B_j =  U_{2j}^2 -  U_{3j}^2, \;\;\;\;\;\;\;\; (\mbox{no sum over } j).
\eea
The coefficients $A^{\mp}$ and $B$ can be written explicitly in terms of mixing
angles and Dirac phase as,
\bea
A_1^{\mp} &=& -c_x\,c_z\left[ c_x\, s_z \left(s_y \mp c_y\right) +
s_x\,\left(c_y \pm s_y\right)\,e^{-i\,\delta}\right],\nn\\
B_1 &=& \left(c_x\,s_y\,s_z + s_x\, c_y\,e^{-i\,\delta}\right)^2 -
\left(-c_x\,c_y\,s_z + s_x\, s_y\,e^{-i\,\delta}\right)^2,  \nn\\
A_2^{\mp} &=& -s_x\,c_z\left[ s_x\, s_z \left(s_y \mp c_y\right) \mp
c_x\,\left(s_y \pm c_y\right)\,e^{-i\,\delta}\right], \nn\\
B_2 &=& \left(-s_x\,s_y\,s_z + c_x\, c_y\,e^{-i\,\delta}\right)^2 -
\left(s_x\,c_y\,s_z + c_x\, s_y\,e^{-i\,\delta}\right)^2, \nn\\
A_3^{\mp} &=& s_z\,c_z\,\left(s_y \mp c_y\right),\nn \\
B_3 &=& c_z^2\,\left(s_y^2 - c_y^2\right).
\label{ABe}
\eea
Provided $\la_3$ is non-vanishing, the equations (\ref{smpt2}) can be treated as
two inhomogeneous linear equations of the ratios $\frac{\la_1}{\la_3}$ and
$\frac{\la_2}{\la_3}$  which can be solved to get,
\bea
\frac{\la_1}{\la_3} &=&
\frac{A_{3}^{\mp}\; B_2-A_{2}^{\mp}\; B_3}
{A_{2}^{\mp}\; B_1-A_{1}^{\mp}\; B_2}, \nn \\
\frac{\la_2}{\la_3} &=& \frac{A_{1}^{\mp}\; B_3-A_{3}^{\mp}\; B_1}
{A_{2}^{\mp}\; B_1-A_{1}^{\mp}\; B_2}.
\label{lams}
\eea
Computing the  mass spectrum, we find that it is always a degenerate one $(m_1=m_2=m_3)$ leading to
vanishing mass-squared differences, which is unacceptable phenomenologically. Explicitly, for the cases (a) to (c)  mentioned in the previous section and
respecting exact $\mu$\,--\,$\tau$ symmetry,  we have all the coefficients  $A^{\mp}$'s and $B$'s vanishing except: $A_3^+ = - A_1^+ = s_z c_z$ (case a),
$A_3^+ = - A_2^+ = s_z c_z$ (case b) and $A_1^- = - A_2^- = -\sqrt{2} s_x c_x e^{-i\d}$ (case c).

\section{The exact $\mu$\,--\,$\tau$ symmetry as a texture for
singular neutrino mass matrix}
One may wonder that our analysis might lead to non trivial results for singular
neutrino mass matrix. Thus, it is crucial to carry the same study for singular
case, and keep in mind that the  viable singular neutrino mass matrices have
to be characterized by vanishing $m_1$ or $m_3$. The vanishing of $m_2$ leading to the
simultaneous vanishing of $m_1$ and $m_2$ is  not at all  phenomenologically
consistent.
%\begin{landscape}
\begin{table}[hbtp]
\begin{center}
\begin{tabular}{|c|c|c|}
\hline
&\multicolumn{2}{c|}{$m_1=0$} \\
\hline
Realization &\multicolumn{2}{c|}{$m_2\over m_3$}    \\
\hline
&&\\
$S_-$ & $\left|{A_3^-\over A_2^-}\right|\approx \sqrt{{1-s_{2y}\over 1+ s_{2y}}}\,{s_z\over s_x\,c_x} + O(s_z^2)$ &
$\left|{B_3\over B_2}\right| \approx {1\over c_x^2} \left( 1 + 2\,t_x\,t_{2y}\,c_\d\, s_z\right) + O(s_z^2)$\\
&&\\
\hline
&&\\
$S_+$ & $\left|{A_3^+\over A_2^+}\right|\approx \sqrt{{1+s_{2y}\over 1- s_{2y}}}\,{s_z\over s_x\,c_x} + O(s_z^2)$ &
$\left|{B_3\over B_2}\right| \approx {1\over c_x^2} \left( 1 + 2\,t_x\,t_{2y}\,c_\d\, s_z\right) + O(s_z^2)$
\\
&& \\
\hline
 & \multicolumn{2}{c|}{$m_3=0$}\\
 \hline
 Realization &  \multicolumn{2}{c|}{$m_2\over m_1$}    \\
 \hline
&&\\
$S_-$ & $\left|{A_1^-\over A_2^-}\right|\approx 1-{\left(1-s_{2y}\right)\,c_\d\,s_z\over c_{2y}\,s_x\,c_x} + O(s_z^2)$ &
$ \left|{B_1\over B_2}\right| \approx t_x^2\,\left(1 + {2\,t_{2y}\,c_\d\,s_z \over s_x c_x}\right) + O(s_z^2)$
\\
&&\\
\hline
&&\\
$S_+$ & $\left|{A_1^+\over A_2^+}\right|\approx 1+{\left(1+s_{2y}\right)\,c_\d\,s_z\over c_{2y}\,s_x\,c_x} + O(s_z^2)$ &
$ \left|{B_1\over B_2}\right| \approx t_x^2\,\left(1 + {2\,t_{2y}\,c_\d\,s_z \over s_x c_x}\right) + O(s_z^2)$\\
&&\\
\hline
\end{tabular}
\end{center}
\caption{\small  The approximate mass ratio formulae for the singular light neutrino mass realizing exact $\mu$ -- $\tau$ symmetry. The
forumlae are calculated in terms of A's or B's coefficients }
\label{singmass}
 \end{table}
 %\end{landscape}

 \subsection{Vanishing $m_1$ singular neutrino mass matrix having exact
$\mu$\,--\,$\tau$ symmetry }
The mass spectrum in this case turns out to be,
 \bea
 m_1 =0, & m_2= \sqrt{\d m^2}, & m_3 = \sqrt{\D m^2 + {\d m^2\over 2}}
 \approx \sqrt{ \D m^2},
 \label{m1spec}
 \eea
 which puts the mass ratio ${m_2\over m_3}$ in the form
 \be
 m_{23}\equiv {m_2\over m_3} = \sqrt{\frac{R_\n}{1+{R_\n\over 2}}}\approx \sqrt{R_\n},
 \label{m23con}
 \ee
where the phenomenologically acceptable value for  $R_\n$ is given in
Table~(\ref{tab1}).
The vanishing of $m_1$ together with imposing the exact $\mu$\,--\,$\tau$ symmetry
as stated in Eqs.(\ref{smpt2}) leads to,
\bea
A_2^{\mp}\, \la_2 \, + A_3^{\mp}\, \la_3 & =& 0,\nn\\
B_2\, \la_2 \, + B_3\, \la_3 & =& 0,
\label{m1emt}
\eea
which gives non trivial solutions,  provided $A_2^{\mp} B_3 - A_3^{\mp} B_2 = 0$, i.e.
 \bea
 m_{23} = \left|{A_3^{\mp}\over A_2^{\mp}}\right|= \left|{B_3\over B_2}\right|, &&
 \s = {1\over 2}\;\mbox{Arg} \left(-{A_3^{\mp}\, m_3\over A_2^{\mp}\, m_2}\right) =
 {1\over 2}\;\mbox{Arg} \left(-{B_3\, m_3\over B_2\, m_2}\right).
 \label{m1zeromt}
 \eea
The Majorana phase $\r$ becomes unphysical,  since $m_1$ vanishes, in this case,
and can be dropped out.

These patterns can be easily shown to be unviable just by comparing the two approximate  expressions obtained for
${m_2\over m_3}$. As an example we consider the case $S_-$ where we have,  as reported in Table~(\ref{singmass}),
\be
\begin{array}{lll}
{\large m_{23} }  & {\large \approx} &
\left\{
\begin{array}{lll}
 & {\large \sqrt{{1-s_{2y}\over 1+ s_{2y}}}\,{s_z\over s_x\,c_x} + O(s_z^2),} & \mbox{from A}{}^{-}\mbox{'s},\\\\
 & {\large {1\over c_x^2} \left( 1 + 2\,t_x\,t_{2y}\,c_\d\, s_z\right) + O(s_z^2),} & \mbox{from B's},
 \end{array}
\right.
\end{array}
\ee
This mass ratio, ${m_2\over m_3}$ should be consistent with the constraint of Eq.~(\ref{m23con}), which means that it should be much less than one. It is hard to satisfy this constraint
because the first expression, obtained from $A^-$'s, starts from $O(s_z)$ and can be  tuned to a small value, while the second one, obtained from $B$'s,
has a leading contribution  (${1\over c_x^2}$) which is greater than one for the admissible range of $\t_x$. To properly tune the second expression, one needs
large negative higher order corrections which can be achieved by choosing negative $c_\d$ and letting $\t_y$ approach ${\pi\over 4}$, but this tends in its turn to diminish
the first expression of the mass ratio more than required. Thus, the two expression cannot be made compatible. A similar reasoning can be applied to the case
$S_+$ to show the incompatibility of the two derived expressions for the mass ratio. Our numerical study confirms this conclusion where
all the phenomenologically acceptable ranges for mixing angles and Dirac phase are scanned, but no solutions could  be found satisfying the mass constraint  expressed in
Eq.~(\ref{m23con})

\subsection{Vanishing $m_3$ singular neutrino mass matrix having exact
$\mu$\,--\,$\tau$  symmetry}
 Along the same lines of the previous subsection, we can treat the case of vanishing $m_3$.
 This time, the mass spectrum is found to be,
 \bea
 m_1 =\sqrt{\D m^2 - {\d m^2\over 2}}, & m_2= \sqrt{\D m^2 + {\d m^2\over 2}}, &
m_3 = 0,
 \label{m3spec}
 \eea
 forcing the mass ratio ${m_2\over m_1}$ to be
 \be
 m_{21}
 \equiv {m_2\over m_1} = \sqrt{{1+{R_\n\over 2}}\over{1-{R_\n\over 2}}} \approx 1 +
{R_\n\over 2} \gtrsim 1.
 \label{m21con}
 \ee
The vanishing of $m_3$ together with imposing exact $\mu$\,--\,$\tau$ symmetry
as stated in Eqs.(\ref{smpt2}) result in the following
equations,
\bea
A_1^{\mp}\, \la_1 \, + A_2^{\mp}\, \la_2 & =& 0,\nn\\
B_1\, \la_1 \, + B_2\, \la_2 & =& 0,
\label{m3emt}
\eea
 which have non trivial solutions as,
 \bea
 m_{21} = \left|{A_1^{\mp}\over A_2^{\mp}}\right| = \left|{B_1\over B_2}\right|, &&
 \r-\s = {1\over 2}\;\mbox{Arg} \left(-{A_2^{\mp}\, m_2\over A_1^{\mp}\, m_1}\right)=
 {1\over 2}\;\mbox{Arg} \left(-{B_2\, m_2\over B_1\, m_1}\right),
 \label{m3zeromt}
 \eea
provided $A_1^{\mp} B_2 - A_2^{\mp} B_1 = 0$. It is clear that the only relevant
physical combination of Majorana phases in such a case is the difference
$\r-\s$. One can use the same reasoning explained in the case of vanishing $m_1$, based on approximate formulae for mass ratios, as
reported in Table~(\ref{singmass}),  to show that the  constraint of Eq.~\ref{m21con} cannot be satisfied, which makes the patterns unviable.
Again, our numerical study based on scanning all phenomenologically acceptable ranges for mixing angles and Dirac phase reveals no solutions found
satisfying the constraint of Eq.~(\ref{m21con}).

Our investigations, which are so far model independent, point out that imposing
exact $\mu$\,--\,$\tau$  symmetry always produces phenomenologically unsatisfactory
 results.
Thus one might find the solace by demanding violation of the exact $\mu$\,--\,$\tau$
 symmetry. In breaking the symmetry, we are going to try the simplest and
minimal ways of breaking.

\section{Deviation from exact $\mu$\,--\,$\tau$  symmetry}
We consider the simplest minimal possible deviation from the exact
$\mu$\,--\,$\tau$  symmetry that can be parameterized by only one parameter.
The relations characterizing these deviations can assume the following two
forms,
\be
M_{\n\,12}\;\left(1+ \chi\right) = \pm M_{\n\,13},\; \mbox{and}\;\; M_{\n\,22} =
M_{\n\,33},
\label{d1chi}
\ee
and
 \be
M_{\n\,12}= \pm M_{\n\,13},\; \mbox{and}\;\; M_{\n\,22}\;\left(1+ \chi\right) =
M_{\n\,33},
\label{d2chi}
\ee
where
$\chi = \left|\chi\right|\,e^{i\,\th}$
is a complex parameter measuring the deviation from exact $\mu$\,--\,$\tau$
symmetry. The absolute value $\left|\chi\right|$
is restricted to fall in the range $\left[0,0.2\right]$, while the phase $\t$
is totally free. The chosen range for $\chi$ is made to ensure a small deviation
that can be treated as a perturbation.

The deviation from exact $\mu$\,--\,$\tau$  symmetry can be treated in an
illuminating way by considering the relations in
Eqs.(\ref{d1chi},\ref{d2chi}) as defining
the following textures
\be
M_{\n\,12}\;\left(1+ \chi\right)  \mp M_{\n\,13}=0,\; \mbox{and}\;\; M_{\n\,22}
- M_{\n\,33}=0,
\label{tex1}
\ee
and
 \be
M_{\n\,12} \mp M_{\n\,13}=0,\; \mbox{and}\;\; M_{\n\,22}\;\left(1+ \chi\right) -
M_{\n\,33}=0.
\label{tex2}
\ee
Following the same procedure as described in section~4, we find that the coefficients
$A$'s and $B$'s corresponding to the textures defined in Eq.(\ref{tex1}) and
Eq.(\ref{tex2}) are respectively,
\bea
\label{AB1}
A^{\mp}_j = U_{1j}\;\left[U_{2j}\,\left(1 + \chi\right) \mp U_{3j}\right],
&\mbox{and}&
B_j =  U_{2j}^2 -  U_{3j}^2, \;\;\;\;\;\;\;\; (\mbox{no sum over } j).
\eea
and
\bea
\label{AB2}
A^{\mp}_j  = U_{1j}\;\left(U_{2j}\, \mp U_{3j}\right), &\mbox{and}&
B_j =  U_{2j}^2\,\left(1 + \chi\right) -  U_{3j}^2, \;\;\;\;\;\;\;\; (\mbox{no
sum over } j).
\eea
Assuming $\la_3 \neq 0$, the resulting $\la$'s ratio are found to be,
\bea
\frac{\la_1}{\la_3} &=&
\frac{A_{3}\; B_2-A_{2}\; B_3}
{A_{2}\; B_1-A_{1}\; B_2}, \nn \\
\frac{\la_2}{\la_3} &=& \frac{A_{1}\; B_3-A_{3}\; B_1}
{A_{2}\; B_1-A_{1}\; B_2},
\label{lam12}
\eea
From these $\la$-ratios, the mass ratios $\left({m _1\over m_3},  {m_2\over
m_3}\right)$ and Majorana phases $\left(\r, \s\right)$ can be determined
in terms of the mixing angles $\left((\t_x, \t_y, \t_z\right)$, the Dirac phase $\d$ and
the complex parameter $\chi$. Thus, we can vary ($\theta_x, \theta_y, \theta_z, \d m^2$)
over their experimentally allowed regions and ($\d, |\chi|, \t$)
in their full range to determine the unknown mass spectra and  Majorana
phases.
We can then confront the whole predictions  with the experimental constraints
given in Table~(\ref{tab1}) and Eq.~(\ref{nosdata}) to find out the admissible
7-dim parameter space
region. For a proper survey of the allowed parameter space, one can  illustrate
graphically all the possible
correlations, at the three levels of $\s$-error, between any two physical
neutrino parameters. We chose to
plot for each pattern and for each type of hierarchy thirty four  correlations at the
3-$\s$ error level involving
the parameters $\left(m_1,m_2, m_3,\t_x,\t_y,\t_z,\r,\s,\d,J,m_{ee},
|\chi|, \t\right)$ and the lowest neutrino mass ({\bf LNM}).
Moreover, for each parameter, one can determine the extremum values it can take
according to the considered
precision level, and we listed in tables these predictions for all the patterns
and for the three $\s$-error
levels.

The resulting mass patterns are found to be classifiable into
three categories:
\begin{itemize}
\item Normal hierarchy: characterized by $m_1 < m_2 < m_3$ and
is denoted by ${\bf N}$ satisfying numerically the bound:
\be
\frac{m_1}{m_3} < \frac{m_2}{m_3} < 0.7
\label{nor}
\ee
\item Inverted hierarchy: characterized
by $m_3 < m_1 < m_2$ and is denoted by ${\bf I}$ satisfying the bound:
\be
\frac{m_2}{m_3} > \frac{m_1}{m_3} > 1.3
\label{inv}
\ee
\item Degenerate hierarchy (meaning quasi- degeneracy): characterized
by $m_1\approx  m_2 \approx m_3$ and is denoted by ${\bf D}$. The corresponding
numeric bound is
taken to be:
\be
0.7 < \frac{m_1}{m_3} < \frac{m_2}{m_3} < 1.3
\label{deg}
\ee
\end{itemize}
Moreover, we studied for each pattern the possibility of having a singular
(non-invertible) mass matrix
characterized by one of the masses ($m_1,
\mbox{and} \; m_3$) being equal to zero (the data prohibits
the
simultaneous vanishing of two masses and thus $m_2$ can not vanish).

\section{Numerical results of various patterns violating exact $\mu$\,--\,$\tau$
 symmetry }
We present now the results of our numerical analysis for the four simplest possible patterns
violating exact $\mu$\,--\,$\tau$  as described in the previous section and
quantified  in Eq.(\ref{tex1}) and Eq.(\ref{tex2}).
The coefficients $A's$  and $B$'s are expressed in Eq.(\ref{AB1}) and
Eq.(\ref{AB2}) according to the pattern under study. Moreover, analytical
expressions of the relevant parameters up to leading order in $s_z$
are provided
in order to get an ``understanding'' of the numerical results. The relevant
parameters include mass ratios, Majorana phases, $R_\n$ parameter, effective
Majorana mass term $\mee$ and
effective electron's neutrino mass $\me$. We stress here that our numerical
analysis is based on the exact formulae and not on the approximate ones.

The large number of correlation figures is organized in plots, at the
3-$\s$-error level, by dividing
 each figure into left and right panels (halves) denoted accordingly by the
letters L and R. Additional
labels (D, N and I) are attached to the plots to indicate the type of hierarchy
(Degenerate, Normal and
Inverted, respectively). Any missing label D, N or I on the figures of
certain pattern means the absence of the corresponding hierarchy type in this
pattern.

We list in tables~(\ref{tab2}) and (\ref{tab3}), and for the three types of
hierarchy and the
three precision levels, the extremum values that the different parameters can
take.
It is noteworthy that our numerical study is based, as was the case in \cite{tex0lash}, on random scanning of
the 7-dim parameter space
composed of $\left(\t_x, \t_y, \t_z, \d, \d m^2, |\chi|\; \mbox{and}\; \t\right)$. This kind of randomness implies
that the reported values in the tables are meant to give only a strong
qualitative indication, in that they might change from one run to another, providing thus a way to check for the stability of the results.
%%%%%%%%%%%%%%%%%%%%%%%%%%%%%%%%%%%%%%%%%%%%%%%%%%%%%%%%%%%%%%%%%%%%%%%%%%%%%%%

%%%%%%%%%%%%%%%%%%%%%%%%%%%%%%%%%%%%%%%%%%%%%%%%%%%%%%%%%%%%%%%%%%%%%%%%%%%%%%%
\subsection{C1: Pattern having $ M_{\n\,12}\,\left(1+ \chi\right)  -
M_{\n\,13}=0,\; \mbox{and}\;\; M_{\n\,22} - M_{\n\,33}=0.$}
In this pattern, C1,the relevant expressions for $A$'s and $B$'s  are
\bea
  A_1 &=& -c_x c_z \left(c_x s_y s_z + s_x c_y e^{-i\,\d}\right)\, \left(1 +
\chi\right) - c_x\,c_z \left(- c_x c_y s_z + s_x s_y e^{-i\,\d}\right), \nn\\
 A_2 & = & s_x c_z \left(- s_x s_y s_z + c_x c_y e^{-i\,\d}\right)\, \left(1 +
\chi\right) + s_x\,c_z \left(s_x c_y s_z + c_x s_y e^{-i\,\d}\right),\nn\\
 A_3 &=& s_z s_y c_z \left(1 + \chi \right) - s_z c_y c_z,\nn\\
 B_1 & =& \left(c_x s_y s_z + s_x c_y e^{-i\,\d}\right)^2 - \left(- c_x c_y s_z
+ s_x s_y e^{-i\,\d}\right)^2, \nn \\
 B_2 & =& \left(-s_x s_y s_z + c_x c_y e^{-i\,\d}\right)^2 - \left(s_x c_y s_z +
c_x s_y e^{-i\,\d}\right)^2, \nn \\
 B_3 &=& s_y^2 c_z^2 - c_y^2 c_z^2,
 \label{abtex1d}
 \eea
leading to mass ratios, up to leading order in $s_z$, as
\bea
 m_{13} \equiv \frac{m_1}{m_3} &\approx&  1 + \frac{2 \, s_\d s_\t
\left|\chi\right| s_z}{t_x T_1},  \nn\\
 m_{23} \equiv \frac{m_2}{m_3} &\approx&  1 - \frac{2\, t_x s_\d s_\t
\left|\chi\right| s_z}{T_1},
\label{mrtex1d}
\eea
where $T_1$ is defined as,
\be
T_1=\left|\chi\right|^2\, c_y^2 + 2\,\left|\chi\right|\,c_\t\,c_y\,\left(c_y + s_y\right) + 1 +s_{2y}. \label{T1}
\ee
While the Majorana phases as,
\bea
\r  &\approx& \d + \frac{s_\d \, s_z\,\left(-s_y c_y \left|\chi\right|^2 + \left|\chi\right|\,c_\t\,\left(c_{2y} - s_{2y}\right) + c_{2y}\right)}{t_x\,T_1}, \nn
\\
\s  &\approx& \d - \frac{s_\d \, t_x\, s_z\,\left(-s_y c_y \left|\chi\right|^2 + \left|\chi\right|\,c_\t\,\left(c_{2y} - s_{2y}\right) + c_{2y}\right)}{T_1}.
\label{phtex1d}
\eea
The parameters $R_\n$, mass ratio square difference $m_{23}^2 - m_{13}^2$,
$\me$ and $\mee$ can be deduced to be,
\bea
 R_\n &\approx& - \frac{8\, s_\d\, \,s_\t \left|\chi\right|
\,s_z}{s_{2x}\, T_1},\nn\\
 m_{23}^2 - m_{13}^2 &\approx&  - \frac{8\, s_\d\, \,s_\t \left|\chi\right|
\,s_z}{s_{2x}\, T_1}, \nn\\
\me &\approx& m_3 \left[ 1 +  \frac{4\, s_\t \, s_\d\, \left|\chi\right|
\,s_z}{t_{2x}\,T_1}\right], \nn\\
 \mee &\approx& m_3 \left[ 1 +  \frac{4\, s_\t \, s_\d\, \left|\chi\right|
\,s_z}{t_{2x}\,T_1}\right].
\label{nostex1d}
\eea

Our expansion in terms of $s_z$ is justified since $s_z$ is typically small for phenomenologically acceptable values where the best fit
for $s_z\approx 0.144$. Therefore, we naively expect that the expansion should work properly but it turns out that there are some subtle points
in this expansion which would invalidate our naive expectation.  To elaborate on this, let us consider the expansion corresponding to the mass ratio $m_{13}$ as,
\be
m_{13}= 1 + \sum_{i=1}^{\infty}\, c_i\left(\t_x,\t_y , \d , \left|\chi\right| , \t\right) s_z^i,
\label{m13exp}
\ee
where $c_i$ is the i$^{th}$-Taylor expansion coefficient depending on $\t_x,\t_y , \d , \left|\chi\right|$ and $\t$. In this pattern, putting  $\t_y$ equal to ${\pi\over 4}$
makes  the spectrum  degenerate $(m_{13}=m_{23}=1)$ irrespective of the values for  $\t_x , \d , \left|\chi\right|$ and $\t$. There are two possible alternatives
to match this finding: in the first one, all the $c_i\left( \t_y={\pi\over 4} \right) $'s are vanishing, whereas in the second one some of the $c_i\left( \t_y={\pi\over 4} \right) $'s are
finite and non-vanishing provided that an infinite number of $c_i\left( \t_y={\pi\over 4} \right) $'s are divergent such that the coefficients recombine in a delicate way to make the sum
$\sum_{i=1}^{\infty}\, c_i\left(\t_x,\t_y={\pi\over 4} , \d , \left|\chi\right| , \t\right) s_z^i$ equaling zero for any $s_z$ \footnote{One can see this simply by noting that in case all the $c_i$'s are bounded then the analyticity of the series forces them to vanish. On the other hand, one can not have a finite number of `unbounded' expansion coefficients, otherwise we could, assuming  without loss of generality two coefficients ($c_{i_1}, c_{i_2}$, $i_1<i_2$) whose limits at $y=y_0=\frac{\pi}{4}$ are divergent, write $c_{i_1}(y) t^{i_1} + c_{i_2}(y) t^{i_2} = g(y,t)$ where $g$ is a well behaved function if the infinite sum of `bounded' terms converge. It suffices then to let $y$, for $t_1 \neq t_2$, approach $y_0$ in the relation $c_{i_2}(y) = \frac{\frac{g(y,t_1)}{t_1^{i_1}}-\frac{g(y,t_2)}{t_2^{i_1}}}{t_{1}^{i_2-i_1}-t_{2}^{i_2-i_1}}$ to reach a contradiction.}.  Explicit calculation reveals that $c_1$ is finite and non vanishing at $\t_y={\pi\over 4}$ as is evident from Eq.(\ref{mrtex1d}), while $c_i$
is divergent  at $\t_y={\pi\over 4}$ for all $i\ge 2$. A similar consideration applies also to the mass ratio $m_{23}$. These divergences, at  $\t_y={\pi\over 4}$,
appearing in the expansion coefficients $c_i$ for mass ratios resurface again in the  expansion coefficients corresponding to $\me$ and $\mee$ but surprisingly enough the
divergences associated with $R_\n$ and $m_{23}^2 - m_{13}^2$ start only from the third order coefficients.  All these subtleties are an artifact of the expansion,
whereas no such problems arise if we use exact formulae. Thus, the formulae due to expansion must be dealt with caution.

\begin{figure}[hbtp]
\centering
\begin{minipage}[l]{0.5\textwidth}
\epsfxsize=8cm
\centerline{\epsfbox{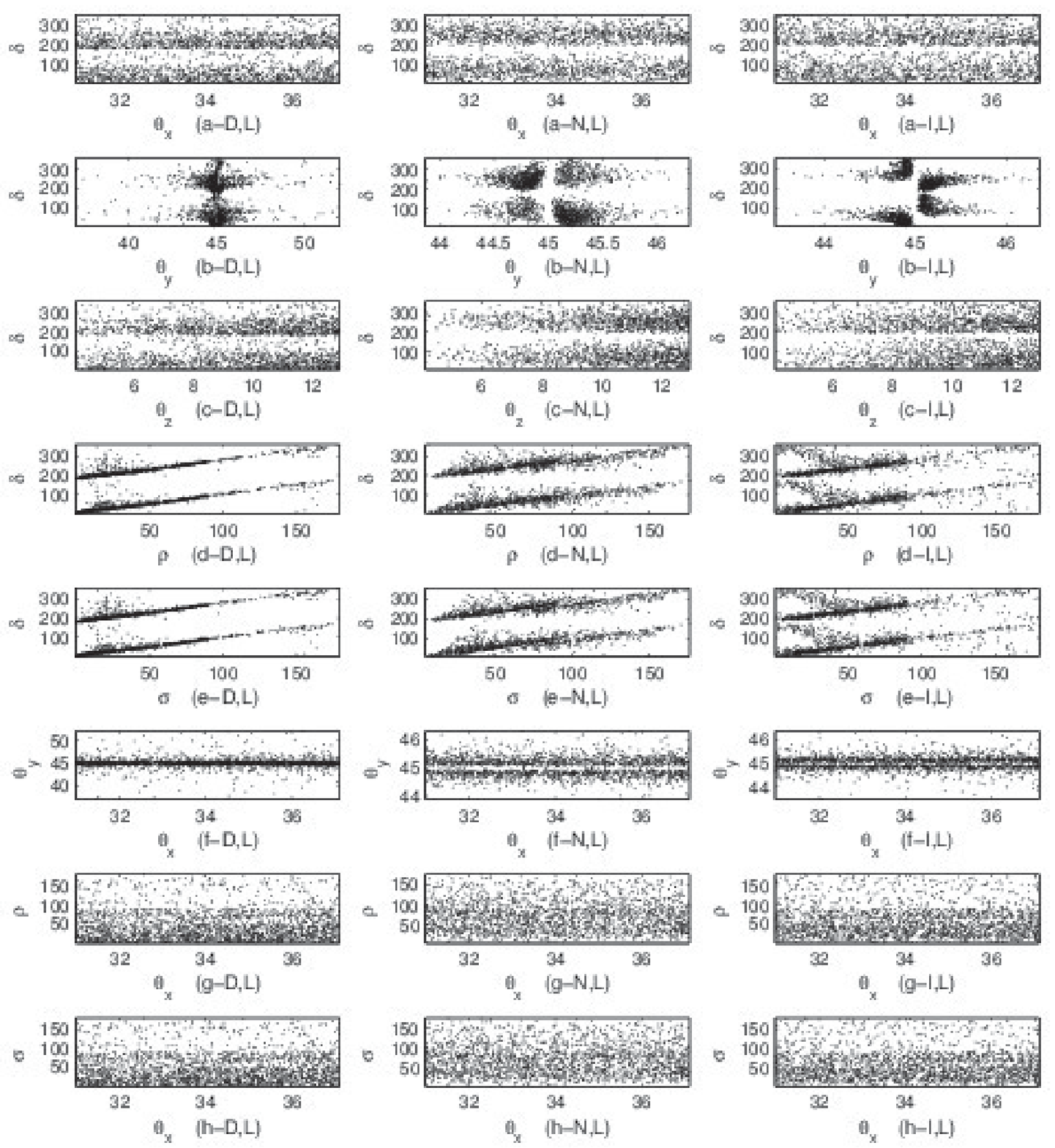}}
\end{minipage}%
\begin{minipage}[r]{0.5\textwidth}
\epsfxsize=8cm
\centerline{\epsfbox{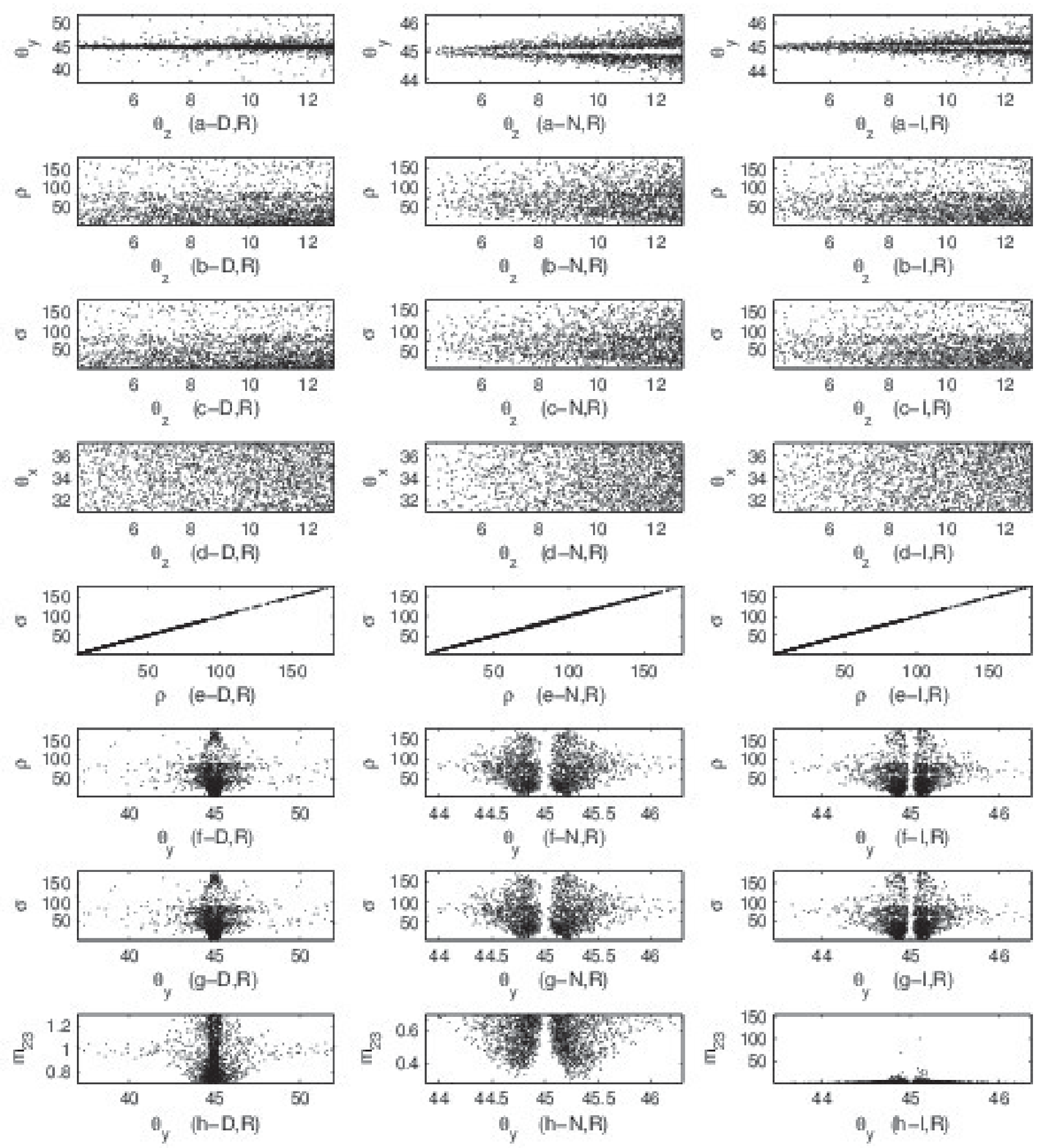}}
\end{minipage}
\vspace{0.5cm}
\caption{{\footnotesize Pattern having $ M_{\n\,12}\,\left(1+ \chi\right)  -
M_{\n\,13}=0,\; \mbox{and}\;\; M_{\n\,22} - M_{\n\,33}=0$: The left panel (the
left three columns) presents correlations of $\delta$ against
mixing angles and Majorana phases ($\r$ and $\s$) and those of $\t_x$ against
$\t_y$, $\r$ and $\s$.
The right panel (the right three columns) shows the correlations of $\t_z$
against $\t_y$, $\r$ ,
$\s$, and $\t_x$ and those of $\r$ against $\s$ and $\t_y$, and also the
correlation of $\t_y$ versus
$\s$ and $m_{23}$.}}
\label{12mfig1}
\end{figure}

All the possible fifteen pair correlations related to
the three mixing angles and the three Majorana and Dirac phases $(\t_x, \t_y,
\t_z, \d, \r, \s)$ are presented in the left and right
panels of Figure \ref{12mfig1}, while the last plot in the right panel is
reserved for the correlation of $m_{23}$ against $\t_y$.

\begin{figure}[hbtp]
\centering
\begin{minipage}[l]{0.5\textwidth}
\epsfxsize=8cm
\centerline{\epsfbox{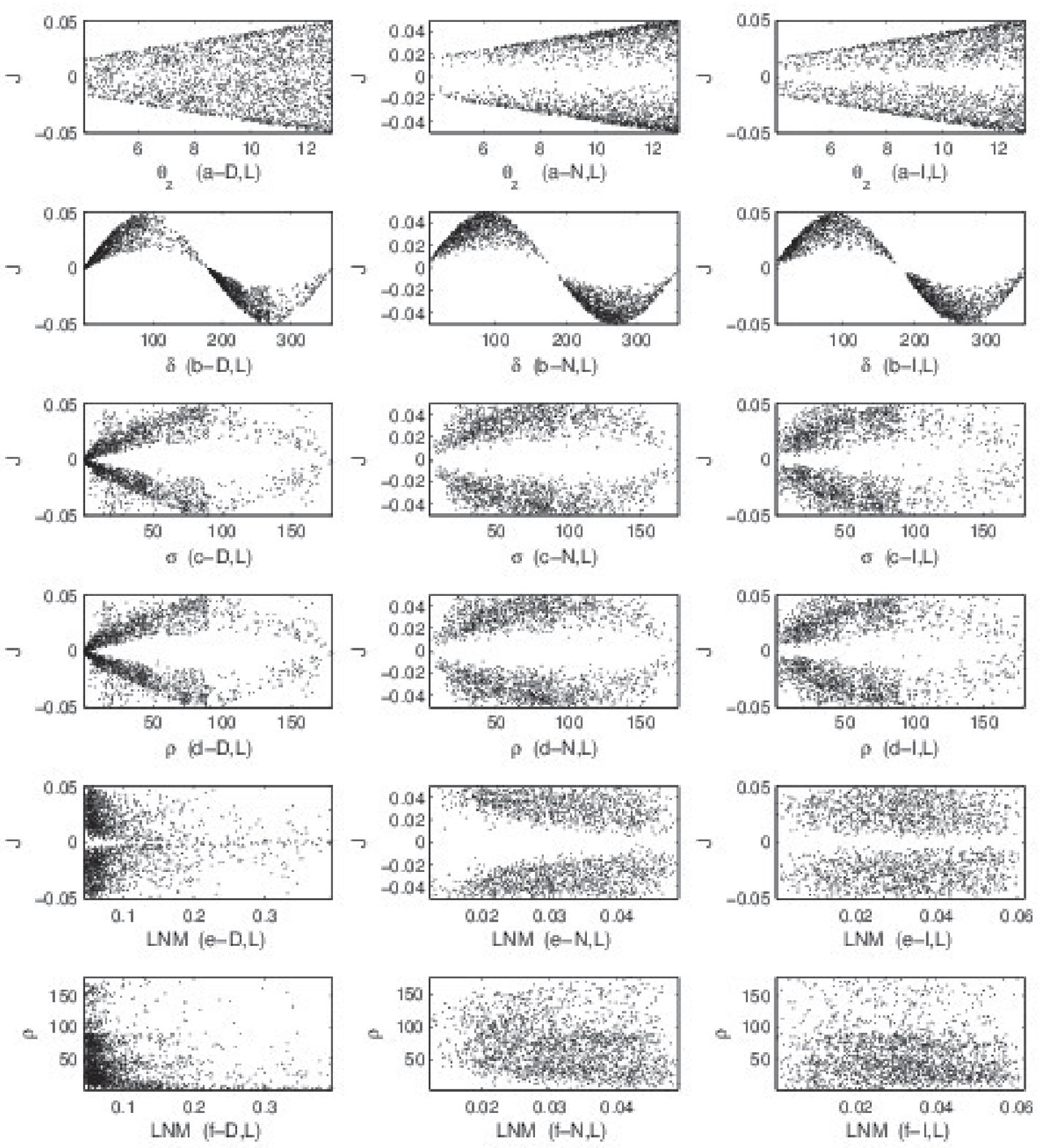}}
\end{minipage}%
\begin{minipage}[r]{0.5\textwidth}
\epsfxsize=8cm
\centerline{\epsfbox{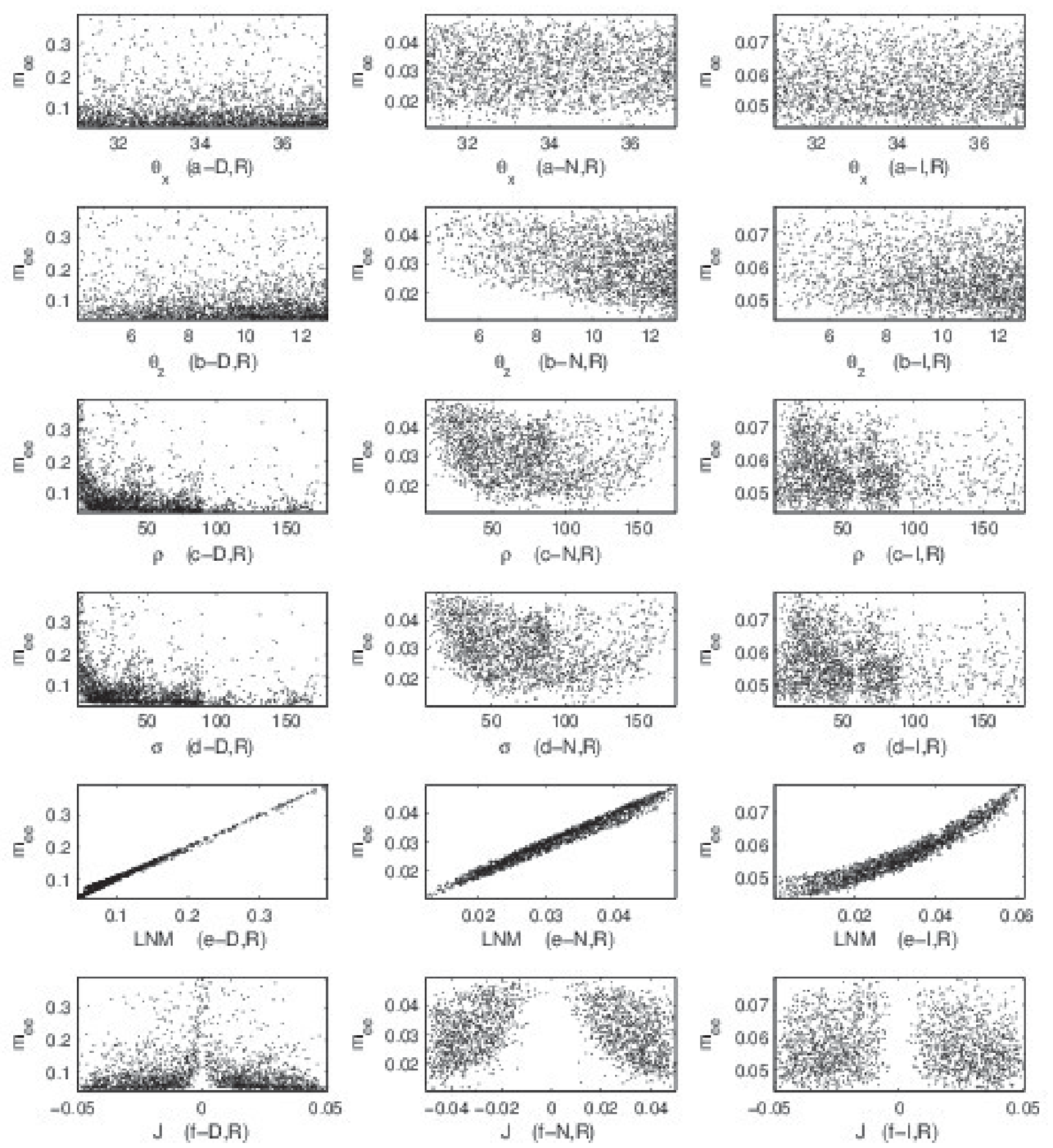}}
\end{minipage}
\vspace{0.5cm}
\caption{{\footnotesize Pattern having $ M_{\n\,12}\,\left(1+ \chi\right)  -
M_{\n\,13}=0,\; \mbox{and}\;\; M_{\n\,22} - M_{\n\,33}=0$: Left panel presents
correlations of $J$ against
$\t_z$, $\d$,  $\s$ , $\r$, and lowest neutrino mass ({\bf LNM}), while the last
one depicts the
correlation of LNM against $\r$. The right panel shows correlations of $m_{ee}$
against $\t_x$,
$\t_z$, $\r$, $\s$, {\bf LNM} and $J$. }}
\label{12mfig2}
\end{figure}

In Fig. \ref{12mfig2}, left panel, we present five correlations of $J$ against
($\t_z,\d, \s, \r$
and {\bf LNM}) and the correlation of $\r$ versus {\bf LNM}. As to the right
panel, we include presentation for the  correlations of $\mee$ against $\t_x$,
$\t_z$, $\r$, $\s$, $\mbox{LNM}$,  and $J$.

\begin{figure}[hbtp]
\centering
\epsfxsize=15cm
\epsfbox{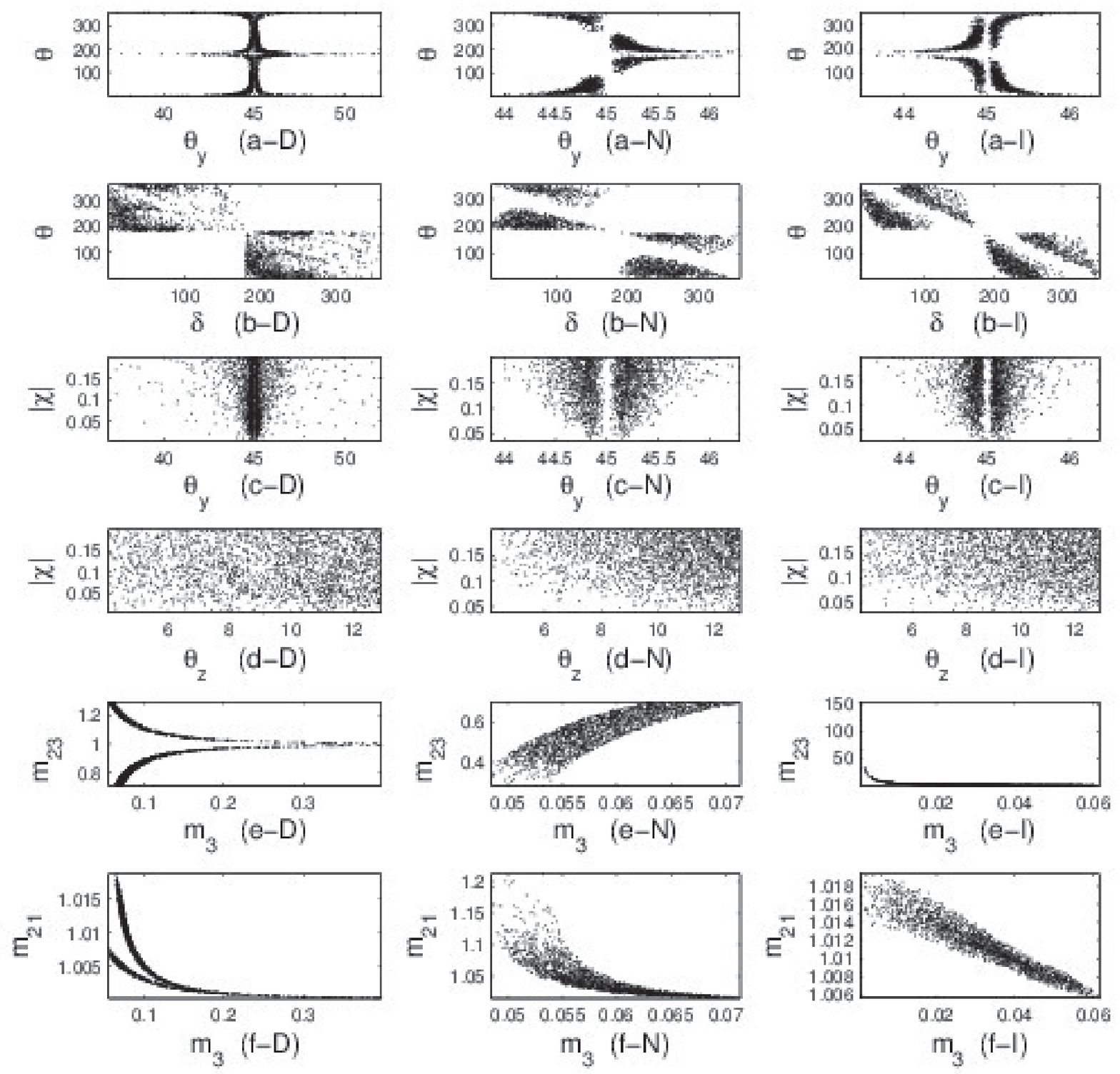}
\caption{{\footnotesize Pattern having $ M_{\n\,12}\,\left(1+ \chi\right)  -
M_{\n\,13}=0,\; \mbox{and}\;\; M_{\n\,22} - M_{\n\,33}=0$: The first two rows
presents the correlations of $\t$ against $\t_y$ and $\d$, while the second two
rows depict those of $\left|\chi\right|$ versus $\t_y$ and $\t_z$. The last two
rows shows the correlations of mass ratios $m_{23}$ and
$m_{21}$ against $m_3$.}}
\label{12mfig3}
\end{figure}

As to Fig. \ref{12mfig3}, and in a similar way, we present correlations for
$\t$ against $\t_y$ and $\d$ and for $\left|\chi\right|$ versus $\t_y$ and
$\t_z$. The correlation of  $m_3$ against $m_{23}$ and
$m_{21}$ are also included. All correlations are exhibited for all
three types of hierarchy and for each type we have thirty four depicted
correlations.
%%%%%%%%%%%%%%%%%%%%%%%%%%%%%%%%%%%%%%%%%%%%%%%%%%%%%%%%%%%%%%%%%%%%%%%%%%%%%%%%%%%%%%%%%%%%%%%%%%%%%%%%%
\clearpage

Before dwelling into examining the correlations provided by the various figures we can infer some restrictions concernning mixing angles and phases
in each pattern just by considering the expression for $R_\n$ as given in Eq.(\ref{nostex1d}). The parameter $R_\n$ must be positive, nonvanishing and at
the $3-\s$ level is restricted to be in the interval $\left[0.0262,0.0397\right]$. This clearly requires nonvanishing values for $s_z$,  $s_\d$, $s_\t$  and $\left|\chi\right|$.
The nonvanishing of $s_z$  means $\t_z \neq 0$ which is phenomenologically favorable,  while vanishing of  $s_\d$, $s_\t$ implies excluding $0$, $\pi$ and $2\pi$
for both $\d$ and $\t$. The nonvanishing of $\left|\chi\right|$ is naturally expected otherwise there would not be  a deviation from exact $\mu$\,--\,$\tau$
 symmetry.  The other required restriction, namely,  $s_\t\, s_\d < 0$ dictates that if $\d$ falls in the first and second quadrants then $\t$ falls in
 third and fourth quadrants and vice versa.  These  conclusions remain valid if  one  used the exact expression
for $R_\n$ instead of the first order expression. Explicit computations of $R_\n$ using its  exact expression tell us that $\t_y$ cannot be exactly  equal to ${\pi\over 4}$ otherwise $R_\n$
would be zero, but nevertheless $\t_y$ can possibly stay very close to ${\pi\over 4}$.

We see in Fig.~\ref{12mfig1}~(plots: a-L $\rightarrow$ c-L, as examples) that
all the experimentally allowed ranges of mixing angles, at $3 \s$
error levels, can be covered in this pattern except for normal and inverted
hierarchy types where $\t_y$ is restricted to be around $45^0$, by at most, plus or minus $1.5^0$.
This restriction on $\t_y$ is a characteristic of the normal and inverted hierarchy type in this pattern.
This characteristic behaviour of $\t_y$ can be understood by expressing the mass ratios, using Eqs.~(\ref{mrtex1d}, \ref{T1} and \ref{nostex1d}), as
\bea
m_{13} &=& 1 -{1\over 2}\, c_x^2\, R_\n + O\left(s_z^2\right), \nn \\
m_{23} &=& 1 + {1\over 2}\, s_x^2\, R_\n + O\left(s_z^2\right),
\label{mrtex1dR}
\eea
where the first order correction is identified consistently with $R_\n$ expressed up to this order. All the remaining higher order corrections to
the mass ratios contribute significantly and in a spiky way in the vicinity of $\t_y={\pi\over 4}$ leading to mass ratios considerably  greater or smaller than
unity. Therefore, to produce the various hierarchy types as marked in Eqs.(\ref{nor}--\ref{deg}), $\t_y$ can take in the degenerate hierarchy type values far from $\pi\over 4$ corresponding to small higher order corrections in Eq.~(\ref{mrtex1dR}) which would keep $m_{13}$ and $m_{23}$ near the value one. However, in order to get normal or inverted hierarchies, the higher order corrections in Eq.~(\ref{mrtex1dR}) should contribute in a noticeably large amount, which could not be happened unless $\t_y$ stays close to $\pi\over 4$,  and this is what the corresponding ranges for $\t_y$
reported in Table~\ref{tab2} confirm. As to the Dirac CP-phase $\d$, the whole range is allowed except the regions around $0$ and $\pi$ whose extensions
depend on the type of hierarchy and the precision level as evident from the
same plots and the reported values in Table~\ref{tab2}. Likewise, the plots (g-L, h-L),
in Figure~\ref{12mfig1}  and the values reported in Table~\ref{tab2} show that the Majorana phases ($\r, \s$) are covering their ranges excluding regions around $0$ and $\pi$.

The plots in Figure~\ref{12mfig1} can reveal  many obvious clear correlations. For example, the plots (a-R) shows that as $\t_z$ decreases $\t_y$ tends
to be very close to $45^0$. The plots (d-L, e-L) show a  sort of distorted linear correlation of $\d$
versus $(\r, \s)$ in all hierarchy types which confirms the relations presented  in Eq.~(\ref{phtex1d}) which give linear relations at zeroth order of $s_z$, while
the found distortion can be attributed to the  higher order corrections.
We may see also also in (plot e-R), a very clear linear correlation between the Majorona phases
($\r, \s$) in all hierarchy types which again confirms  the relations presented  in Eq.~(\ref{phtex1d})
which at zeroth order produces the linear relation $\r\approx \s$.

The Figure~\ref{12mfig2}~(plots: a-L ,b-L) shows that the correlations ($J,\t_z$) and ($J,\d$) have each a specific geometrical
shape irrespective  of the hierarchy type.
In fact,  Eq.~(\ref{jg}) indicates that the correlation
($J,\d$) can be seen as a superposition of many sinusoidal graphs in $\d$, the `positive'
amplitudes of which are determined by the acceptable mixing angles, whereas the ($J,\t_z$) correlation
is a superposition of straight-lines in $s_z \sim
\t_z$, for small $\t_z$, the slopes of which are positive or negative according to
the sign of $s_\d$.  The resulting shape for ($J,\t_z$) correlation being
trapezoidal rather than isosceles is due to the exclusion of zero and its vicinity for $\t_z$
considering the latest oscillation data. The unfilled region in the plots originates from the disallowed region of
$\d$ around $0$ and $\pi$, which would have led, if allowed, to zero $J$.

The left panel of Figure~\ref{12mfig2} (plots: c-L, d-L), unveils a correlation
of  $J$ versus $(\r, \s)$
 which is a direct consequence of the `linear' correlations of  $\d$ against
$(\r, \s)$ and of the `geometrical' correlation of ($J, \d$). The two correlations
 concerning the {\bf LNM} (plots: e-L, f-L)
reveals that as the {\bf LNM} increases the parameter space becomes more
restricted. This seems to be a general tendency in all the patterns, where the {\bf
LNM} can reach in the degenerate case values higher than in the normal and inverted hierarchies.

To gain more insight about  the correlattions involving $\mee$ as defined in Eq.~(\ref{mee}), we work out approximate formulae for
$\mee$ corresponding to different hierarchy types. It is helpful in deriving these approximate formulae to
realize that $\r \approx \s$  and $m_1\approx m_2$  in all hierarchy types as is evident respectively from
Fig.~\ref{12mfig1} (plots: e-R) and  Fig.~\ref{12mfig3} (plots: f), and also to realize that the normal hierarchy is moderate
(meaning $m_3$ is of the same order as $m_1$)  while the inverted one is acute as can be inferred from  Fig.~\ref{12mfig3} (plots: e-N, e-I).
Thus, the resulting formulae  are,
\bea
\mee &\approx& m_1 \left(1-2\,s_z^2\, c_z^2\, s_\s^2\right)\;\;\;\; \mbox{For normal and degenerate cases},\nn\\
\mee & \approx& m_1 \left(1-s_z^2\right)\;\;\;\;\;\;\;\;\;\;\;\;\;\;\; \mbox{For inverted  case}.
\label{meeatex1d}
\eea
The correlations of $\mee$ against $(\t_x, \t_z, \r, \s)$ as depicted in the right panel  of Fig.~\ref{12mfig2} (plots: a-R --d-R)
can be understood by exploiting the approximate expression for $\mee$ in conjunction with the correlations found between $\t_z$ and $\left(\t_x, \r , \s\right)$.
The totality of correlations of $\mee$ presented in  the right panel  of Fig.~\ref{12mfig2} indicate that the increase
of $\mee$ would on the whole constrain the allowed parameter space. We note also a general trend of
increasing $\mee$ with increasing LNM in all cases of hierarchy (plots e-R). The values of $\mee$
can not reach the zero-limit in all types of hierarchy, as is evident from
the graphs or explicitly from the corresponding covered range in Table~\ref{tab2}. Another point concerning $\mee$ is that its
scale is triggered by the scale of $m_1$ ($\approx m_2$) as is evident from both the approximate formula in Eq.~(\ref{meeatex1d}) and the
corresponding covered range in Table~\ref{tab2}.

The plots in Fig.~\ref{12mfig3} (plots: b) disclose a clear correlation between $\t$ and $\d$ which is in accordance with what was derived before
in that ($s_\t s_\d <0$). The plots also reveal that there are disallowed regions for both $\t$ and $\d$, which must definitely contain domains around
$0$ and $\pi$ besides other possible additional areas. The disallowed regions can be also checked with the help of Tables~(\ref{tab2}--{\ref{tab3}) where one
additionally finds that the regions around $0$ and $\pi$ tend to be  shrunk for the degenerate case. The plots (c) in Fig.~\ref{12mfig3}
show that as $\t_y$ deviates slightly from ${\pi\over 4}$ then $\left|\chi\right|$ tends to increase.

For the mass spectrum, we see from Fig.~\ref{12mfig3} (plots: e, f) that the normal
hierarchy is mild in that the mass ratios do not reach extreme
values. In contrast, the inverted hierarchy can be acute in that the mass ratio
$m_{23}$ can reach values up to  $O(10^2)$. The values of $m_1$ and $m_2$ are nearly equal in all hierarchy types. We also see that if $m_3$ is large enough then only
the degenerate case with $m_1 \sim m_2$ can be phenomenologically acceptable.

\subsection{C2:  Pattern having $ M_{\n\,12}\,\left(1+ \chi\right)  +
M_{\n\,13}=0,\; \mbox{and}\;\; M_{\n\,22} - M_{\n\,33}=0.$}
In this pattern, C2,the relevant expressions for $A$'s and $B$'s  are
\bea
  A_1 &=& -c_x c_z \left(c_x s_y s_z + s_x c_y e^{-i\,\d}\right)\, \left(1 +
\chi\right) + c_x\,c_z \left(- c_x c_y s_z + s_x s_y e^{-i\,\d}\right), \nn\\
 A_2 & = & s_x c_z \left(- s_x s_y s_z + c_x c_y e^{-i\,\d}\right)\, \left(1 +
\chi\right) - s_x\,c_z \left(s_x c_y s_z + c_x s_y e^{-i\,\d}\right),\nn\\
 A_3 &=& s_z s_y c_z \left(1 + \chi \right) + s_z c_y c_z,\nn\\
 B_1 & =& \left(c_x s_y s_z + s_x c_y e^{-i\,\d}\right)^2 - \left(- c_x c_y s_z
+ s_x s_y e^{-i\,\d}\right)^2, \nn \\
 B_2 & =& \left(-s_x s_y s_z + c_x c_y e^{-i\,\d}\right)^2 - \left(s_x c_y s_z +
c_x s_y e^{-i\,\d}\right)^2, \nn \\
 B_3 &=& s_y^2 c_z^2 - c_y^2 c_z^2,
 \label{abtex2d}
 \eea
leading to mass ratios, up to leading order in $s_z$, as
\bea
  m_{13}  &\approx&  1 - \frac{2 \, s_\d s_\t
\left|\chi\right| s_z}{t_x T_2},  \nn\\
 m_{23} &\approx&  1 + \frac{2\, t_x s_\d s_\t
\left|\chi\right| s_z}{T_2},
\label{mrtex2d}
\eea
where $T_2$ is defined as,
\be
T_2=\left|\chi\right|^2\, c_y^2 + 2\,\left|\chi\right|\,c_\t\,c_y\,\left(c_y - s_y\right) + 1 - s_{2y} \label{T2}.
\ee
The Majorana phases are given by
\bea
\r  &\approx& \d - \frac{s_\d \, s_z\,\left(s_y c_y \left|\chi\right|^2 + \left|\chi\right|\,c_\t\,\left(c_{2y} + s_{2y}\right) + c_{2y}\right)}{t_x\,T_2}, \nn
\\
\s  &\approx& \d + \frac{s_\d \, t_x\, s_z\,\left(s_y c_y \left|\chi\right|^2 + \left|\chi\right|\,c_\t\,\left(c_{2y} + s_{2y}\right) + c_{2y}\right)}{T_2}.
\label{phtex2d}
\eea
The parameters $R_\n$, mass ratio square difference $m_{23}^2 - m_{13}^2$,
$\me$ and $\mee$ can be deduced to be,
\bea
R_\n &\approx& \frac{8\, s_\d\, \,s_\t \left|\chi\right|
\,s_z}{s_{2x}\, T_2},\nn\\
 m_{23}^2 - m_{13}^2 &\approx&  \frac{8\, s_\d\, \,s_\t \left|\chi\right|
\,s_z}{s_{2x}\, T_2}, \nn\\
\me &\approx& m_3 \left[ 1 -  \frac{4\, s_\t \, s_\d\, \left|\chi\right|
\,s_z}{t_{2x}\,T_2}\right], \nn\\
 \mee &\approx& m_3 \left[ 1 - \frac{4\, s_\t \, s_\d\, \left|\chi\right|
\,s_z}{t_{2x}\,T_2}\right].
\label{nostex2d}
\eea
One can notice that the all results concerning this pattern, C2, can be derived from those of the previous one, C1, by simply making the substitutions
$s_y \rightarrow - s_y$ and $\d \rightarrow \d + \pi$. Unfortunately, the found relation cannot be used in practice to derive the predictions of
one pattern from the other because the mapping $s_y  \rightarrow -s_y$ takes $\t_y$ from a physically admissible region to a forbidden one. However, one can also verify that the
two patterns have the same properties regarding divergences for the expansion coefficients of the mass ratios.

The approximate expression for $R_\n$ in Eq.~(\ref{nostex2d}) provides us with similar restrictions like those of the previous
pattern C1, except that both $\d$ and $\t$ should now fall in same upper or lower semicircles. Once again the derived restriction remains unchanged when
using the exact expression for $R_\n$.

We plot the corresponding correlations in Figures (\ref{12pfig1}, \ref{12pfig2}
and \ref{12pfig3}) with the same conventions as before.
In contrast to the C1 case, we see here that the
mixing angle $(\t_y)$ can cover a wider range in the normal and inverted hierarchy cases instead of being confined around
$\t_y= {\pi\over 4}$. In the normal hierarchy case $\t_y$ falls in the interval $[41^o - 50^o]$, while it almost covers all the admissible range in
the inverted case. In the degenerate case, however, there is no restriction on $\t_y$, as it was in the C1 pattern.
 Another contrasting feature is the range of $\t_z$ in the normal hierarchy type, where it is now restricted to be less than $10^o$, whereas
 it can, similarly to the C1 pattern, cover all its allowed range in the inverted and degenerate cases.

We can understand the  behaviour of $\t_y$, compared to that of the previous pattern C1,  by expressing the mass ratios, from Eqs.~(\ref{mrtex2d},\ref{T2}) and (\ref{nostex2d}), as
\bea
m_{13} &=& 1 -{1\over 2}\, c_x^2\, R_\n + O\left(s_z^2\right), \nn \\
m_{23} &=& 1 + {1\over 2}\, s_x^2\, R_\n + O\left(s_z^2\right),
\label{mrtex2dR}
\eea
where the first order correction is identified consistently with $R_\n$ expressed up to this order, and thus representing a small quantity. In contrast to the situation in the pattern C1, the remaining higher-order corrections in
the mass ratios can be tuned to have a significant contribution in the vicinity of any $\t_y$ depending on the other combinations of mixing angles and phases, which would lead to
mass ratios considerably  greater or smaller than unity. Therefore the various hierarchy types as marked in Eqs.(\ref{nor}--\ref{deg}) can be generated
for almost all  $\t_y$ in its allowed range, and the values of $\t_y$
reported in Table~\ref{tab2} confirm this. As to the Dirac CP-phase $\d$, the whole range is allowed except the regions around $0$ and $\pi$ whose extensions
depend on the type of hierarchy and the precision level as is evident from the
corresponding plots and from the reported values in Table~\ref{tab2}.

The plots in Figure~\ref{12pfig1} can disclose  many obvious clear correlations. For example, the plots (a-R) show, in normal and inverted hierarchy cases,
that as $\t_z$ decreases $\t_y$ tends to be spread over its admissible range  while the contrary occurs when $\t_z$ increases. The plots (d-L, e-L) do not show a
simple correlation of $\d$ versus $(\r, \s)$ in the various hierarchy types which would have been consistent with the zeroth order linear relation given  in Eq.~(\ref{phtex2d}).
In fact, the  higher order corrections bring a severe distortion that invalidate the zeroth order linear relation even at the approximate level.
These higher order corrections do not work in the same manner for both $\r$ and $\s$, so they do not cancel out upon subtraction
producing ambiguous correlation between $\r$ and $\s$, as depicted in the (plot e-R), contrasted with the simple linearity in the previous pattern C1. The absence of linear relations
among the phases ($\d,\r,\s$) forbids the allowed region of Majorana phases to be straightforwardly determined from that of the Dirac phase ($\d$), as can be figured out
looking at  the corresponding allowed values in Table~\ref{tab2}.

The special `sinusoidal' and `trapezoidal' shapes of $J$ versus $\d$ and $\t_z$ remain intact
(Fig.~\ref{12pfig2}, plots: a-L, b-L), and as before  the unfilled region in the trapezoidal shaped plots is attributed to the disallowed region for
$\d$ around $0$ and $\pi$. The usual correlations of $J$ versus $\r$ and $\s$ (Fig.~\ref{12pfig2} plots: c-L, d-L) emerge from those of
$\d$ versus  $\r$ and $\s$. The two correlations concerning the {\bf LNM} (plots: e-L, f-L)
indicate that as the {\bf LNM} increases (say, larger than $0.1$ ev) the parameter space becomes more
restricted. This seems to represent an inclination in all the patterns, where the {\bf
LNM} can reach in the degenerate case values higher than the other hierarchies.

The correlations involving $\mee$ can be made more transparent by deriving an approximate formula for $\mee$ capturing the essential observed features for
all kinds of hierarchies in this specific pattern C2 which are: first,  the equality of $m_1$ and $m_2$ as is clear
in Fig.~\ref{12pfig3} (plots: f); second, the mild hierarchy in both normal and inverted cases as is evident from Fig.~\ref{12pfig3} (plots: e-N, e-I).
Thus, one can deduce from Eq.~(\ref{mee}) that $\mee$ is approximated by
\be
\mee  \approx  m_1\, c_z^2\, \sqrt{\left[1- s_{2x}^2\, \sin^2\left(\r-\s\right)\right]}.
\label{meeatex2d}
\ee
Now, the correlations of $\mee$ against $(\t_x, \t_z, \r, \s)$ as displayed in the right panel  of Figure~\ref{12pfig2}~(plots: a-R --d-R)
can be comprehended by invoking the approximate expression for $\mee$ in conjunction with the pair correlations found amidst  $\t_x,\, \t_z,\, \r$ and  $\s$.
The whole correlations of $\mee$ presented in  the right panel  of Figure~\ref{12pfig2} point out that the increase
of $\mee$ would generally constrain the allowed parameter space. We note also a general tendency of
increasing $\mee$ with increasing LNM in all cases of hierarchy (plots e-R). The values of $\mee$
can not attain the zero-limit in all types of hierarchy, as is evident from
the graphs or explicitly from the corresponding covered range in Table~\ref{tab2}. Another point concerning $\mee$ is that its
scale is triggered by the scale of $m_1$ ($\approx m_2$) as is evident from both the approximate formula in Eq.~(\ref{meeatex2d}) and
the corresponding covered range stated in Table~\ref{tab2}.

The plots in Fig.~\ref{12pfig3} (plots: b) shows both that $\t$ and $\d$ must lie in the same upper or lower semicircle which confirms our inference based
on the approximate formula for $R_\n$ in Eq.~(\ref{nostex2d}). The plots also reveal that there are disallowed regions for both $\t$ and $\d$,
which definitely should contain regions around
$0$ and $\pi$ besides other possible additional regions. The disallowed regions can be also checked with the help of Tables~(\ref{tab2}--\ref{tab3}) where one
can additionally find that the forbidden regions around $0$ and $\pi$ tend to be  shrunk for the degenerate case and that the allowed range for $\t$ is very limited
in normal and inverted hierarchy. The  Figure~\ref{12pfig3}~(plots: c,d)
shows that $\left|\chi\right|$ tends to increase in normal and inverted heirarchies as $\t_y$ deviates from ${\pi\over 4}$ or as $\t_z$ increases.

For the mass spectrum, we see from Fig.~\ref{12pfig3} (plots: e)  that all hierarchy types are characterized by nearly equal
values of $m_1$ and $m_2$. Moreover, Fig.~\ref{12pfig3} (plots: f) reveals that both normal
and inverted hierarchies are of moderate type in that the mass ratios $m_{23}$  does not reach extremely low nor high
values. We also see that if $m_3$ is large enough then only
the degenerate case with $m_1 \sim m_2$ can be compatible with data.
\begin{figure}[hbtp]
\centering
\begin{minipage}[l]{0.5\textwidth}
\epsfxsize=8cm
\centerline{\epsfbox{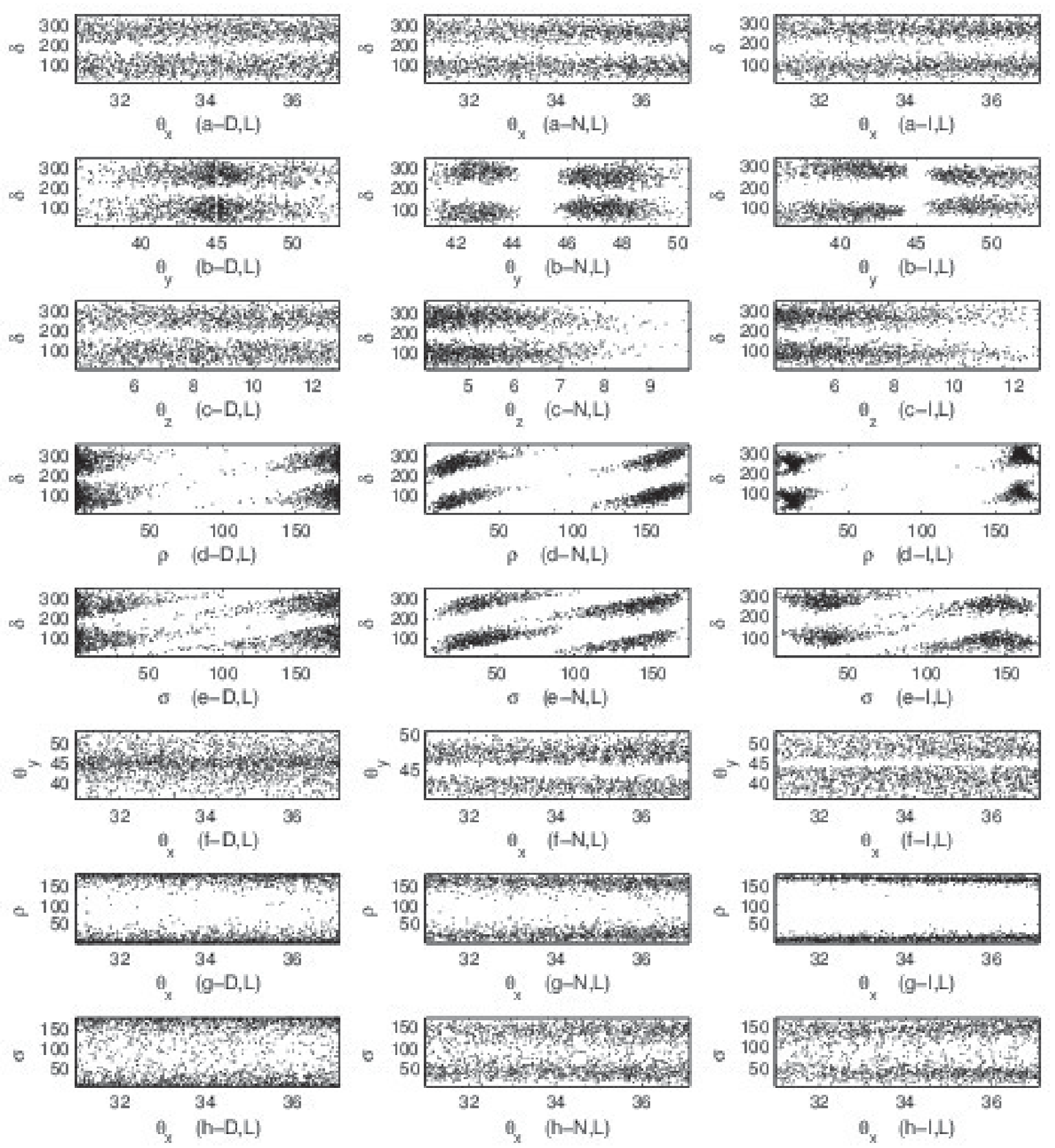}}
\end{minipage}%
\begin{minipage}[r]{0.5\textwidth}
\epsfxsize=8cm
\centerline{\epsfbox{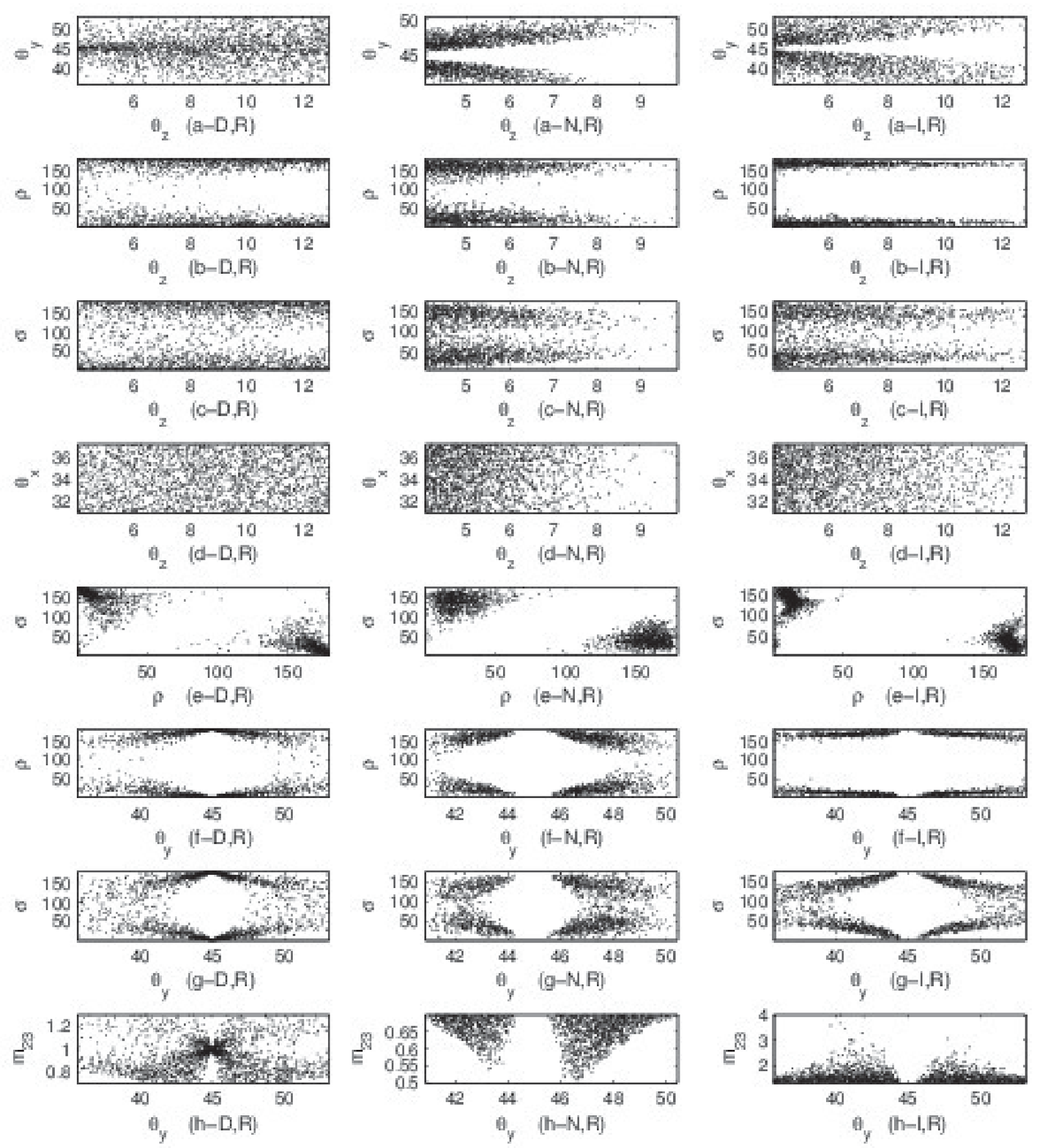}}
\end{minipage}
\vspace{0.5cm}
\caption{{\footnotesize Pattern having $ M_{\n\,12}\,\left(1+ \chi\right)  +
M_{\n\,13}=0,\; \mbox{and}\;\; M_{\n\,22} - M_{\n\,33}=0$: The left panel (the
left three columns) presents correlations of $\delta$ against
mixing angles and Majorana phases ($\r$ and $\s$) and those of $\t_x$ against
$\t_y$, $\r$ and $\s$.
The right panel (the right three columns) shows the correlations of $\t_z$
against $\t_y$, $\r$ ,
$\s$, and $\t_x$ and those of $\r$ against $\s$ and $\t_y$, and also the
correlation of $\t_y$ versus
$\s$ and $m_{23}$.}}
\label{12pfig1}
\end{figure}
%%%%%%%%%%%%%%%%%%%%%%%%%%%%%%%%%%%%%%%%%%%%%%%%%%%%%%%%%%%%%%%%%%%%%%%%%%%%%%%%%%%%%%%%%%%%%%%%%%%%%%%%%%
%%%%%%%%%%%%%%%%%%%%%%%%%%%%%%%%%%%%%%%%%%%%%%%%%%%%%%%%%%%%%%%%%%%%%%%%%%%%%%%%%%%%%%%%%%%%%%%%%%%%%
\begin{figure}[hbtp]
\centering
\begin{minipage}[l]{0.5\textwidth}
\epsfxsize=8cm
\centerline{\epsfbox{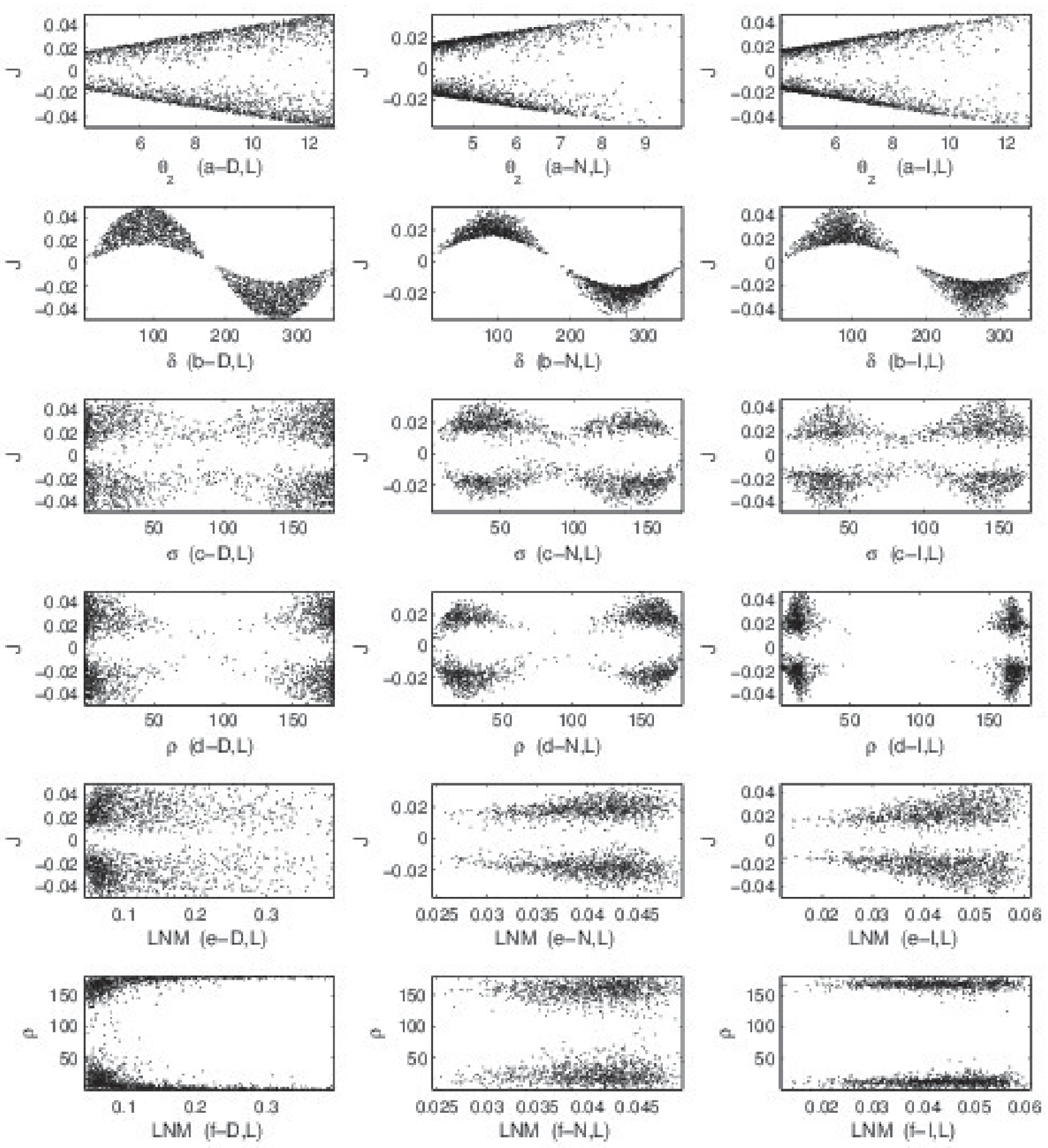}}
\end{minipage}%
\begin{minipage}[r]{0.5\textwidth}
\epsfxsize=8cm
\centerline{\epsfbox{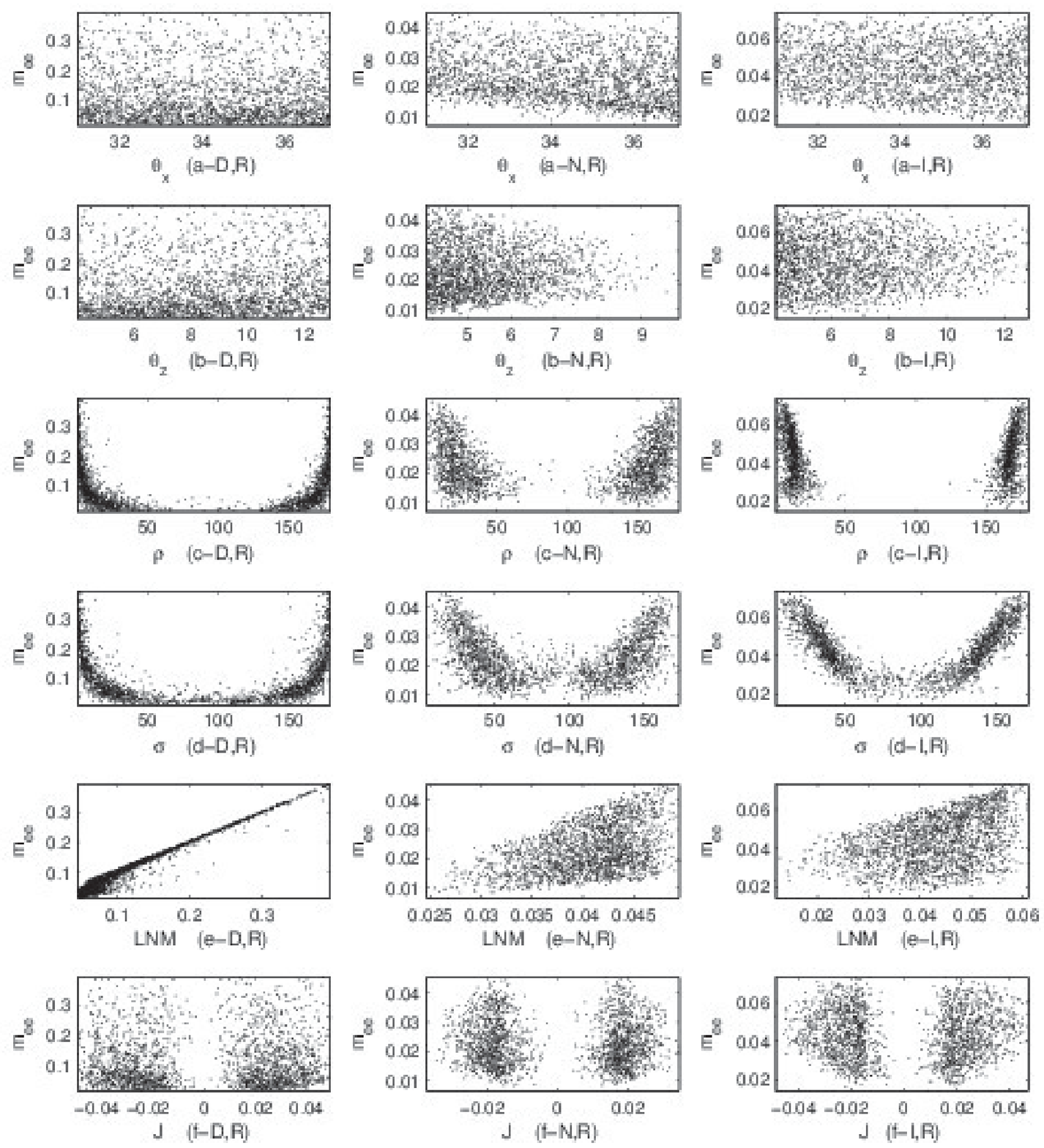}}
\end{minipage}
\vspace{0.5cm}
\caption{{\footnotesize Pattern having $ M_{\n\,12}\,\left(1+ \chi\right)  +
M_{\n\,13}=0,\; \mbox{and}\;\; M_{\n\,22} - M_{\n\,33}=0$: Left panel presents
correlations of $J$ against
$\t_z$, $\d$,  $\s$ , $\r$, and lowest neutrino mass ({\bf LNM}), while the last
one depicts the
correlation of LNM against $\r$. The right panel shows correlations of $m_{ee}$
against $\t_x$,
$\t_z$, $\r$, $\s$, {\bf LNM} and $J$. }}
\label{12pfig2}
\end{figure}
%%%%%%%%%%%%%%%%%%%%%%%%%%%%%%%%%%%%%%%%%%%%%%%%%%%%%%%%%%%%%%%%%%%%%%%%%%%%%%%%%%%%%%%%%%%%%%%%%%%%%%%%%%%%%5
\clearpage
\begin{figure}[hbtp]
\centering
\epsfxsize=15cm
\epsfbox{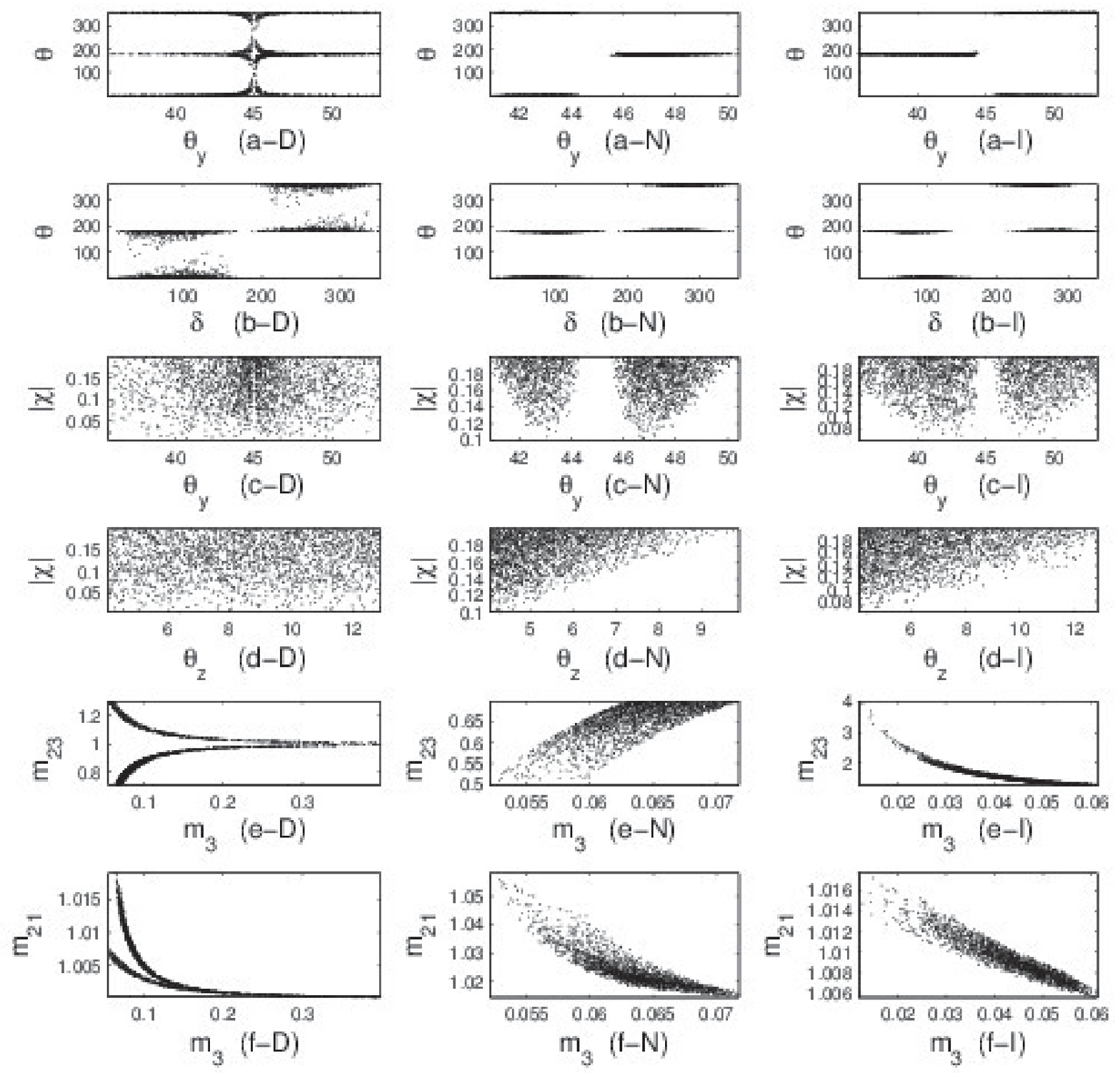}
\caption{{\footnotesize Pattern having $ M_{\n\,12}\,\left(1+ \chi\right)  +
M_{\n\,13}=0,\; \mbox{and}\;\; M_{\n\,22} - M_{\n\,33}=0$: The first two rows
presents the correlations of $\t$ against $\t_y$ and $\d$, while the second two
rows depict those of $\left|\chi\right|$ versus $\t_y$ and $\t_z$. The last two
rows shows the correlations of mass ratios $m_{23}$ and
$m_{21}$ against $m_3$.}}
\label{12pfig3}
\end{figure}
%%%%%%%%%%%%%%%%%%%%%%%%%%%%%%%%%%%%%%%%%%%%%%%%%%%%%%%%%%%%%%%%%%%%%%%%%%%%%%%%%%%%%%%%%%
%%%%%%%%%%%%%%%%%C3 Pattern M12 - M13 =0,  M22(1+chi) -M33=0
%%%%%%%%%%%%%%%%%%%%%%%%%%%%%%%%%%%%%%%%%%%%%%%%%%%%%%%%%%%%%%%%%%%%%%%%%%%%%%%%%%%%%%%%%%
\subsection{C3: Pattern having $ M_{\n\,12} -
M_{\n\,13}=0,\; \mbox{and}\;\; M_{\n\,22}\,\left(1+ \chi\right)  - M_{\n\,33}=0.$}
In this pattern, the relevant expressions for $A$'s and $B$'s  are
\bea
  A_1 &=& -c_x c_z \left(c_x s_y s_z + s_x c_y e^{-i\,\d}\right) - c_x\,c_z \left(- c_x c_y s_z + s_x s_y e^{-i\,\d}\right), \nn\\
 A_2 & = & s_x c_z \left(- s_x s_y s_z + c_x c_y e^{-i\,\d}\right) + s_x\,c_z \left(s_x c_y s_z + c_x s_y e^{-i\,\d}\right),\nn\\
 A_3 &=& s_z c_z\,\left(s_y - c_y\right) ,\nn\\
 B_1 & =& \left(c_x s_y s_z + s_x c_y e^{-i\,\d}\right)^2\,\left(1+ \chi\right)  - \left(- c_x c_y s_z
+ s_x s_y e^{-i\,\d}\right)^2, \nn \\
 B_2 & =& \left(-s_x s_y s_z + c_x c_y e^{-i\,\d}\right)^2\,\left(1+ \chi\right) - \left(s_x c_y s_z +
c_x s_y e^{-i\,\d}\right)^2, \nn \\
 B_3 &=& s_y^2 c_z^2\,\left(1+ \chi\right)  - c_y^2 c_z^2,
 \label{abtex3d}
 \eea
leading to mass ratios, up to leading order in $s_z$, as
\bea
  m_{13}  &\approx& \sqrt{{T_3\over T_4}}\,\left[ 1 - {\left|\chi\right|\, c_{2y}\, \left(-c_\d \,s_y^2 \left|\chi\right| + c_{2y}\, c_{\d-\t}\right)\,s_z
  \over t_x\, \left( 1 +s_{2y}\right)\, T_3} \right] + O(s_z^2),  \nn\\
 m_{23} &\approx&  \sqrt{{T_3\over T_4}}\,\left[ 1 + {\left|\chi\right|\, c_{2y}\, t_x \left(-c_\d\, s_y^2 \,\left|\chi\right| + c_{2y}\, c_{\d-\t}\right)\,s_z
  \over \left( 1 +s_{2y}\right) T_3} \right] + O(s_z^2),
\label{mrtex3d}
\eea
where $T_3$ and $T_4$ are  defined as,
\bea
T_3=\left|\chi\right|^2\, s_y^4 - 2\,\left|\chi\right|\,c_\t\,s_y^2\,c_{2y} + c_{2y}^2, \nn \\
T_4=\left|\chi\right|^2\, c_y^4 + 2\,\left|\chi\right|\,c_\t\,c_y^2\,c_{2y} + c_{2y}^2,
\label{deft34}
\eea
While the Majorana phases as,
\bea
\r  &\approx& {1\over 2}\,\arctan{
\left[
{\left|\chi\right|^2\,c_y^2\,s_y^2\,s_{2\d}  - \left|\chi\right|\,c_{2y}\,\left(2\, c_y^2\,s_{2\d}\,c_\t - s_{2\,\d + \t}\right)
- s_{2\d}\,c_{2y}^2 \over
\left|\chi\right|^2\,c_y^2\,s_y^2\,c_{2\d}  - \left|\chi\right|\,c_{2y}\,\left(2 c_y^2\,c_{2\d}\,c_\t - c_{2\,\d + \t}\right)
- c_{2\d}\,c_{2y}^2 }\right]} + O\left(s_z\right), \nn \\
&\approx& \d\hspace{3cm} \mbox{for small enough}\; \left|\chi\right|; \left|\chi\right| \le 0.2 , \nn
\\
\s  &\approx& {1\over 2}\,\arctan{
\left[
{\left|\chi\right|^2\,c_y^2\,s_y^2\,s_{2\d}  - \left|\chi\right|\,c_{2y}\,\left(2 c_y^2\,s_{2\d}\,c_\t - s_{2\,\d + \t}\right) - s_{2\d}\,c_{2y}^2 \over
\left|\chi\right|^2\,c_y^2\,s_y^2\,c_{2\d}  - \left|\chi\right|\,c_{2y}\,\left(2 c_y^2\,c_{2\d}\,c_\t - c_{2\,\d + \t}\right)
- c_{2\d}\,c_{2y}^2 }\right]} + O\left(s_z\right), \nn \\
&\approx& \d\hspace{3cm}\mbox{for small enough}\; \left|\chi\right|; \left|\chi\right| \le 0.2.
\label{phtex3d}
\eea

The parameters $R_\n$, mass ratio square difference $m_{23}^2 - m_{13}^2$,
$\me$ and $\mee$ can be deduced to be,
\bea
R_\n &\approx& {2 \,\left|\chi\right|\,c_{2y}\,\left(-c_\d\,s_y^2 \,\left|\chi\right| + c_{2y}\, c_{\d-\t}\right)\,s_z \over s_x\, c_x\, \left(1+s_{2y}\right)\, T_4} +
O\left(s_z^2\right),\nn\\
 m_{23}^2 - m_{13}^2 &\approx&  {2 \,\left|\chi\right|\,c_{2y}\,\left(-c_\d\,s_y^2 \,\left|\chi\right| + c_{2y}\, c_{\d-\t}\right)\,s_z
 \over s_x\, c_x\, \left(1+s_{2y}\right)\, T_4} +
O\left(s_z^2\right), \nn\\
\me &\approx& m_3\, \sqrt{{T_3 \over T_4}} \left[ 1 +  {2\, s_z\,\left|\chi\right|\,c_{2y}\,\left( \left|\chi\right|\, s_y^2 \, c_\d - c_{2y}\,c_{\d - \t}\right)
\over t_{2x}\,\left(1 + s_{2y}\right)\, T_3} \right] + O\left(s_z^2\right), \nn\\
 \mee &\approx& m_3\, \sqrt{{T_3 \over T_4}} \left[ 1 +  {2\, s_z\,\,\left|\chi\right|\,c_{2y}\,\left( \left|\chi\right|\, s_y^2 \, c_\d - c_{2y}\,c_{\d - \t}\right)
\over t_{2x} \left(1 + s_{2y}\right)\, T_3} \right] + O\left(s_z^2\right).
\label{nostex3d}
\eea
It is worthy to mention that the expansions in terms of $s_z$ for this pattern are well behaved in the sense that the expansion coefficients appearing in the
mass ratio expressions are not divergent for certain values of the mixing angles as it is the case in the C1 and C2 patterns. Therefore, the expansion can be reliably used
as a perturbative expansion in which higher order terms have negligible contribution compared to the lower ones. In this pattern, it remains forbidden for
$\t_z$ or the difference $(\t_y - {\pi\over 4})$ to vanish otherwise, as exact computations show, we would have degeneracy for $m_1$ and $m_2$ leading to vanishing $R_\n$. In contrast,
the phases $\d$ (Dirac phase) and $\t$ can attain the values zero or $\pi$ without implying vanishing $R_\n$. These findings can be easily deduced using the
approximate formula for $R_\n$ as given in Eq.~(\ref{nostex3d}). The complete degeneracy ($m_1 = m_2 =m_3$) is achieved when
$\t_y={\pi\over 4}$ and $\d ={\pi\over 2}$ which can only be checked using the exact complicated formulae for $m_{13}$ and $m_{23}$. At this particular value,
$( \t_y={\pi\over 4}, \d ={\pi\over 2})$, the zeroth order expansion coefficient, of say $m_{13}\sqrt{T4/T3}$, assumes the value of one, while the other remaining coefficients are
checked to be vanishing. The positivity of $R_\n$ and the constraint to lie within the interval $\left[0.0262,0.0397\right]$ (at $3-\s$ level)
imposes a complicated relation between $\d$ and $\t$ rather than the simple constraint of belonging to alternate (identical) semicircles in the cases C1 (C2).

The phenomenology of this pattern has many features in common with that of the pattern C1 in terms of correlations and allowed values for the parameters
as can checked from the corresponding Figs.-(\ref{22mfig1}--\ref{22mfig3}) versus (\ref{12mfig1}--\ref{12mfig3})-  and Tables~(\ref{tab2}--\ref{tab3}).
Thus, we shall not repeat the same discussions and descriptions. Rather, we mention few dissimilarities: first, the mixing angle $\t_y$ is
allowed to cover all of its admissible range even in the cases of inverted and normal hierarchies; second, the correlation between $\d$ and $\t$
is not as simple as that of belonging to opposite semicircles in the pattern C1, where the  $R_\n$'s expression allows interpreting it.

\begin{figure}[hbtp]
\centering
\begin{minipage}[l]{0.5\textwidth}
\epsfxsize=8cm
\centerline{\epsfbox{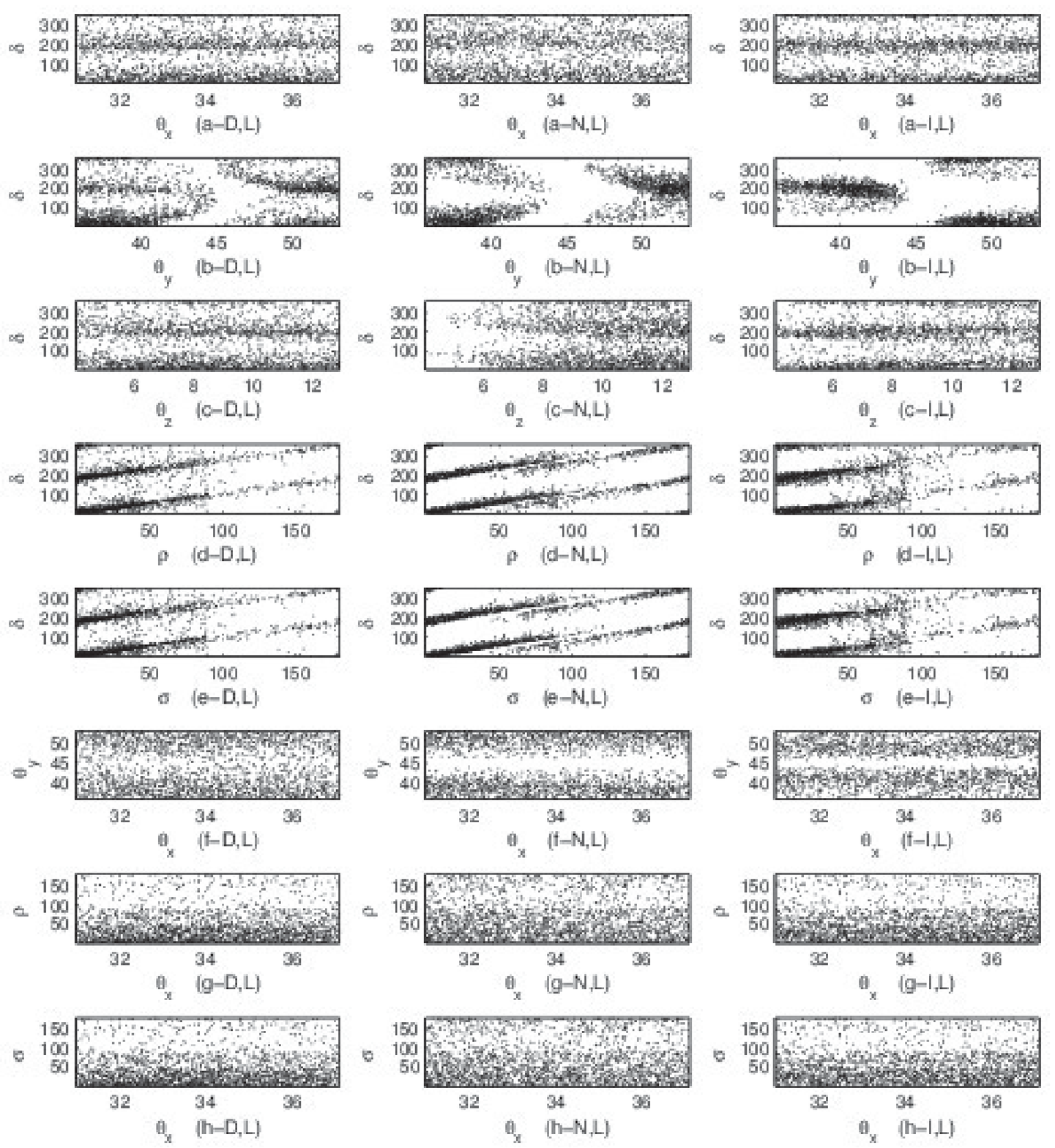}}
\end{minipage}%
\begin{minipage}[r]{0.5\textwidth}
\epsfxsize=8cm
\centerline{\epsfbox{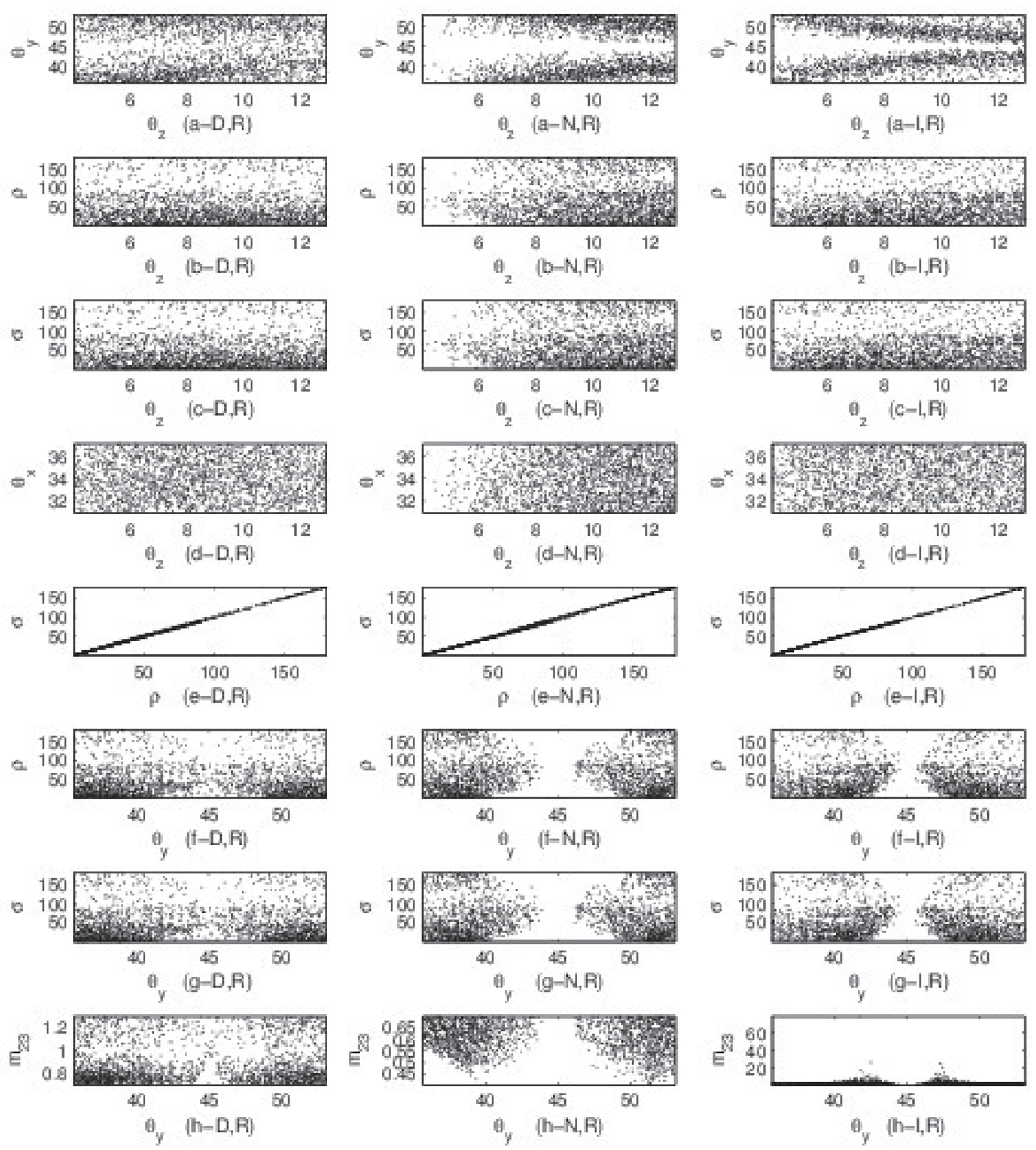}}
\end{minipage}
\vspace{0.5cm}
\caption{{\footnotesize Pattern having $ M_{\n\,12} -
M_{\n\,13}=0,\; \mbox{and}\;\; M_{\n\,22}\,\left(1+ \chi\right)- M_{\n\,33}=0$: The left panel (the
left three columns) presents correlations of $\delta$ against
mixing angles and Majorana phases ($\r$ and $\s$) and those of $\t_x$ against
$\t_y$, $\r$ and $\s$.
The right panel (the right three columns) shows the correlations of $\t_z$
against $\t_y$, $\r$ ,
$\s$, and $\t_x$ and those of $\r$ against $\s$ and $\t_y$, and also the
correlation of $\t_y$ versus
$\s$ and $m_{23}$.}}
\label{22mfig1}
\end{figure}

\begin{figure}[hbtp]
\centering
\begin{minipage}[l]{0.5\textwidth}
\epsfxsize=8cm
\centerline{\epsfbox{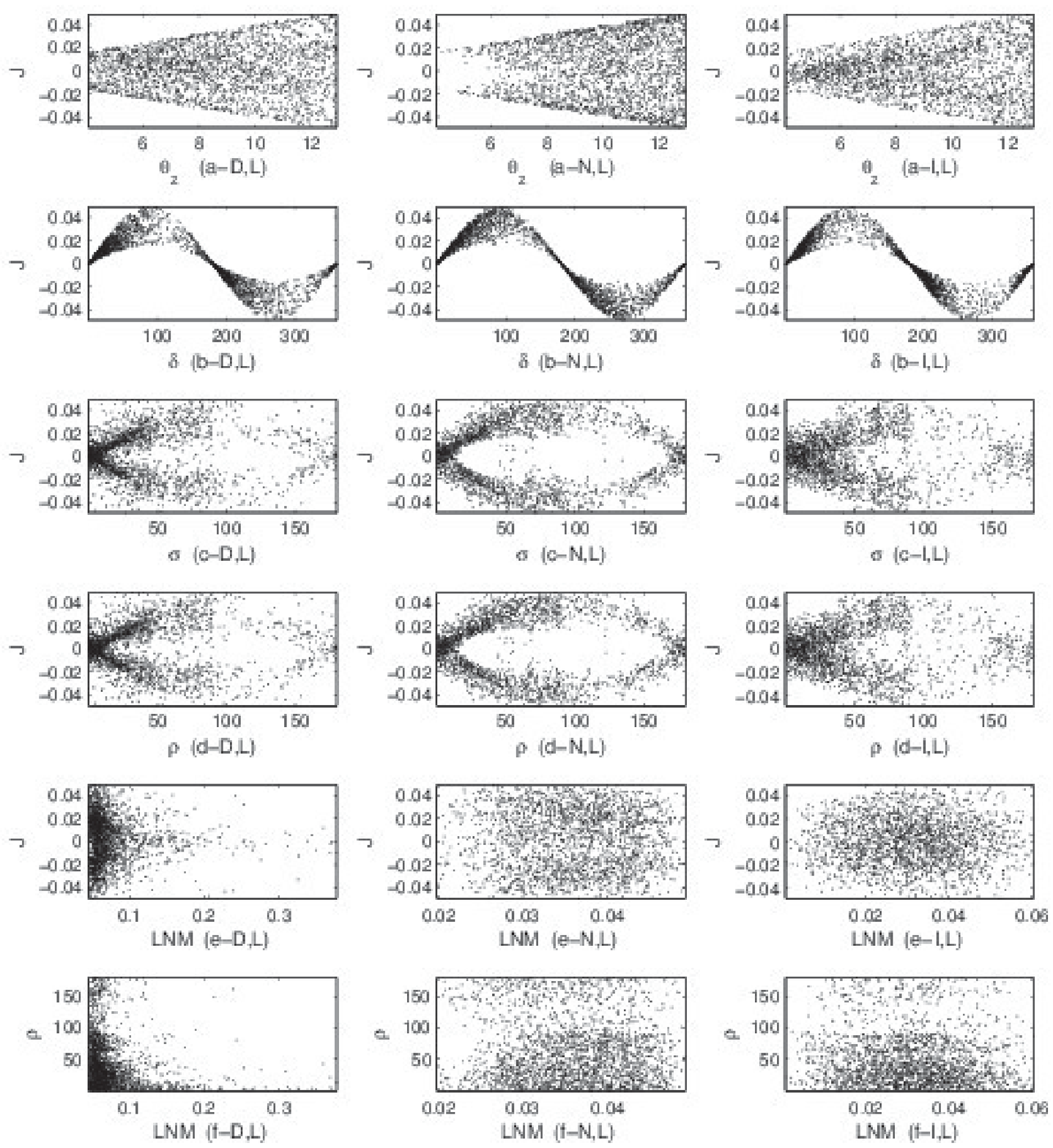}}
\end{minipage}%
\begin{minipage}[r]{0.5\textwidth}
\epsfxsize=8cm
\centerline{\epsfbox{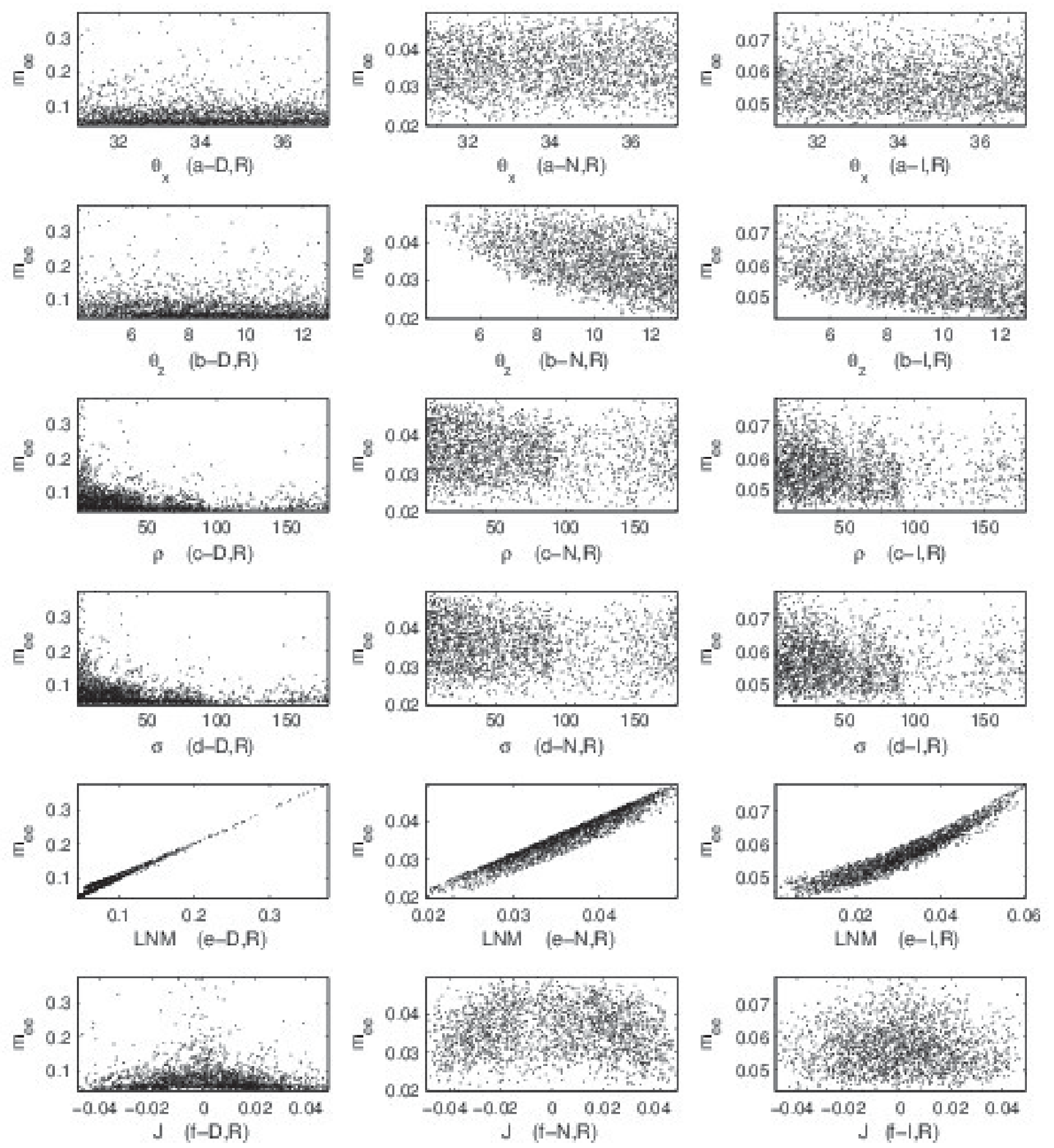}}
\end{minipage}
\vspace{0.5cm}
\caption{{\footnotesize Pattern having $ M_{\n\,12}  -
M_{\n\,13}=0,\; \mbox{and}\;\; M_{\n\,22}\,\left(1+ \chi\right) - M_{\n\,33}=0$: Left panel presents
correlations of $J$ against
$\t_z$, $\d$,  $\s$ , $\r$, and lowest neutrino mass ({\bf LNM}), while the last
one depicts the
correlation of LNM against $\r$. The right panel shows correlations of $m_{ee}$
against $\t_x$,
$\t_z$, $\r$, $\s$, {\bf LNM} and $J$. }}
\label{22mfig2}
\end{figure}
%%%%%%%%%%%%%%%%%%%%%%%%%%%%%%%%%%%%%%%%%%%%%%%%%%%%%%%%%%%%%%%%%%%%%%%%%%%%%%%%%%%%%%%%%%%%%%%%%%%%%%%%%%5
\clearpage
\begin{figure}[hbtp]
\centering
\epsfxsize=15cm
\epsfbox{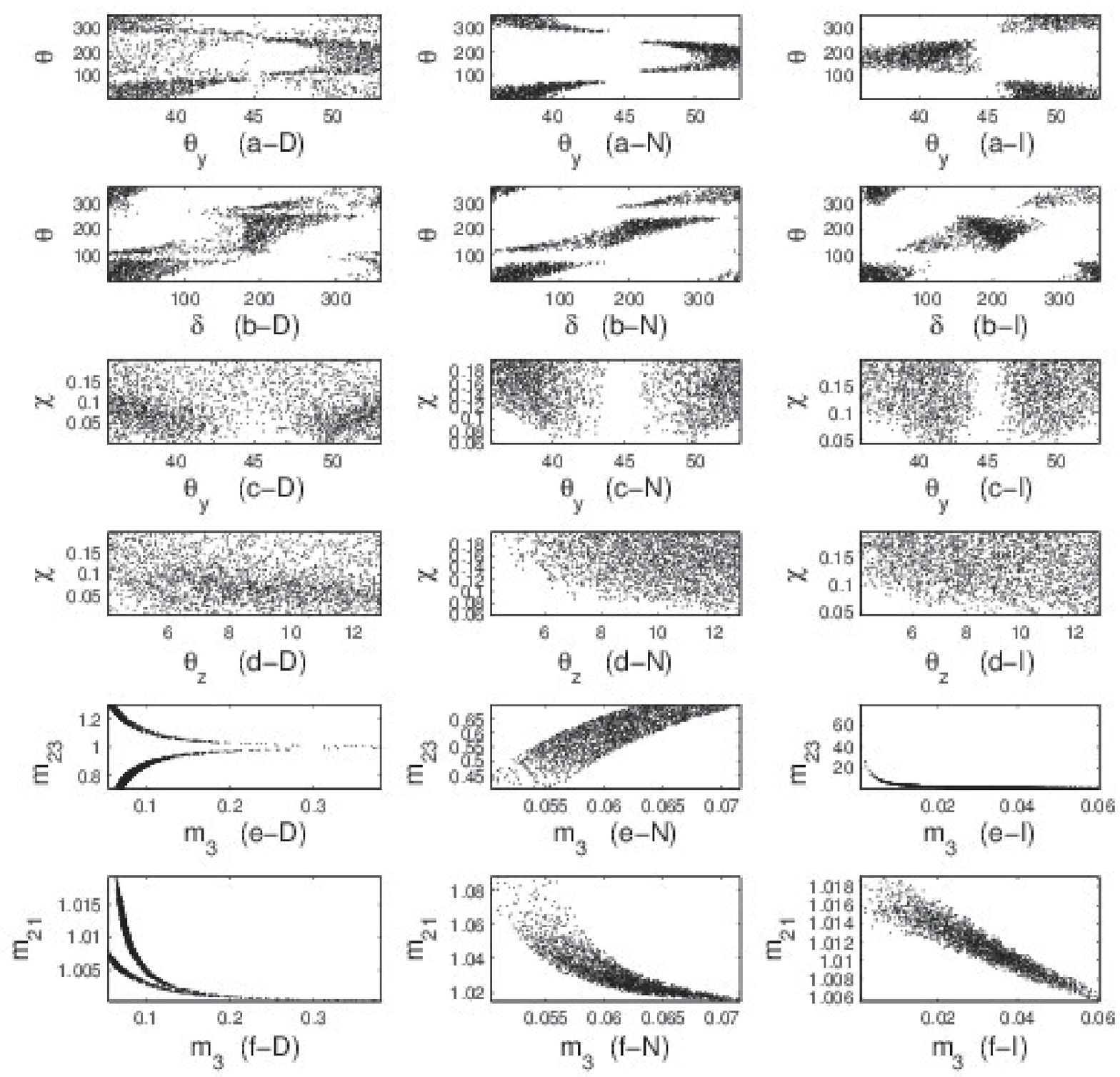}
\caption{{\footnotesize Pattern having $ M_{\n\,12}  -
M_{\n\,13}=0,\; \mbox{and}\;\; M_{\n\,22}\,\left(1+ \chi\right) - M_{\n\,33}=0$: The first two rows
presents the correlations of $\t$ against $\t_y$ and $\d$, while the second two
rows depict those of $\left|\chi\right|$ versus $\t_y$ and $\t_z$. The last two
rows shows the correlations of mass ratios $m_{23}$ and
$m_{21}$ against $m_3$.}}
\label{22mfig3}
\end{figure}
%%%%%%%%%%%%%%%%%%%%%%%%%%%%%%%%%%%%%%%%%%%%%%%%%%%%%%%%%%%%%%%%%%%%%%%%%%%%%%%%%%%%%%%%%%
%%%%%%%%%%%%%%%%% C4 Pattern M12 + M13 =0,  M22(1+chi) -M33=0
%%%%%%%%%%%%%%%%%%%%%%%%%%%%%%%%%%%%%%%%%%%%%%%%%%%%%%%%%%%%%%%%%%%%%%%%%%%%%%%%%%%%%%%%%%
\subsection{ C4: Pattern having $ M_{\n\,12} +
M_{\n\,13}=0,\; \mbox{and}\;\; M_{\n\,22}\,\left(1+ \chi\right)  - M_{\n\,33}=0.$}
In this pattern, the relevant expressions for $A$'s and $B$'s  are
\bea
  A_1 &=& -c_x c_z \left(c_x s_y s_z + s_x c_y e^{-i\,\d}\right) + c_x\,c_z \left(- c_x c_y s_z + s_x s_y e^{-i\,\d}\right), \nn\\
 A_2 & = & s_x c_z \left(- s_x s_y s_z + c_x c_y e^{-i\,\d}\right) - s_x\,c_z \left(s_x c_y s_z + c_x s_y e^{-i\,\d}\right),\nn\\
 A_3 &=& s_z c_z\,\left(s_y + c_y\right) ,\nn\\
 B_1 & =& \left(c_x s_y s_z + s_x c_y e^{-i\,\d}\right)^2\,\left(1+ \chi\right)  - \left(- c_x c_y s_z
+ s_x s_y e^{-i\,\d}\right)^2, \nn \\
 B_2 & =& \left(-s_x s_y s_z + c_x c_y e^{-i\,\d}\right)^2\,\left(1+ \chi\right) - \left(s_x c_y s_z +
c_x s_y e^{-i\,\d}\right)^2, \nn \\
 B_3 &=& s_y^2 c_z^2\,\left(1+ \chi\right)  - c_y^2 c_z^2,
 \label{abtex4d}
 \eea
leading to mass ratios, up to leading order in $s_z$, as
\bea
  m_{13}  &\approx& \sqrt{{T_3\over T_4}}\,\left[ 1 + {\left|\chi\right|\, c_{2y}\, \left(-c_\d \,s_y^2 \left|\chi\right| + c_{2y}\, c_{\d-\t}\right)\,s_z
  \over t_x\, \left( 1 - s_{2y}\right)\, T_3} \right] + O(s_z^2),  \nn\\
 m_{23} &\approx&  \sqrt{{T_3\over T_4}}\,\left[ 1 - {\left|\chi\right|\, c_{2y}\, t_x \left(-c_\d\, s_y^2 \,\left|\chi\right| + c_{2y}\, c_{\d-\t}\right)\,s_z
  \over \left( 1 - s_{2y}\right) T_3} \right] + O(s_z^2).
\label{mrtex4d}
\eea
While the Majorana phases as,
\bea
\r  &\approx& {1\over 2}\,\arctan{
\left[
{\left|\chi\right|^2\,c_y^2\,s_y^2\,s_{2\d}  - \left|\chi\right|\,c_{2y}\,\left(2\, c_y^2\,s_{2\d}\,c_\t - s_{2\,\d + \t}\right)
- s_{2\d}\,c_{2y}^2 \over
\left|\chi\right|^2\,c_y^2\,s_y^2\,c_{2\d}  - \left|\chi\right|\,c_{2y}\,\left(2 c_y^2\,c_{2\d}\,c_\t - c_{2\,\d + \t}\right)
- c_{2\d}\,c_{2y}^2 }\right]} + O\left(s_z\right), \nn
\\
\s  &\approx& {1\over 2}\,\arctan{
\left[
{\left|\chi\right|^2\,c_y^2\,s_y^2\,s_{2\d}  - \left|\chi\right|\,c_{2y}\,\left(2 c_y^2\,s_{2\d}\,c_\t - s_{2\,\d + \t}\right)
- s_{2\d}\,c_{2y}^2 \over
\left|\chi\right|^2\,c_y^2\,s_y^2\,c_{2\d}  - \left|\chi\right|\,c_{2y}\,\left(2 c_y^2\,c_{2\d}\,c_\t - c_{2\,\d + \t}\right)
- c_{2\d}\,c_{2y}^2 }\right]} + O\left(s_z\right).
\label{phtex4d}
\eea

The parameters $R_\n$, mass ratio square difference $m_{23}^2 - m_{13}^2$,
$\me$ and $\mee$ can be deduced to be,
\bea
R_\n &\approx& {2 \,\left|\chi\right|\,c_{2y}\,\left(+c_\d\,s_y^2 \,\left|\chi\right| - c_{2y}\, c_{\d-\t}\right)\,s_z \over s_x\, c_x\, \left(1-s_{2y}\right)\, T_4} +
O\left(s_z^2\right),\nn\\
 m_{23}^2 - m_{13}^2 &\approx&  {2 \,\left|\chi\right|\,c_{2y}\,\left(+c_\d\,s_y^2 \,\left|\chi\right| - c_{2y}\, c_{\d-\t}\right)\,s_z
 \over s_x\, c_x\, \left(1 - s_{2y}\right)\, T_4} +
O\left(s_z^2\right), \nn\\
\me &\approx& m_3\, \sqrt{{T_3 \over T_4}} \left[ 1 -  {2\, s_z\,\left|\chi\right|\,c_{2y}\,\left( \left|\chi\right|\, s_y^2 \, c_\d - c_{2y}\,c_{\d - \t}\right)
\over t_{2x}\,\left(1 - s_{2y}\right)\, T_3} \right] + O\left(s_z^2\right), \nn\\
 \mee &\approx& m_3\, \sqrt{{T_3 \over T_4}} \left[ 1 -  {2\, s_z\,\,\left|\chi\right|\,c_{2y}\,\left( \left|\chi\right|\, s_y^2 \, c_\d - c_{2y}\,c_{\d - \t}\right)
\over t_{2x} \left(1 - s_{2y}\right)\, T_3} \right] + O\left(s_z^2\right).
\label{nostex4d}
\eea
Once again, and as it was for the two patterns C1 and C2, one can find the same interrelations between C3  and C4 where the results (formulae) of C4 can be
derived from those of  C3, by simply making the substitutions $s_y \rightarrow - s_y$ and $\d \rightarrow \d + \pi$. Another time, the found relations cannot be
used in a useful way to derive the predictions of one pattern from the other because the mapping $s_y  \rightarrow -s_y$ does not keep the physically admissible
region of  $\t_y$  invariant. Furthermore, we are ill-fated that the properties regarding boundedness of the expansion coefficients of the mass ratios are mapped
so that the bounded coefficient at $(\t_y = {\pi\over 4}, \d ={\pi\over 2})$ in the pattern C3 may become divergent in the case of  C4. This becomes clear by looking at the expressions in Eq.~(\ref{mrtex4d}), where the
zeroth order expansion coefficient, for say $m_{13} \sqrt{T_4/T_3}$,  assumes the value one, and the first order coefficient is convergent at  $(\t_y ={\pi\over 4}, \d={\pi \over 2})$
, whereas all higher order expansion
coefficients are divergent at this point while they were vanishing in the C3 pattern. This finding is consistent with the infinite number of divergent terms
summing up to a smooth function as was discussed in Section~(7.1).
The divergence for $R_\n$ expansion is starting from the second order coefficient in harmony with the corresponding behaviour in the patterns C1 and C2.
Using the exact expression of $R_\n$ corresponding to this pattern shows that the mixing angle $\t_y$ is allowed to be exactly ${\pi \over 4}$ without forcing $R_\n$ to vanish.
The phases $\d$ and $\t$ can assume also any arbitrary values, but we should note that the point $(\t_y ={\pi\over 4}, \d={\pi \over 2})$  causes the exact form of $R_\n$ to be null. It is obvious that vanishing $\t_z$ leads also to vanishing $R_\n$,
but this choice is already  excluded by data. As was the case in the C3 pattern,  the correlation between $\d$ and $\t$ that emerges from the positivity of $R_\n$ and its allowed
range cannot, due to the complicated expression of $R_\n$ that involves complicated dependence on phases
even at the approximate level, be  described in a simple manner. We stress again that the expansion should be dealt and interpreted with caution in case
of divergent coefficients and cannot be reliably used as perturbative expansion. Thus to avoid these kinds of problems, our numerical results are based on exact expressions that
do not suffer from divergences.

We checked when we spanned the parameter space that the normal hierarchy could accommodate the data only at the $3-\s$ error level, whereas the inverted hierarchy could do
it at the $2-3\, \s$  error levels, and the  degenerate hierarchy could survive at all error levels.  The figures (\ref{22pfig1}, \ref{22pfig2}  and \ref{22pfig3})
show the corresponding correlation plots, with the same conventions as in the previous patterns. The appearance of the normal hierarchy  only at the $3-\s$ error level makes
it so special, and it turns out to be quite restrictive in the sense that the mixing angle $\t_y$ is severely bounded to be around
two possible values, namely,  $36^0$ or $52^0$, whereas $\t_z$ has only one narrow band close to $4^0$, while the Dirac phase $\d$ covers almost all its range excluding the
region\, $]158^0 - 188.4^0[$. Moreover, in this normal hierarchy case the parameter $\chi$, parameterizing the deviation from exact $\mu$--$\tau$ symmetry, cannot assume
an arbitrary value in its prescribed range:  $\left| \chi \right|$ must be in the range $[0.16-0.2]$, whereas  the phase $\t$ can cover all its allowable range excluding
the region $]19.47^0 - 139.9^0[\, \bigcup\, ]217.4^0 - 340.8^0[$.

Once again, there is a close resemblance between the pattern C4 and C2 in terms of correlations and allowed values for the parameters,
as can be checked respectively from the corresponding Figs.-(\ref{22pfig1}--\ref{22pfig3}) versus (\ref{12pfig1}--\ref{12pfig3})-
and Tables~(\ref{tab2}--\ref{tab3}). Therefore it is not necessary to repeat the same discussions and descriptions but rather we focus on
the few dissimilarities: First, the mixing angle $\t_y$ is allowed to cover all of its admissible range in the inverted hierarchy type, and in particular the
value ${\pi \over 4}$  which is excluded with its small neighborhood in the pattern C2; second, the Dirac phase $\d$ is allowed to cover all of
its ranges in the inverted and degenerate hierarchy types without any exclusion as was the case in the pattern C2 concerning the  values $(0, \,\mbox{and}\; \,\pi)$ together with their neighborhoods;
third, the mixing angle $\t_z$ tends to have a far more restrictive range in case of the pattern C4 compared to that of C2; fourth,  the normal hierarchy case for the pattern C4,
as explained above,  represents
an exceptional situation, which was not the case in the pattern C2. The figures depicting the  correlations for the two patterns C2 and C4 look, more or less,  similar
provided the loose restrictions on $\t_y$ and $\d$ associated with the pattern C4 are taken into consideration.

\begin{figure}[hbtp]
\centering
\begin{minipage}[l]{0.5\textwidth}
\epsfxsize=8cm
\centerline{\epsfbox{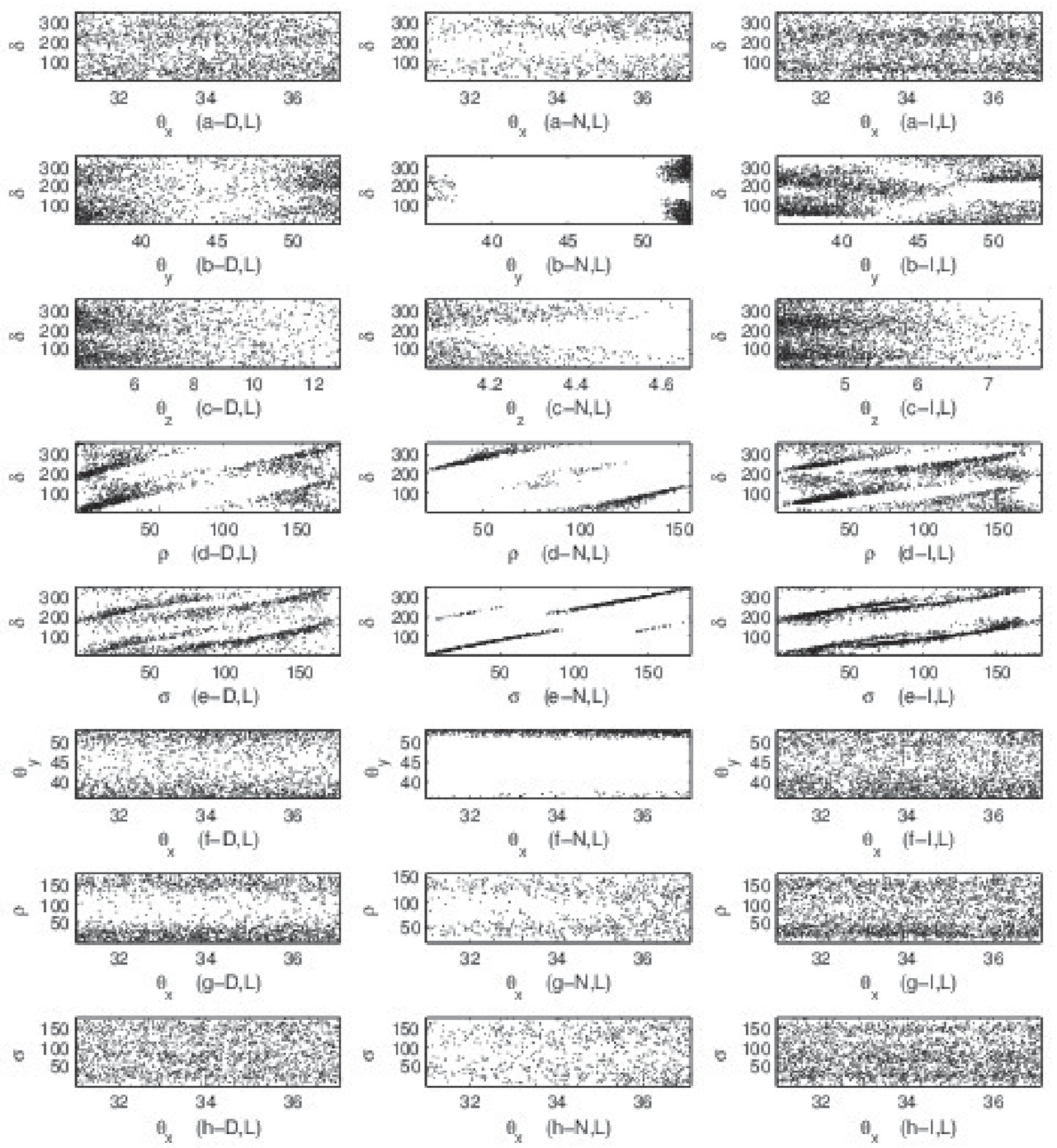}}
\end{minipage}%
\begin{minipage}[r]{0.5\textwidth}
\epsfxsize=8cm
\centerline{\epsfbox{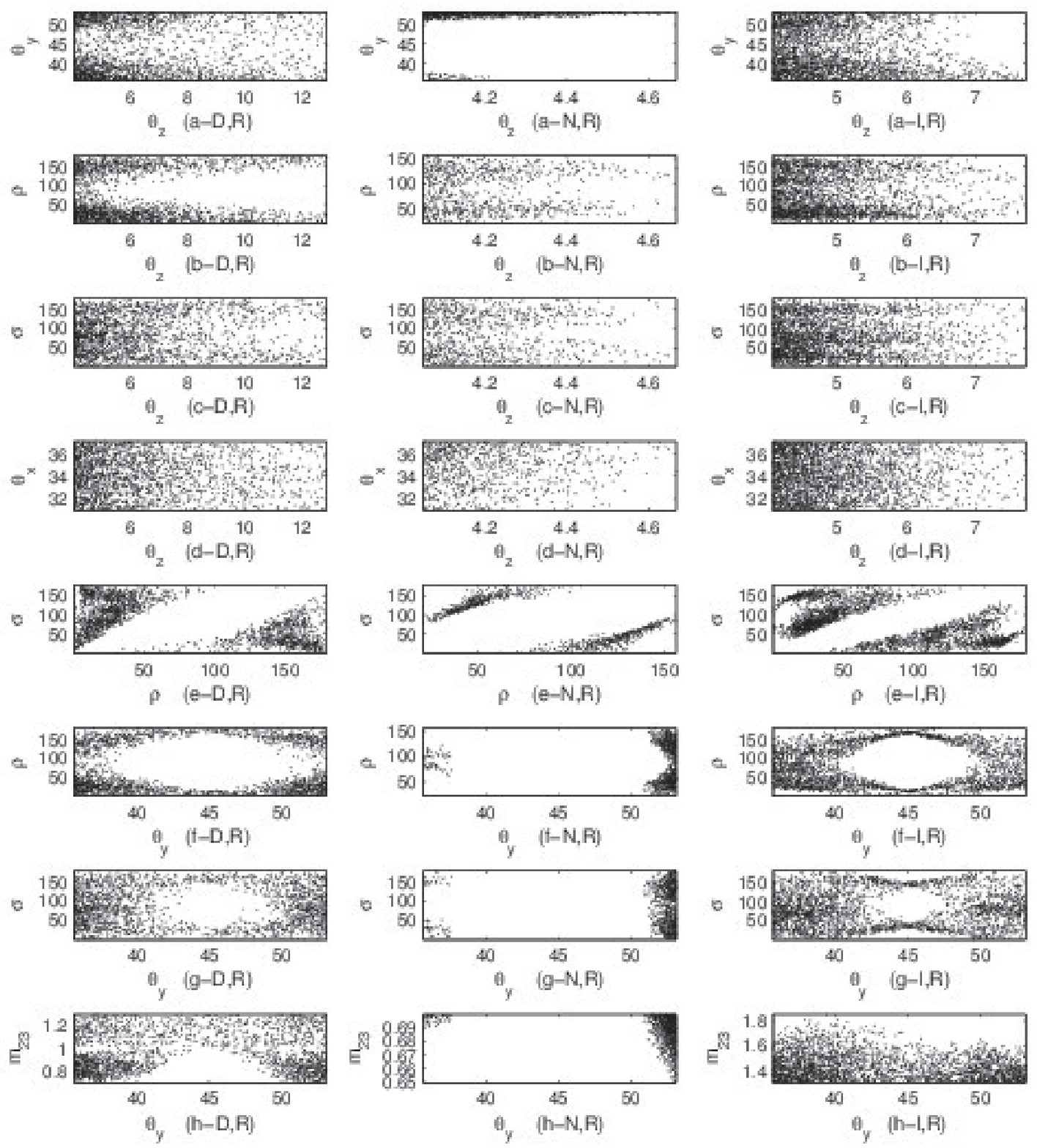}}
\end{minipage}
\vspace{0.5cm}
\caption{{\footnotesize Pattern having $ M_{\n\,12} +
M_{\n\,13}=0,\; \mbox{and}\;\; M_{\n\,22}\,\left(1+ \chi\right)- M_{\n\,33}=0$: The left panel (the
left three columns) presents correlations of $\delta$ against
mixing angles and Majorana phases ($\r$ and $\s$) and those of $\t_x$ against
$\t_y$, $\r$ and $\s$.
The right panel (the right three columns) shows the correlations of $\t_z$
against $\t_y$, $\r$ ,
$\s$, and $\t_x$ and those of $\r$ against $\s$ and $\t_y$, and also the
correlation of $\t_y$ versus
$\s$ and $m_{23}$.}}
\label{22pfig1}
\end{figure}

\begin{figure}[hbtp]
\centering
\begin{minipage}[l]{0.5\textwidth}
\epsfxsize=8cm
\centerline{\epsfbox{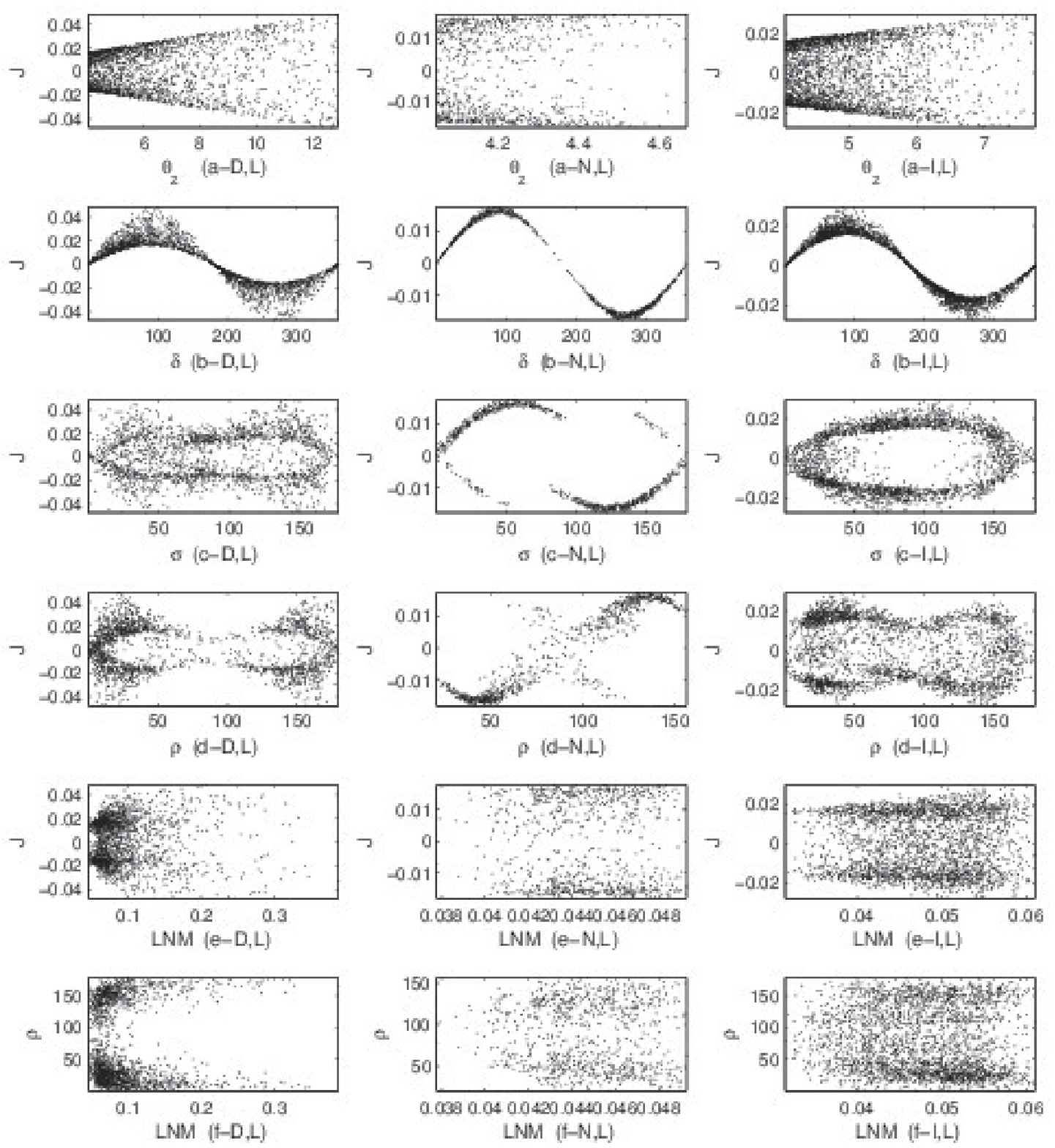}}
\end{minipage}%
\begin{minipage}[r]{0.5\textwidth}
\epsfxsize=8cm
\centerline{\epsfbox{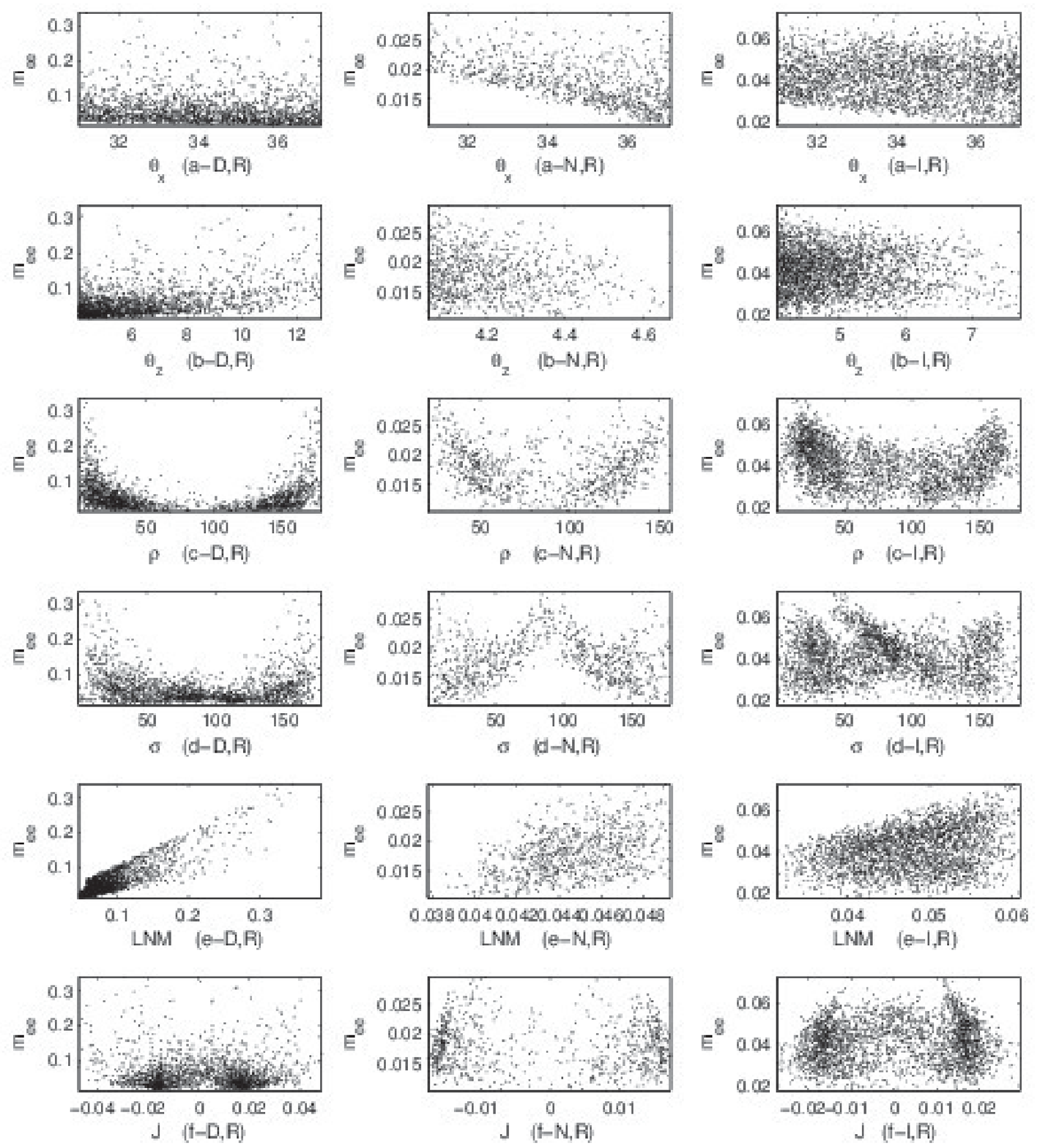}}
\end{minipage}
\vspace{0.5cm}
\caption{{\footnotesize Pattern having $ M_{\n\,12}  +
M_{\n\,13}=0,\; \mbox{and}\;\; M_{\n\,22}\,\left(1+ \chi\right) - M_{\n\,33}=0$: Left panel presents
correlations of $J$ against
$\t_z$, $\d$,  $\s$ , $\r$, and lowest neutrino mass ({\bf LNM}), while the last
one depicts the
correlation of LNM against $\r$. The right panel shows correlations of $m_{ee}$
against $\t_x$,
$\t_z$, $\r$, $\s$, {\bf LNM} and $J$. }}
\label{22pfig2}
\end{figure}
\clearpage
\begin{figure}[hbtp]
\centering
\epsfxsize=15cm
\epsfbox{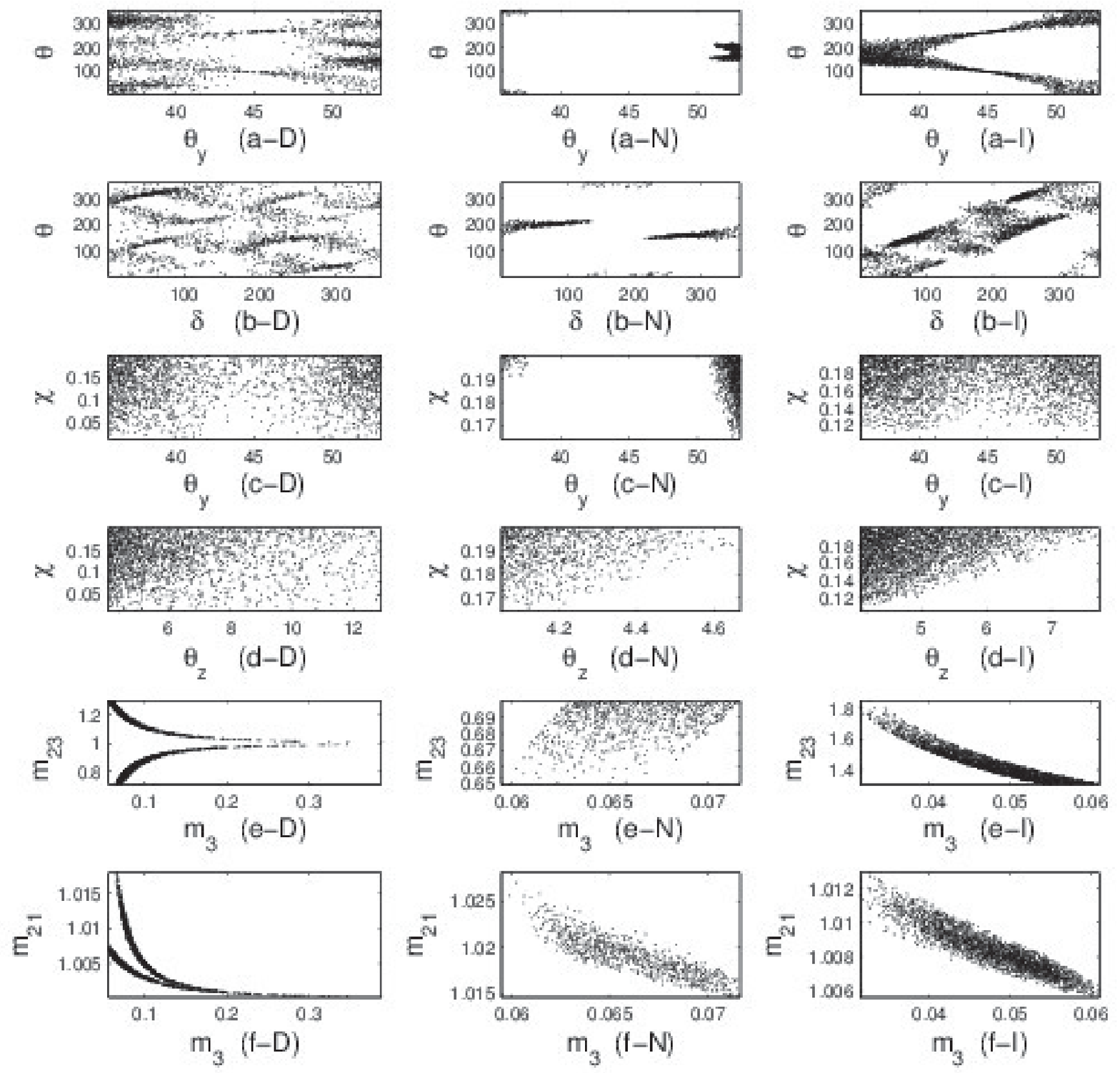}
\caption{{\footnotesize Pattern having $ M_{\n\,12}  +
M_{\n\,13}=0,\; \mbox{and}\;\; M_{\n\,22}\,\left(1+ \chi\right) - M_{\n\,33}=0$: The first two rows
presents the correlations of $\t$ against $\t_y$ and $\d$, while the second two
rows depict those of $\left|\chi\right|$ versus $\t_y$ and $\t_z$. The last two
rows shows the correlations of mass ratios $m_{23}$ and
$m_{21}$ against $m_3$.}}
\label{22pfig3}
\end{figure}
%%%%%%%%%%%%%%%%%%%%%%%%%%%%%%%%%%%%%%%%%%%%%%%%%%%%%%%%%%%%%%%%%
%%%%%%%%%%%%%%%%%%%%%%%%%%%%%%%%%%%%%%%%%%%%%%%%%%%%%%%%%%%%%%%%%
%%%%%%%%%%%%%%%%%%%%%%%%%%%%%%%%%%%%%%%%%%%%%%%%%%%%%%%%%
\begin{landscape}
\begin{table}[h]
 \begin{center}
\scalebox{0.7}{
{\tiny
 \begin{tabular}{c|c|c|c|c|c|c|c|c|c|c|c|c}
%%%%%%%%%%%%%%%%%%%%%%%%%%%%%%%%%%%%%%%%%%%%%%%%%%%%%%%%%%%%%%%%%%%%%%%%%%%%%%%
%%%%%%%%%%%%%%%%%%%%%%%%%%%%%%%%%%%%%%%%%%%%%%%Model M12 (1+x)- M13, M22 - M33%%%%%%%%%
%%%%%%%%%%%%%%%%%%%%%%%%%%%%%%%%%%%%%%%%%%%%%%%%%%%%%%%%%%%%%%%%%%%%%%%%%%%%%%%
 \hline
 \hline
\multicolumn{13}{c}{\mbox{Pattern:} $M_{\n\,12}\,\left(1+ \chi\right)  -
M_{\n\,13}=0,\; \mbox{and}\;\; M_{\n\,22} - M_{\n\,33}=0$} \\
\hline
 \mbox{quantity} & $\t_x$ & $\t_y$& $\t_z$ & $m_1$ & $m_2$& $m_3$ & $\r$ &
$\s$ & $\d$ & $\me$
 & $\mee$ & $J$\\
 \hline
 \multicolumn{13}{c}{\mbox{Degenerate  Hierarchy}} \\
 \cline{1-13}
 $1\, \sig$ &$32.96 - 35.00$ & $38.77 - 44.99$ & $ 7.71 - 10.30$ & $0.0470 -
0.3975$ & $0.0478 - 0.3976$ &
  $0.0583 - 0.3971$& $0.1910-  177.00$ & $0.1915 - 176.96$&$[0.3653
-176.6]\bigcup [180.8 -358.27]$ & $0.0479 - 0.3975$
  &$0.0448 - 0.3939 $ & $-0.0402 - 0.0402$ \\
 \hline
 $2\, \sig$ & $31.95 - 36.09$ & $36.88 - 50.77$& $6.29 - 11.68$ & $0.0463 -
0.3942$ & $0.0471 - 0.3943$ &
 $0.0568 - 0.3970$&$ 0.2341 -  178.17$ & $0.2670 - 178.16$ & $[0.7333
-173.3]\bigcup [180.3- 357.99]$& $0.0470 - 0.3942$ &
 $0.0429 - 0.3938$ & $-0.0459 - 0.0444$  \\
 \hline
 $3\, \sig$ &$30.98 - 37.11$ &$36.96 -52.01$ & $4.08 -12.92$ & $0.0457- 0.3947$
& $0.0465- 0.3948$ &
  $0.0557 - 0.3975$& $0.1981 - 179.55 $ & $0.2046 - 179.46$ & $[0.1882 -176.7]
\bigcup [180.6 - 359.79]$ &
  $0.0463-  0.3949$ & $0.0411 - 0.3947$ & $-0.0502 -  0.0506$ \\
 \hline
 %%%%%%%%%%%%%%%%%%%%%%%%%%%%%%%%%%%%%%%%%%%%%%%%%%%%%%%%%%%%%%%%%
 \multicolumn{13}{c}{\mbox{Normal  Hierarchy}} \\
 \cline{1-13}
 $1\, \sig$ &$32.96 - 35.00$ & $44.29 - 44.96$ & $ 7.71 - 10.30$ & $0.0163-
0.0471$ & $0.0186 - 0.0479$ &
  $ 0.0510 - 0.0686$& $9.71-  167.30$ & $9.77 - 167.1$&$[14.44 - 167.1]\bigcup
[188 - 354.00]$ & $0.019 - 0.0481$
  &$0.0151 - 0.0476 $ & $[-0.0406 - -0.0041] \bigcup [0.0079  - 0.0404]$ \\
 \hline
 $2\, \sig$ & $31.95 - 36.09$ &$ [44.03 - 44.95] \bigcup [45.05 - 46.07]$ &
$6.29 - 11.68 $ &$0.0129 -0.0483$ &$0.0155 - 0.0491$
 &$ 0.0497 - 0.0703$ & $7.36 - 171.71$ & $ 7.32 - 171.56$ & $[3.22 -
166.9]\bigcup [188.2- 346.46]$& $0.0166 - 0.0496$  &
 $0.0115 - 0.0485$ & $[-0.0456 - -0.0061] \bigcup [0.0021 -  0.0457]$  \\
 \hline
 $3\, \sig$ &$30.98- 37.10$ & $[43.87 -44.98]\bigcup [45.04 - 46.30]$ & $4.11
-12.92$& $0.0124-0.0490$ & $0.0151 - 0.0498$ &
  $ 0.0485 - 0.0714$& $4.48 - 175.92$ & $4.88 -175.83$ &$[8.71 -173.5]\bigcup
[190.1-357.69]$ & $0.0168 - 0.050$ &
  $ 0.0107 -  0.0496$ & $[-0.0504 - -0.0019] \bigcup [0.0053 -  0.050]$ \\
 \hline
 %%%%%%%%%%%%%%%%%%%%%%%%%%%%%%%%%%%%%%%%%%%%%%%%%%%%%%%%%%%%%%%%%%%%%%%%
 \multicolumn{13}{c}{\mbox{Inverted  Hierarchy}} \\
 \cline{1-13}
 $1\, \sig$ &$32.96 - 35.00$ & $43.89 - 44.97$ & $7.71 - 10.30$ & $0.0463
-0.0783$ &
 $0.0471 - 0.0787$ & $7.4\times 10^{-4} - 0.0602$ & $0.2721 - 179.84$ & $0.0356
- 179.49$ & $[2.87-117.4] \bigcup [235.6-357.6]$ &
  $0.0459 - 0.0779$& $ 0.0452 - 0.0779$ & $  [-0.0403- -0.0017]\bigcup [0.0020 -
0.0402]$ \\
 \hline
 $2\, \sig$ &$31.95  - 36.08$ & $[43.57 -  44.97]\bigcup [45.04 - 46.13]$ &
$6.29 -11.68$ & $0.0466-  0.0783$ &
 $0.0474 -  0.0788$ & $8.48\times 10^{-4}- 0.0601$ & $0.0617 -179.40$ & $0.0771-
179.81$ & $[7.11-174.1] \bigcup [185.7-356.34]$ &$ 0.0461 - 0.0780$& $ 0.0453 -
0.0775$ & $ [-0.0453 - -0.0020] \bigcup [0.0033 - 0.0456]$  \\
 \hline
 $3\, \sig$ & $30.98 -37.11$ & $[43.46 -44.98] \bigcup [45.02 - 46.35]$ & $4.05
-12.92$ & $0.0452- 0.0802$ &
 $ 0.0460 - 0.0806$ & $3.2\times 10^{-4} -0.0617$ & $0.8583 -179.39$ & $0.5892
-179.73$ & $[6.60-172.4] \bigcup [188.7-352.83]$ &
  $0.0445 -  0.0796$ & $ 0.0436 - 0.0784$& $ [-0.0501 - -0.0036] \bigcup [0.0036
- 0.0504]$  \\
 \hline
%%%%%%%%%%%%%%%%%%%%%%%%%%%%%%%%%%%%%%%%%%%%%%%%%%%%%%%%%%%%%%%%%%%%%%%%%%%%%%%
%%%%%%%%%%%%%%%%%%%%%%%%%%%%%%%%%%%%%%%%%%%%%%%%Model M12 (1+x)+ M13, M22 - M33%%
%%%%%%%%%%%%%%%%%%%%%%%%%%%%%%%%%%%%%%%%%%%%%%%%%%%%%%%%%%%%%%%%%%%%%%%%%%%%%%%
 \hline
\multicolumn{13}{c}{\mbox{Pattern} $M_{\n\,12}\,\left(1+ \chi\right)  +
M_{\n\,13}=0,\; \mbox{and}\;\; M_{\n\,22} - M_{\n\,33}=0$} \\
\hline
 \mbox{quantity} & $\th_x$ & $\th_y$& $\th_z$ & $m_1$ & $m_2$& $m_3$ & $\r$ &
$\sig$ & $\d$ & $\me$
 & $\mee$ & $J$\\
 \hline
 \multicolumn{13}{c}{\mbox{Degenerate  Hierarchy}} \\
 \cline{1-13}
 $1\, \sig$ &$ 32.96 -35.00$ & $38.65 -45.91$ & $7.71 -10.30$ & $0.0475  -
0.3950$ & $0.0483 - 0.3951$ &
  $0.0579 - 0.3979$& $[0.1509 - 40.42] \bigcup [136.9 - 179.95]$ & $0.5750  -
179.29$&$[2.83 - 164.7] \bigcup [199.5 - 356.94]$ & $0.0482 - 0.3950$
  &$0.0193 - 0.3947$ & $-0.0396 - 0.0406$ \\
 \hline
 $2\, \sig$ & $31.95  - 36.10$ & $36.87 - 50.77$& $6.29 - 11.68$ & $0.0471 -
0.3959$ & $0.0479 - 0.3960$ &
 $0.0579 - 0.3927$&$[0.0045 - 88.53]\bigcup [111.5 - 179.95]$ & $0.5585 -
179.45$ & $[1.96 - 174.7]\bigcup [189.9 - 352.1]$& $0.0477 - 0.3958$ &
 $0.0155 - 0.3958$ & $[-0.0453 - -0.004] \bigcup [0.001 - 0.0448]$  \\
 \hline
 $3\, \sig$ &$30.98 - 37.11$ &$35.67  - 53.10$ & $4.05 - 12.92$ & $0.0454 -
0.3947$ & $0.0462 - 0.3948$ &
  $0.0554 - 0.3980$& $[0.0064 - 93.2] \bigcup [99.53 - 179.90]$ & $0.6741 -
179.36$ & $[4.71 -167.8] \bigcup[188 - 350.9698]$ &
  $ 0.0459 - 0.3949$ & $ 0.0148 - 0.3941$ & $[-0.0496 - -0.0034] \bigcup [
0.0019 - 0.0492]$ \\
 \hline
 %%%%%%%%%%%%%%%%%%%%%%%%%%%%%%%%%%%%%%%%%%%%%%%%%%%%%%%%%%%%%%%%%
 \multicolumn{13}{c}{\mbox{Normal  Hierarchy}} \\
 \cline{1-13}
 $1\, \sig$ &$32.98 - 34.99$& $40.85 - 42.05$ &$7.71 - 8.16$ &$0.0444 - 0.0474$
&$0.0452 - 0.0482$ &
  $0.0655 - 0.0689$& $[5.01 - 23.17]\bigcup [156.9 - 177.81]$ &$[41.78 -
74.46]\bigcup [100.3 - 137.8190]$ & $[15.28 -78.84] \bigcup [279 - 352.81]$ &$
0.0451 - 0.0481$ &
  $0.0175 - 0.0298$ & $[-0.0304 - -0.0123] \bigcup  [0.008 -0.0302]$  \\
 \hline
 $2\, \sig$ & $31.95 - 36.09$ &$[40.70 - 43.12]\bigcup [46.45 - 50.31]$ & $6.29
- 9.89$ &$0.0345 - 0.0485$ &$0.0356 - 0.0493$
 &$0.0586 - 0.0704$ & $[0.1354 - 59.76]\bigcup [121.2 - 179.89]$ & $ 18.58 -
162.63$ & $[12.42 - 177.6]\bigcup [185.9 -345.12]$& $0.0353 - 0.0493$ &
 $0.0120  - 0.0406$ & $[-0.0346 - -0.003]\bigcup [0.005 - 0.0371]$  \\
 \hline
 $3\, \sig$ &$ 30.98-  37.11$ & $[40.88 -44.26] \bigcup [45.52 - 50.43]$ & $4.05
- 9.87$& $0.0246 - 0.0495$ & $0.0260 - 0.0502$ &
  $0.0521  - 0.0718$& $[0.0144 - 89.4] \bigcup [112.4 - 179.44]$ & $4.49 -
173.68$ &$[10.79 - 167.3] \bigcup [187.6 - 353.32]$ & $ 0.0253 - 0.0500$ &
  $0.0067 - 0.0453$ & $[-0.0374 - -0.0024]\bigcup [0.0045 - 0.0347]$ \\
 \hline
 %%%%%%%%%%%%%%%%%%%%%%%%%%%%%%%%%%%%%%%%%%%%%%%%%%%%%%%%%%%%%%%%%%%%%%%%
 \multicolumn{13}{c}{\mbox{Inverted  Hierarchy}} \\
 \cline{1-13}
 $1\, \sig$ & $32.96 - 35.00$  &$38.65 - 43.46$ &$7.71 - 10.30$ &$0.0551 -
0.0784$ &$0.0558 - 0.0789$&
 $0.0294 - 0.0603$ &$[3.54 - 19.71] \bigcup [160.2 - 176.77]$ &$[14.78 - 69.1]
\bigcup [109.7 165.11]$ &$[18.51 - 121.2] \bigcup  [236.7 -343.59]$ &$0.0550 -
0.0781$ &$0.0289  - 0.0717$ &$[-0.0400 - -0.01] \bigcup [0.01 - 0.0398]$  \\
 \hline
 $2\, \sig$ &$31.95- 36.09$ & $[36.89 - 43.81] \bigcup [46.3 - 50.77]$ & $6.29 -
11.67$ & $0.0526 - 0.0784$ &
 $0.0534  - 0.0788$ & $0.0248  - 0.0602$ & $[0.4268 - 28.53] \bigcup [153.5 -
177.64]$ & $9.58 - 168.19$ &$[6.25 - 157.6] \bigcup [196.6 - 347.4449]$ &$0.0526
- 0.0779$& $ 0.0199 - 0.0724$ & $[-0.0439 - -0.007]\bigcup [0.003 - 0.0438]$  \\
 \hline
 $3\, \sig$ & $30.98 -37.11$ & $[35.7 - 44.39] \bigcup [45.57  - 53.13]$ & $4.05
- 12.84$ & $0.0468 - 0.0797$ &
 $0.0476 - 0.0802$ & $0.0118  - 0.0612$ & $[0.1457 - 48.96] \bigcup [137.7  -
179.93]$ & $4.47 - 170.68$ &$[4.56 - 162.9] \bigcup [190.2 - 341.24]$ &
  $0.0469 - 0.0798$ & $0.0145 - 0.0730$& $ [-0.0488 - -0.0038] \bigcup [0.0019 -
0.0470]$  \\
 \hline
 %%%%%%%%%%%%%%%%%%%%%%%%%%%%%%%%%%%%%%%%%%%%%%%%%%%%%%%%%%%%%%%%%%%%%%%%%%%%%%%
%%%%%%%%%%%%%%%%%%%%%%%%%%%%%%%%%%%%%%%%%%%%%%%%Model  M_12 - M_13=0, M_22 (1+ chi) -M_33=0%%%%%
%%%%%%%%%%%%%%%%%%%%%%%%%%%%%%%%%%%%%%%%%%%%%%%%%%%%%%%%%%%%%%%%%%%%%%%%%%%%%%%
 \hline
\multicolumn{13}{c}{\mbox{Pattern:} $M_{\n\,12}\, - M_{\n\,13}=0,\;
\mbox{and}\;\; M_{\n\,22}\,\left(1+ \chi\right)  - M_{\n\,33}=0$} \\
\hline
 \mbox{quantity} & $\t_x$ & $\t_y$& $\t_z$ & $m_1$ & $m_2$& $m_3$ & $\r$ &
$\s$ & $\d$ & $\me$
 & $\mee$ & $J$\\
 \hline
 \multicolumn{13}{c}{\mbox{Degenerate  Hierarchy}} \\
 \cline{1-13}
 $1\, \sig$ &$32.96 -35$ & $38.65 - 44.848$ & $ 7.71 - 10.30$ & $0.0472 - 0.3790$
&$ 0.0480 - 0.3791$ &
  $ 0.0579 - 0.3822$& $0.0149 - 179.30$ & $0.0169 - 179.29$ &$0.0484 - 359.94$ &
$ 0.0480 -  0.3791$
  &$ 0.0447 - 0.3718$ & $ -0.0398 -  0.0398$ \\
 \hline
 $2\, \sig$ & $31.95 - 36.09$ & $[36.87 - 44.88] \bigcup [45.13 - 50.77]$& $6.29 -11.68$ & $0.0465 - 0.3951$
 &$ 0.0473 - 0.3952$ &
 $0.0574 - 0.3921$&$ 0.0305 - 179.84$ & $ 0.0546 - 179.84$ & $ 0.0702 - 359.88$&
$ 0.0472  - 0.3950$ &
 $ 0.0435  -  0.3949$ & $-0.0442 - 0.0447$  \\
 \hline
 $3\, \sig$ &$ 30.98 - 37.11$ & $[35.67 -44.93] \bigcup [45.08 - 53.1295]$ & $ 4.06 - 12.92$& $ 0.0453 - 0.3777$ & $ 0.0462 - 0.3778$ &
  $ 0.0556  -  0.3810$& $0.0191 - 180$ & $ 0.0192 - 180$ &$0.0257 - 359.86$ &
$ 0.0463  - 0.3779$ &
  $  0.0421 -  0.3761$ & $ -0.0488 - 0.0487$ \\
 \hline
 %%%%%%%%%%%%%%%%%%%%%%%%%%%%%%%%%%%%%%%%%%%%%%%%%%%%%%%%%%%%%%%%%
 \multicolumn{13}{c}{\mbox{Normal  Hierarchy}} \\
 \cline{1-13}
 $1\, \sig$ &$32.96 - 35$& $ 38.65 - 43.72$ &$7.72 - 10.30$ &$ 0.0259 - 0.0473$ &$ 0.0272 - 0.0481$ &
  $0.0550 - 0.0689$& $ 0.2156 - 179.98$ &$ 0.0009 - 179.97$ & $ [0.1287 - 172.7] \bigcup [193.3 - 359.1856]$ &$0.0277 - 0.0481$ &
  $ 0.0256  - 0.0479$ & $ -0.0393 - 0.0399$  \\
 \hline
 $2\, \sig$ &$31.95 - 36.09$& $[36.88 -44.04]  \bigcup [46.1 - 50.77]$ &$6.30 -11.68$ &$ 0.0223  - 0.0481$ &$ 0.0239 - 0.0489$ &
  $0.0531 - 0.0701$& $ 0.0070 - 179.98$ &$0.0522 - 179.95$ & $ [0.0141 -171.5] \bigcup [181.7 - 359.94]$ &$0.0247  - 0.0493$ &
  $ 0.0216 - 0.0490$ & $ -0.0452  - 0.0456$  \\
 \hline
 $3\, \sig$ &$ 30.98 - 37.11$& $[35.67 - 43.87]\bigcup [46.17 - 53.12]$ &$4.08 -  12.92$ &$ 0.0198 - 0.0492$ &$ 0.0216 - 0.0500$ &
  $ 0.0503 - 0.0715$& $ 0.0615 - 180$ &$0.0269 -  179.99$ & $0.1590 - 359.92$ &$0.0228 - 0.0503$ &
  $0.0199  - 0.0498$ & $-0.0493 - 0.0492$  \\
 \hline
 %%%%%%%%%%%%%%%%%%%%%%%%%%%%%%%%%%%%%%%%%%%%%%%%%%%%%%%%%%%%%%%%%%%%%%%%
 \multicolumn{13}{c}{\mbox{Inverted  Hierarchy}} \\
 \cline{1-13}
 $1\, \sig$ &$32.96 - 35.00$ & $38.65 - 44.36$ & $7.71 - 10.30$ & $0.0464 - 0.0776$ &
 $ 0.0472  - 0.0781$ & $ 0.0008  - 0.0592$ & $  0.0224 - 179.77$ & $0.0452 - 179.73$
& $59.89 -  281.52$ & $0.0461 - 0.0774$& $ 0.0455  - 0.0773$ & $ -0.0390 - 0.0396$  \\
 \hline
 $2\, \sig$ &$31.95 - 36.09$ & $[36.89 - 44.32] \bigcup [45.64 - 50.77]$ & $ 6.30 - 11.68$ & $ 0.0463 - 0.0777$ &
 $ 0.0471 - 0.0782$ & $ 0.0019 -   0.0598$ & $ 0.1213 - 179.93$ & $ 0.0469 - 179.96$ &
$0.1177 - 359.95$ &$ 0.0458  - 0.0776$& $ 0.0448  -  0.0775$ & $-0.0439  - 0.0445$  \\
 \hline
 $3\, \sig$ & $30.98 - 37.10$ & $[35.70 - 44.48] \bigcup [45.63 - 53.13]$ & $4.05 -12.92$ & $0.0453 - 0.0790$ &
 $ 0.0462  - 0.0794$ & $ 0.0006 - 0.0604$ & $ 0.0132 - 179.99$ & $ 0.0244 - 179.84$ &
$ 0.0079 -  359.91$ &$0.0448 - 0.0788$ & $ 0.0436 - 0.0782$& $ -0.0486 -   0.0492$  \\
 \hline
%%%%%%%%%%%%%%%%%%%%%%%%%%%%%%%%%%%%%%%%%%%%%%%%%%%%%%%%%%%%%%%%%%%%%%%%%%%%%%%%
%%%%%%%%%%%%%%%%%%%%%%%%%%%%%%%%%%%%%%%%%%%%%%%Model  M_12 + M_13=0, M_22 (1+ chi) -M_33=0%%%%%%%%%%%
%%%%%%%%%%%%%%%%%%%%%%%%%%%%%%%%%%%%%%%%%%%%%%%%%%%%%%%%%%%%%%%%%%%%%%%%%%%%%%%
 \hline
\multicolumn{13}{c}{\mbox{Pattern:} $M_{\n\,12}\, + M_{\n\,13}=0,\;
\mbox{and}\;\; M_{\n\,22}\,\left(1+ \chi\right)  - M_{\n\,33}=0$} \\
\hline
 \mbox{quantity} & $\th_x$ & $\th_y$& $\th_z$ & $m_1$ & $m_2$& $m_3$ & $\r$ &
$\sig$ & $\d$ & $\me$
 & $\mee$ & $J$\\
 \hline
 \multicolumn{13}{c}{\mbox{Degenerate  Hierarchy}} \\
 \cline{1-13}
 $1\, \sig$ &$32.96 -35$ & $38.65- 44.98$ & $7.71 -10.30$ & $0.0757 - 0.3966$
&$  0.0755  - 0.3965$ & $ 0.0293  - 0.3962$& $[0.0844- 40.68] \bigcup [135.6 - 179.67]$ & $1.29 - 177.65$ &$0.2316 -  359.73$ &
$0.0745 - 0.3954$
  &$0.0483 - 0.3617$ & $  -0.0397  -  0.0395$ \\
 \hline
 $2\, \sig$ & $31.95 - 36.09$ & $36.87 - 50.77$& $6.29 - 11.68$ & $ 0.0658 - 0.3955$ &$ 0.0664 - 0.3956$ &
 $0.0574 - 0.3926$&$ [0.0111 - 63.82] \bigcup [113  - 179.59]$ & $  1.38 - 176.99$ & $ 0.4530  - 359.73$&
$ 0.0663 - 0.3955$ &
 $0.0231 - 0.3628$ & $ -0.0446 - 0.0443$  \\
 \hline
 $3\, \sig$ &$30.99 -  37.10$ & $35.67 - 53.13$ & $4.05 -12.90$& $ 0.0456 - 0.3902$ & $ 0.0464 - 0.3903$ &
  $ 0.0561 - 0.3875$& $0.1229 - 179.71$ & $  0.1735 - 177.30$ &$0.4436 - 359.90$ &
$ 0.0460 - 0.3902$ &
  $  0.0138 - 0.3377$ & $  -0.0474 - 0.0484$ \\
 \hline
 %%%%%%%%%%%%%%%%%%%%%%%%%%%%%%%%%%%%%%%%%%%%%%%%%%%%%%%%%%%%%%%%%
 \multicolumn{13}{c}{\mbox{Normal  Hierarchy}} \\
 \cline{1-13}
 $1\, \sig$ &$\times$& $\times$ &$\times$ &$\times$ &$\times$ &
  $\times$& $\times$ &$\times$ & $\times$ &$\times$ &
  $\times$ & $\times$  \\
 \hline
 $2\, \sig$ &$\times$& $\times$ &$\times$ &$\times$ &$\times$ &
  $\times$& $\times$ &$\times$ & $\times$ &$\times$ &
  $\times$ & $\times$  \\
 \hline
 $3\, \sig$ &$ 30.99 -  37.11$& $ [35.68 - 37.61] \bigcup [50.89 - 53.13]$ &$4.05 - 4.67$ &$ 0.0378  - 0.0493$ &$0.0388  - 0.0501$ &
  $0.0595 - 0.0717$& $ 20.85 -156.35$ &$0.3957 - 178.58$ & $[0.7022 - 158] \bigcup [188.4 - 358.43]$ &$ 0.0383 - 0.0497$ &
  $0.0104 - 0.0297$ & $ -0.0179 - 0.0176$  \\
 \hline
 %%%%%%%%%%%%%%%%%%%%%%%%%%%%%%%%%%%%%%%%%%%%%%%%%%%%%%%%%%%%%%%%%%%%%%%%
 \multicolumn{13}{c}{\mbox{Inverted  Hierarchy}} \\
 \cline{1-13}
 $1\, \sig$ &$\times$& $\times$ &$\times$ &$\times$ &$\times$ &
  $\times$& $\times$ &$\times$ & $\times$ &$\times$ &
  $\times$ & $\times$  \\
 \hline
 $2\, \sig$ &$31.95 - 36.09$ & $ 36.87 - 50.77$ & $ 6.29 - 8.07$ & $ 0.0652 - 0.0787$ &
 $ 0.0657 - 0.0792$ & $ 0.0458 - 0.0605$ & $ [0.1822 - 90.22] \bigcup [94.89 - 178.27]$ & $ 0.0532 - 179.63$ &
$0.1149 - 354.39$ & $0.0651  - 0.0787$& $ 0.0203  -  0.0635$ & $-0.0291  -  0.0295$  \\
 \hline
 $3\, \sig$ & $30.98 - 37.11$ & $35.68 -53.12$ & $4.05 - 7.73$ & $ 0.0554 - 0.0795$ &
 $ 0.0561 - 0.0800$ & $0.0314 - 0.0611$ & $0.1136 - 179.52$ & $0.0084 - 179.96$ &
$ 0.2029 - 359.92$ & $ 0.0555 - 0.0796$ & $0.0175 -0.0725$& $ -0.0274 - 0.0295$  \\
 \hline
 \end{tabular}
 }}
 \end{center}
 \caption{\small  The various prediction for the patterns of
  violating exact $\mu$--$\tau$ symmetry. All
the angles (masses) are
  evaluated in degrees ($eV$).}
  \label{tab2}
 \end{table}
\end{landscape}

%%%%%%%%%%%%%%%%%%%%%%%%%%%%%%%%%%%%%%%%%%%%%%%%%%%%%%%%%%%%%%%%%%%%%%%%%%%%%%
%%%%%%%%%%%%%%%%%%Second TAble%%%%%%%%%%%%%%%%%%%%%%%%%%%%%%%%%%%%%%%%%%%
%%%%%%%%%%%%%%%%%%Second Table%%%%%%%%%%%%%%%%%%%%%%%%%%%%%%%%%%%%%%%%%%%
%\begin{landscape}
\begin{table}[h]
 \begin{center}
\scalebox{1}{
{\tiny
 \begin{tabular}{c|c|c|c|c|c}
 \hline
 \hline
\multicolumn{6}{c}{$\mbox{Pattern:} M_{\n\,12}\,\left(1+ \chi\right)  - M_{\n\,13}=0,\; \mbox{and}\;\; M_{\n\,22} - M_{\n\,33}=0$} \\
\hline
\multicolumn{3}{c|}{ $\left|\chi\right|$} & \multicolumn{3}{c}{ $\t$} \\
\cline{1-6}
$1\, \s$ & $2\, \s$ & $3\, \s$ &  $1\, \s$  & $2\, \s$ & $3\, \s$  \\
\hline
\multicolumn{6}{c}{\mbox{Degenerate  Hierarchy}} \\
\cline{1-6}
$0.0023 - 0.2$ & $0.0030 - 0.2$ & $0.0047 - 0.2$ & $0.85 - 359.6$ & $0.75 - 359.12$ & $0.82 - 359.2$ \\
 \hline
 \multicolumn{6}{c}{\mbox{Normal   Hierarchy}} \\
 \cline{1-6}
 $0.0398 - 0.2$ & $0.0434 - 0.2$ & $0.0378 -0.2$ & $\left[3.43 - 91\right] \cup \left[269.9 - 351.77\right]$ &
 $\left[11.74 - 90.57\right] \cup \left[101 - 172.5\right] \cup$& $\left[8.73 - 89.7\right] \cup \left[104 - 176\right]\cup$
\\
   &    &  &  & $\left[188.4 - 263.4\right] \cup \left[277.5 - 358.13\right]$ & $\left[186 -262\right] \cup \left[273.5 - 352.6\right]$ \\
 \hline
\multicolumn{6}{c}{\mbox{Inverted   Hierarchy}} \\
 \cline{1-6}
 $0.0309 - 0.2 $ & $0.020 - 0.2 $& $0.0276  -0.2$ & $\left[13.51 - 1272.2\right] \cup \left[188.1 - 349.7\right]$ &
 $\left[8.48 - 172.9\right] \cup \left[188 - 349 \right]$ &  $\left[8.55 - 172.7\right] \cup \left[185.8 - 350.3\right]$  \\
 %%%%%%%%%%%%%%%%%%%%%%%%%%%%%%%%%%%%%%%%%%%%%%%%%%%%%%%%%%%%%%%%%%%%%%%%%%%%%%%%%%%%%%%%%%%%%%%%%%%%%%%%%%%%%%%%
\hline
\hline
\multicolumn{6}{c}{$\mbox{Pattern:} M_{\n\,12}\,\left(1+ \chi\right)  + M_{\n\,13}=0,\; \mbox{and}\;\; M_{\n\,22} - M_{\n\,33}=0$} \\
\hline
\multicolumn{3}{c|}{ $\left|\chi\right|$} & \multicolumn{3}{c}{ $\t$} \\
\cline{1-6}
$1\, \s$ & $2\, \s$ & $3\, \s$ &  $1\, \s$  & $2\, \s$ & $3\, \s$  \\
\hline
\multicolumn{6}{c}{\mbox{Degenerate  Hierarchy}} \\
\cline{1-6}
$0.0066 - 0.2$ & $0.0085 -0.2  $ & $ 0.0049 - 0.2 $ & $\left[0.075 - 50.24\right] \cup \left[59.11 - 77.88\right] \cup$ &
$\left[0.20 - 76.48\right] \cup \left[108.4 - 242.2\right] \cup $ & $\left[0.1 - 70.91\right] \cup \left[85.6 - 254.1\right] \cup $
 \\
  &     &   &  $\left[129.4 - 233.3\right] \cup \left[288.4 - 359.9\right]$ & $\left[300.8 - 359.78\right]$ & $\left[276.5 - 359.74\right]$ \\
 \hline
 \multicolumn{6}{c}{\mbox{Normal   Hierarchy}} \\
 \cline{1-6}
 $0.1889 - 0.2$ & $0.14 - 0.2$ & $0.1 -0.2$ & $\left[0.48 - 2.17\right] \cup \left[357.7 - 359.8\right]$ &
 $\left[0.43 - 3.1\right] \cup \left[176.7 - 183.4\right] \cup$& $\left[0.33 - 4.9\right] \cup \left[173.2 - 185\right]\cup$
\\
   &    &  &  &  $\left[356.5 -359.54\right] $ & $\left[355.1 - 359.75\right]$  \\
 \hline
\multicolumn{6}{c}{\mbox{Inverted   Hierarchy}} \\
 \cline{1-6}
 $0.0992 - 0.2 $ & $0.0814 - 0.2$ & $0.06 - 0.2$ & $\left[ 176.08 - 179.6\right] \cup \left[180.4 - 184.12\right]$ &
 $\left[0.57 - 4.98\right] \cup \left[175.6 - 184.4 \right] \cup$ &  $\left[0.33- 6.6\right] \cup \left[174.3 - 186.8\right] \cup$
 \\
    &            &     &  & $\left[355.6 - 359.65\right]$ & $ \left[353.8 - 359.87\right]$\\
%%%%%%%%%%%%%%%%%%%%%%%%%%%%%%%%%%%%%%%%%%%%%%%%%%%%%%%%%%%%%%%%%%%%%%%%%%%%%%%%%%%%%%%%%%%%%%%%%%%%%%%%%%%%%%%%
\hline
\hline
\multicolumn{6}{c}{$\mbox{Pattern:} M_{\n\,12} -  M_{\n\,13}=0,\; \mbox{and}\;\; M_{\n\,22}\,\left(1+ \chi\right) - M_{\n\,33}=0$} \\
\hline
\multicolumn{3}{c|}{ $\left|\chi\right|$} & \multicolumn{3}{c}{ $\t$} \\
\cline{1-6}
$1\, \s$ & $2\, \s$ & $3\, \s$ &  $1\, \s$  & $2\, \s$ & $3\, \s$  \\
\hline
\multicolumn{6}{c}{\mbox{Degenerate  Hierarchy}} \\
\cline{1-6}
$0.0017 - 0.2$ & $0.0016 -0.2  $ & $ 0.0025 - 0.2 $ & $0.30 - 359.75$ &
$0.08 - 359.84$ & $0.15 - 359.65 $
 \\
\hline
 \multicolumn{6}{c}{\mbox{Normal   Hierarchy}} \\
 \cline{1-6}
 $0.0729 - 0.2$ & $0.0658 - 0.2$ & $0.0587 -0.2$ & $\left[0.17 - 74.5\right] \cup \left[285.6 - 359.12\right]$ &
 $\left[0.40 - 74.9\right] \cup \left[114.3 - 247.4\right] \cup$& $\left[0.08 - 72.7\right] \cup \left[112.4 - 246.9\right]\cup$
\\
   &    &  &  &  $\left[286.3 -359.68\right] $ & $\left[285.4 - 359.54\right]$  \\
 \hline
\multicolumn{6}{c}{\mbox{Inverted   Hierarchy}} \\
 \cline{1-6}
 $0.0393 - 0.2 $ & $0.0345 - 0.2$ & $0.0426 - 0.2$ & $110.1 - 251.33$ &
 $\left[0.13 - 77.84\right] \cup \left[112.5 - 250.5 \right] \cup$ &  $\left[0.21- 76.57\right] \cup \left[108.3 - 247.6\right] \cup$
 \\
    &            &     &  & $\left[283.3 - 359.87\right]$ & $ \left[283.7 - 359.87\right]$\\
 %%%%%%%%%%%%%%%%%%%%%%%%%%%%%%%%%%%%%%%%%%%%%%%%%%%%%%%%%%%%%%%%%%%%%%%%%%%%%%%%%%%%%%%%%%%%%%%%%%%%%
 \hline
\hline
\multicolumn{6}{c}{$\mbox{Pattern:} M_{\n\,12} + M_{\n\,13}=0,\; \mbox{and}\;\; M_{\n\,22}\,\left(1+ \chi\right) - M_{\n\,33}=0$} \\
\hline
\multicolumn{3}{c|}{ $\left|\chi\right|$} & \multicolumn{3}{c}{ $\t$} \\
\cline{1-6}
$1\, \s$ & $2\, \s$ & $3\, \s$ &  $1\, \s$  & $2\, \s$ & $3\, \s$  \\
\hline
\multicolumn{6}{c}{\mbox{Degenerate  Hierarchy}} \\
\cline{1-6}
$0.0112 - 0.2$ & $0.0079 -0.2  $ & $ 0.0115 - 0.2 $ & $1.53 - 359.94$ &
$0.61 - 358.42$ & $0.82 - 359.45 $
 \\
\hline
 \multicolumn{6}{c}{\mbox{Normal   Hierarchy}} \\
 \cline{1-6}
 $\times$ & $\times$ & $0.16 -0.2$ & $\times$ & $\times$& $\left[0.12 - 19.47\right] \cup \left[139.9 - 217.4\right]\cup$
\\
   &    &  &  &   & $\left[340.8 - 359.9\right]$  \\
 \hline
\multicolumn{6}{c}{\mbox{Inverted   Hierarchy}} \\
 \cline{1-6}
 $\times $ & $0.1572 - 0.2$ & $0.1047 - 0.2$ & $\times$ &
 $ 45.7 - 310.22  $ &  $ 0.03 - 360$
 \\
 \hline
 \end{tabular}
 }}
 \end{center}
  \caption{\small The allowed values for $\left|\chi\right|$  (pure number) and $\t$  for the patterns  of
  violating exact $\mu$--$\tau$ symmetry. All
the angles  are evaluated in degrees.}
\label{tab3}
 \end{table}
%\end{landscape}

\section{Singular patterns violating exact $\mu$--$\tau$ symmetry}
As was the case in the exact symmetry, the violation of exact $\mu$--$\tau$ symmetry does not allow for singular neutrino mass matrix.
The same analysis and arguments against the viability of the singular patterns having exact $\mu$--$\tau$ symmetry in section~(5) can be carried out here to
show the inviability of the various singular deformed patterns. The numerical study based on scanning all acceptable ranges for the mixing angles and the Dirac phase $\d$
assures the absence of any solution satisfying the mass ratio constraints as expressed in Eq.~(\ref{m23con}) and Eq.~(\ref{m21con}). All the relevant formulae
for mass ratios are collected in Table~(\ref{singmass2}) in order to ease judging the inviability of patterns. The $T_3$ and $T_4$ present in the formulae
are the ones defined before in Eq.~(\ref{deft34}), while $T_5$ is introduced as
\be
T_5 =  \left|\chi\right|^2\, c_y^2\,c_\d + \left|\chi\right|\,\left[c_\d\,c_\t\left( 4\,c_y^2 -1 \right) + s_\t\,s_\d\right] + 2\,c_\d\,c_{2y}.
\label{deft5}
\ee
%%%%%%%%%%%%%%%%%%%%%%%%%%%%%%%%%%%%%%%%%%%%%%%%%%%%%%%%%%%%%%%%%%%%%%%%%%%%%%%%%%%%%%%%%%%%
\clearpage
\begin{table}[hbtp]
\begin{center}
\scalebox{0.95}{
\begin{tabular}{|c|c|c|}
\hline
&\multicolumn{2}{c|}{$m_1=0$} \\
\hline
Pattern &\multicolumn{2}{c|}{$m_2\over m_3$}    \\
\hline
&&\\
C1 & $\left|{A_3\over A_2}\right|\approx \sqrt{{\left|\chi\right|^2\, s_y^2 + 2\,\left|\chi\right|\,c_\t\,s_y\left(s_y - c_y\right) + 1 - s_{2y}
\over \left|\chi\right|^2\, c_y^2 + 2\,\left|\chi\right|\,c_\t\,s_y\left(s_y + c_y\right) + 1 + s_{2y}}}\,{s_z\over s_x\,c_x} + O(s_z^2)$ &
$\left|{B_3\over B_2}\right| \approx {1\over c_x^2} \left( 1 + 2\,t_x\,t_{2y}\,c_\d\, s_z\right) + O(s_z^2)$\\
&&\\
\hline
&&\\
C2 & $\left|{A_3\over A_2}\right|\approx \sqrt{{\left|\chi\right|^2\, s_y^2 + 2\,\left|\chi\right|\,c_\t\,s_y\left(s_y + c_y\right) + 1 + s_{2y}
\over \left|\chi\right|^2\, c_y^2 + 2\,\left|\chi\right|\,c_\t\,s_y\left(c_y - s_y\right) + 1 - s_{2y}}}\,{s_z\over s_x\,c_x} + O(s_z^2)$ &
$\left|{B_3\over B_2}\right| \approx {1\over c_x^2} \left( 1 + 2\,t_x\,t_{2y}\,c_\d\, s_z\right) + O(s_z^2)$\\
&&\\
\hline
&&\\
C3 & $ \left|{A_3\over A_2}\right|\approx \sqrt{{1-s_{2y}\over 1 + s_{2y}}}\, {s_z\over s_x\,c_x} + O(s_z^2)$ &
$ \left|{B_3\over B_2}\right| \approx {1\over c_x^2}\,\sqrt{{T_3\over T_4}}\left( 1 + {t_x\,s_{2y}\,T_5\,s_z\over T_4}\right) + O(s_z^2)$\\
&&\\
\hline
&&\\
C4 & $ \left|{A_3\over A_2}\right|\approx \sqrt{{1+s_{2y}\over 1 - s_{2y}}}\, {s_z\over s_x\,c_x} + O(s_z^2)$ &
$ \left|{B_3\over B_2}\right| \approx {1\over c_x^2}\,\sqrt{{T_3\over T_4}}\left( 1 + {t_x\,s_{2y}\,T_5\,s_z\over T_4}\right) + O(s_z^2)$\\
&& \\
\hline
 & \multicolumn{2}{c|}{$m_3=0$}\\
 \hline
 Pattern &  \multicolumn{2}{c|}{$m_2\over m_1$}    \\
 \hline
&&\\
C1 & $\left|{A_1\over A_2}\right|\approx 1+ {\left|\chi\right|^2\, s_y\,c_y\,c_\d +
\left|\chi\right|\,\left[c_\d\,c_\t\left(s_{2y} - c_{2y}\right) -s_\t\,s_\d\right] - c_\d\,c_{2y}
\over \left|\chi\right|^2\, c_y^2 + 2\,\left|\chi\right|\,c_\t\,c_y\left(s_y + c_y\right) + 1 + s_{2y}}\,{s_z\over s_x\,c_x} + O(s_z^2)$ &
$ \left|{B_1\over B_2}\right| \approx t_x^2\,\left(1 + {2\,t_{2y}\,c_\d\,s_z \over s_x c_x}\right) + O(s_z^2)$\\
&&\\
\hline
&&\\
C2 & $\left|{A_1\over A_2}\right|\approx 1+ {\left|\chi\right|^2\, s_y\,c_y\,c_\d +
\left|\chi\right|\,\left[c_\d\,c_\t\left(s_{2y} + c_{2y}\right) + s_\t\,s_\d\right] + c_\d\,c_{2y}
\over \left|\chi\right|^2\, c_y^2 + 2\,\left|\chi\right|\,c_\t\,c_y\left(c_y - s_y\right) + 1 - s_{2y}}\,{s_z\over s_x\,c_x} + O(s_z^2)$ &
$ \left|{B_1\over B_2}\right| \approx t_x^2\,\left(1 + {2\,t_{2y}\,c_\d\,s_z \over s_x c_x}\right) + O(s_z^2)$\\
&&\\
\hline
&&\\
C3 & $\left|{A_1\over A_2}\right|\approx 1- {\left(1-s_{2y}\right)\,c_\d\,s_z
\over c_{2y}\,s_x\,c_x} + O(s_z^2)$ &
$ \left|{B_1\over B_2}\right| \approx t_x^2\,\left(1 + {T_5\,s_{2y}\,s_z \over T_4 s_x c_x}\right) + O(s_z^2)$\\
&&\\
\hline
&&\\
C4 & $\left|{A_1\over A_2}\right|\approx 1+ {\left(1+ s_{2y}\right)\,c_\d\,s_z
\over c_{2y}\,s_x\,c_x} + O(s_z^2)$ &
$ \left|{B_1\over B_2}\right| \approx t_x^2\,\left(1 + {T_5\,s_{2y}\,s_z \over T_4 s_x c_x}\right) + O(s_z^2)$\\
&&\\
\hline
\end{tabular}
}
\end{center}
\caption{\small  The approximate mass ratio formulae for the singular light neutrino mass violating exact $\mu$ -- $\tau$ symmetry. The
forumlae are calculated in terms of A's and B's coefficients }
\label{singmass2}
 \end{table}

%%%%%%%%%%%%%%%%%%%%%%%%%%%%%%%%%
\section{Exact $\m$-$\tau$ symmetry and realizations of the perturbed textures}

We study now in detail how the perturbed textures can arise assuming an exact $\m$-$\tau$ symmetry at the Lagrangian level but at the expense of introducing
new matter fields and symmetries.
To fix the ideas, let's take the $C1$ pattern put in the form:
\bea
M_\n & =&
\left(
\begin {array}{ccc}
A &B& B(1+\chi)\\
B& C& D\\
B(1+\chi)&D&C
\end {array}
\right).
\label{C1pattern}
\eea
The exact $\m$-$\tau$ symmetry (the $S$ symmetry) corresponding to this pattern is given by the matrix
\bea
S & =&
\left(
\begin {array}{ccc}
1 &0& 0\\
0& 0& 1\\
0&1&0
\end {array}
\right)
\label{S}
\eea
in that we have $S^2=1$ and
\bea
\label{FI1}
\left\{ \left(M=M^{\mbox{\textsc{t}}}\right)\,\wedge \left[ S^{\mbox{\textsc{t}}} \cdot M \cdot S=M \right] \right\}
& \Leftrightarrow & { \left[ \exists\, A,B,C,D: M = \pmatrix{A & B & B
\cr B & C &  D\cr
B &  D & C}
\right]},
\eea
We shall need also the following relations:
\bea
\label{FI2}
\left\{ \left(M=M^{\mbox{\textsc{t}}}\right)\,\wedge \left[ S^{\mbox{\textsc{t}}} \cdot M \cdot S=-M \right] \right\}
& \Leftrightarrow & { \left[ \exists\,B,C: M = \pmatrix{0 & B & -B
\cr B & C &  0\cr
-B &  0 & -C}
\right]},
\eea
\bea
\label{FI1g}
 \left[ S^{\mbox{\textsc{t}}} \cdot M \cdot S=M \right]
& \Leftrightarrow & { \left[ \exists\, A,B,C,D: M = \pmatrix{A & B & B
\cr E & C &  D\cr
E &  D & C}
\right]},
\eea
\bea
\label{FI2g}
 \left[ S^{\mbox{\textsc{t}}} \cdot M \cdot S=-M \right]
& \Leftrightarrow & { \left[ \exists\, B,C,D: M = \pmatrix{0 & B & -B
\cr E & C &  D \cr
-E &  -D & -C}
\right]},
\eea
\bea
\label{FI3}
 \left[  S \cdot M = M \right]
 &\Leftrightarrow&  { \left[ \exists\, A,B,C, D, E, F: M = \pmatrix{A & B & C
\cr D & E &  F\cr
D &  E & F}
\right]}
\eea
We shall achieve the texture of Eq. \ref{C1pattern} using both types II and I of the seesaw mechanism.
\subsection{Type II-seesaw}
In the type II seesaw \cite{seesaw2} mechanism, we show now how one can reach the desired form by assuming a flavor symmetry of the form $S \times Z_2$ and by having three Higgs triplets
for the neutrino mass matrix and three Higgs doublets for the charged lepton mass matrix.
\subsubsection{Matter content and symmetries}
First, we extend the SM
by
introducing three $SU(2)_L$ scalar triplets $H_a$, $(a=1,2,3)$,
\be
H_a\equiv \left[H_a^{++}, H_a^+, H_a^0\right].
\ee
In addition to the $S$ symmetry, we introduce another $Z_2$ symmetry, and we assume the following transformations:

\bea L \stackrel{S}{\longrightarrow} S L &,& L \stackrel{Z_2}{\longrightarrow} \mbox{diag}(1,-1,-1) L \label{typeIIL}\\
H \stackrel{S}{\longrightarrow} \mbox{diag}(1,1,-1) H &,& H \stackrel{Z_2}{\longrightarrow} \mbox{diag}(1,-1,-1) H \label{typeIIH}
\eea
where the $L^\textsc{t} =  (L_1,L_2,L_3), H^\textsc{t} =  (H_1,H_2,H_3)$ with $L_i$'s,$(i=1,2,3)$ are the components of the $i^{th}$-family LH lepton doublets (we shall adopt this notation of `vectors' in flavor space even for other fields, like $l^c$, $\n_R$
 and $\phi, \ldots$).
Note that the assignments of $L_2,L_3$ should be the same under $Z_2$ as the $S$ symmetry interchanges them, otherwise the factor subgroups $S$ and $Z_2$ do not commute.
For this reason, the $S$-charges of $H_2,H_3$ are allowed to be different because $Z_2$ acts on $H$ diagonally.
There will be also the RH charged lepton singlets and the Higgs fields responsible for the charged lepton mass matrix.

\subsubsection{Neutrino mass matrix}
The Yukawa interaction relevant for neutrino mass has the form,
\be
\mathcal{L}_{H,L} = \sum_{i,j=1}^3\sum_{a=1}^3\,G_{ij}^a\,\left[ H_a^0 \n_{Li}^T\, \mathcal{C}\, \n_{Lj} + H_a^+ \left(\n_{Li}^T\, \mathcal{C}\, l_{Lj}
+ l_{Lj}^T \,\mathcal{C}\, \n_{Li}\right) + H_a^{++} l_{Li}^T \,\mathcal{C}\, l_{Lj}\right],
\ee
where $G_{ij}^a$ are Yaukawa coupling constants, the indices $i,j$ are flavor ones, and $\mathcal{C}$ is the charge conjugation matrix.

The field $H_a^0$ can get a small vacuum expectation value (vev), $\langle H_a^0\rangle_0 = v_a$ leading to a Majorana neutrino mass
matrix,
\be
M_{\n\;ij} =  \sum_{a=1}^3\,G_{ij}^a\,\langle H_a^0\rangle_0.
\ee
The smallness of the vev $\langle H_a^0\rangle_0$ is due to the largeness of the triplet scalar mass scale\cite{seesaw2}.

The bilinear  of $\n_{Li}\,  \n_{Lj}$ relevant for Majorana mass matrix
transforms, via Eq. \ref{typeIIL}, under $Z_2$ as:
 \bea
 \nu_{Li}\, \nu_{Lj} \stackrel{Z_2}{\sim} B = \left(
\begin {array}{ccc}
1&-1 & -1\\
-1 & 1 & 1\\
-1 &1 &1
\end {array}
\right),&\mbox{meaning:}&
\nu_{Li}\, \nu_{Lj}  \stackrel{Z_2}{\longrightarrow} Z_2(\nu_{Li}\, \nu_{Lj} )  = B_{ij} \nu_{Li}\, \nu_{Lj}  (\mbox{no sum})
\label{biseesaw2}
\eea

Thus we have:
\bea \left. \begin{array}{c} S^{\mbox{\textsc{t}}} G^1 S = G^1, {G^1}^\textsc{t} = G^1 \\ G^1_{ij} Z_2(H_1)Z_2(\nu_{Li}\, \nu_{Lj}) = G^1_{ij} H_1 \nu_{Li}\, \nu_{Lj} (\mbox{no sum})\end{array} \right\}
 &\stackrel{\mbox{Eqs.}\ref{FI1},\ref{typeIIH},\ref{biseesaw2}}{\Longrightarrow}& G^1 = \left(
\begin {array}{ccc}
A^1 &0& 0\\
0& C^1& D^1\\
0&D^1&C^1
\end {array} \right)
 \label{G1}\eea
\bea \left. \begin{array}{c} S^{\mbox{\textsc{t}}} G^2 S = G^2 ,  {G^2}^\textsc{t} = G^2\\ G^2_{ij} Z_2(H_2)Z_2(\nu_{Li}\, \nu_{Lj}) = G^2_{ij} H_2 \nu_{Li}\, \nu_{Lj}(\mbox{no sum}) \end{array} \right\}
 &\stackrel{\mbox{Eqs.}\ref{FI1},\ref{typeIIH},\ref{biseesaw2}}{\Longrightarrow}& G^2 = \left(
\begin {array}{ccc}
0 &B^2& B^2\\
B^2& 0& 0\\
B^2&0&0
\end {array} \right)
 \label{G2}\eea
The two Higgs fields $H_1, H_2$ generate the unperturbed texture, whereas the perturbation is generated by the field $H_3$:
\bea \left. \begin{array}{c} S^{\mbox{\textsc{t}}} G^3 S = -G^3, {G^3}^\textsc{t} = G^3 \\ G^3_{ij} Z_2(H_3)Z_2(\nu_{Li}\, \nu_{Lj}) = G^3_{ij} H_3 \nu_{Li}\, \nu_{Lj} (\mbox{no sum})\end{array} \right\}
 &\stackrel{\mbox{Eqs.}\ref{FI2},\ref{typeIIH},\ref{biseesaw2}}{\Longrightarrow}& G^3 = \left(
\begin {array}{ccc}
0 &B^3& -B^3\\
B^3& 0& 0\\
-B^3&0&0
\end {array} \right)
 \label{G3}\eea
The mass matrix we get is of the form: \bea M_\n & =&
\left(
\begin {array}{ccc}
v_1 A^1 &v_2 B^2+v_3 B^3& v_2 B^2 - v_3 B^3 \\
v_2 B^2+v_3 B^3& v_1 C^1& v_1 D^1\\
v_2 B^2 - v_3 B^3& v_1 D^1& v_1 C^1
\end {array}
\right).
\label{C1patternform}
\eea
Thus if the Yukawa couplings are all of the same order while the vevs satisfy $v_2 \gg v_3 $ we get the desired form of the pattern $C1$ (Eq. \ref{C1pattern})
with $\chi = \frac{-2v_3 B^3 }{v_2 B^2 +v_3B^3}$ .

\subsubsection{Charged lepton mass matrix -- flavor basis}
We need here to extend the symmetry to the charged lepton sector and arrange the couplings in order to be in the `flavor basis' where the charged lepton mass matrix is diagonal.
For this we present  three possible options.
\begin{enumerate}
\item {\bf Just the SM Higgs}

We have the usual Yukawa coupling term
\bea
 \label{L1}
 {\cal{L}}_1 &=& Y_{ij} \overline{L}_i \Phi  l^c_j \,
 \eea
We assume the SM Higgs $\Phi$ is singlet under the flavor symmetry.
\bea \Phi \stackrel{S}{\longrightarrow} \Phi &,& \Phi \stackrel{Z_2}{\longrightarrow} \Phi \label{typeIISMPhi}\eea
and present two scenarios for the RH charged lepton singlets $l^c_j$ transformation under $S\times Z_2$ as follows.

\begin{itemize}

  \item {\bf $l^c_j$ transforms similarly as $L$}

We assume: \bea l^c \stackrel{S}{\longrightarrow} S l^c &,& l^c \stackrel{Z_2}{\longrightarrow} \mbox{diag}(1,-1,-1) l^c \label{typeIISMlc1}
\eea
We get via Eqs. (\ref{typeIIL},\ref{typeIISMPhi} and \ref{typeIISMlc1}) then: \bea S^{\mbox{\textsc{t}}} Y S = Y &,& \overline{L}_i  l^c_j
 \stackrel{Z_2}{\sim} \left(
\begin {array}{ccc}
1&-1 & -1\\
-1 & 1 & 1\\
-1 &1 &1
\end {array}
\right)
\label{typeIISMlc1invariance}
\eea
which would lead, upon acquiring a vev $v$ for the SM Higgs, to a charged lepton mass matrix of the form (see Eqs. \ref{FI1g}, \ref{typeIISMlc1invariance}):
\bea
M_l  = v
\left(
\begin {array}{ccc}
A &0& 0\\
0& C& D\\
0&D&C
\end {array}
\right) &\Rightarrow&
M_l M_l^\dagger = v^2 \left(
\begin {array}{ccc}
|A|^2 &0& 0\\
0& |C|^2+|D|^2& 2\Re (CD^*)\\
0&2\Re (CD^*)&|C|^2+|D|^2
\end {array} \right).
\label{ChargedMasstypeIISMlc1}
\eea
Thus we need to perform a rotation across the $1^{st}$-axis by an angle $\t_y=\pi/4$ in order to diagonalize the squared charged lepton mass matrix and be in the flavor basis.
Thus, this option is not interestsing since it spoils the neutrino mixing predictions carried out in the flavor basis.

  \item {\bf $l^c_j$ is singlet under flavor symmetry}

We assume: \bea l^c \stackrel{S}{\longrightarrow} l^c &,& l^c \stackrel{Z_2}{\longrightarrow}  l^c \label{typeIISMlc2}
\eea

We get via Eqs. (\ref{typeIIL},\ref{typeIISMPhi} and \ref{typeIISMlc2}) then: \bea S Y  = Y &,& \overline{L}_i  l^c_j
 \stackrel{Z_2}{\sim} \left(
\begin {array}{ccc}
1&1 & 1\\
-1 & -1 & -1\\
-1 &-1 &-1
\end {array}
\right)
\label{typeIISMlc2invariance}
\eea
which would lead, upon acquiring a vev $v$ for the SM Higgs, to a charged lepton mass matrix of the form (see Eqs. \ref{FI3}, \ref{typeIISMlc2invariance}):
\bea
M_l  = v
\left(
\begin {array}{ccc}
A &B& C\\
0& 0& 0\\
0&0&0
\end {array}
\right) &\Rightarrow&
M_l M_l^\dagger = v^2 \left(
\begin {array}{ccc}
|A|^2+|B|^2+|C|^2 &0& 0\\
0& 0& 0\\
0&0&0
\end {array} \right).
\label{ChargedMasstypeIISMlc1}
\eea
The squared mass matrix is diagonal, but it predicts two vanishing eigen masses for the $2^{\mbox{nd}}$ and $3^{\mbox{rd}}$ families which is not acceptable experimentally.

\end{itemize}

\item {\bf Three SM-like Higgs doublets}

We extend the SM to include three
scalar doublets $\phi_k$  playing the role of the ordinary
SM-Higgs field. The Lagrangian reponsible for the charged lepton mass is given by:
\bea
 \label{L2}
 {\cal{L}}_2 &=& f^j_{ik} \overline{L}_i \phi_k  l^c_j \,
 \eea
We assume the Higgs fields $\phi_k$, $k=1,2,3$ transform as $L_i$ under $S\times Z_2$:
\bea \phi \stackrel{S}{\longrightarrow} S \phi &,& \phi \stackrel{Z_2}{\longrightarrow} \mbox{diag} (1,-1,-1) \phi \label{typeII3Higssphi}\eea
Equally, the RH charged leptons are supposed to transform as singlets under $S$:
\bea l^c \stackrel{S}{\longrightarrow}  l^c  \label{typeII3HigssSlc}\eea
whereas we
present two scenarios for their transformations under $Z_2$ as follows.
\begin{itemize}

  \item {\bf $l^c_j$ transforms similarly as $L$ under $Z_2$}

We assume \bea  l^c \stackrel{Z_2}{\longrightarrow} \mbox{diag}(1,-1,-1) l^c \label{typeII3HiggsZ2lc1}
\eea
We get via Eqs. (\ref{typeIIL},\ref{typeII3Higssphi}, \ref{typeII3HigssSlc} and \ref{typeII3HiggsZ2lc1}) then:
\bea S^{\mbox{\textsc{t}}} f^{(j)} S = f^{(j)} &,& \overline{L}_i  \phi_k
 \stackrel{Z_2}{\sim} \left(
\begin {array}{ccc}
1&-1 & -1\\
-1 & 1 & 1\\
-1 &1 &1
\end {array}
\right)
\label{typeII3Higgslc1invariance}
\eea
where $f^{(j)}$ is the matrix whose $(i,k)^{th}$-entry is the Yukawa coupling $f^j_{ik}$. Then,  Eqs. (\ref{FI1g}, \ref{typeII3HiggsZ2lc1} and \ref{typeII3Higgslc1invariance}) lead
to the following forms of the Yukawa coupling matrices:
\bea
f^{(1)}  =
\left(
\begin {array}{ccc}
A^1 &0& 0\\
0& C^1& D^1\\
0&D^1&C^1
\end {array}\right) , f^{(2)}  =
\left(
\begin {array}{ccc}
0 &B^2& B^2\\
E^2& 0& 0\\
E^2&0&0
\end {array}\right), f^{(3)}  =
\left(
\begin {array}{ccc}
0 &B^3& B^3\\
E^3& 0& 0\\
E^3&0&0
\end {array}\right)\eea
If there is cute hierarchy in the vevs: $v_3 \gg v_1, v_2$, say, we get, for real entries, a charged lepton mass matrix of the form

\bea
M_l  &=& v_3
\left(
\begin {array}{ccc}
0 &B^2& B^3\\
D^1& 0& 0\\
C^1&0&0
\end {array}
\right)
\label{ChargedMasstypeIISMlc1}
\eea
We see that this choice of $Z_2$-charge assignments for the RH lepton singlets leads to one vanishing mass, which is excluded by experiment. Thus we turn to the other choice which
would prove capable of producing the charged lepton mass spectrum.
\item{\bf $l^c_j$ transforms differently from $L$ under $Z_2$}

We assume \bea  l^c \stackrel{Z_2}{\longrightarrow} \mbox{diag}(1,1,-1) l^c \label{typeII3HiggsZ2lc2}
\eea
We get the same Eq. (\ref{typeII3Higgslc1invariance}), but Eq. (\ref{typeII3HiggsZ2lc2}) leads now to:
\bea
f^{(1)}  =
\left(
\begin {array}{ccc}
A^1 &0& 0\\
0& C^1& D^1\\
0&D^1&C^1
\end {array}\right) , f^{(2)}  =
\left(
\begin {array}{ccc}
A^2 &0& 0\\
0& C^2& D^2\\
0&D^2&C^2
\end {array}\right), f^{(3)}  =
\left(
\begin {array}{ccc}
0 &B^3& B^3\\
E^3& 0& 0\\
E^3&0&0
\end {array}\right)\eea
The hierarchy ($v_3 \gg v_1, v_2$) would now lead to the following form for the charged lepton mass matrix:

\bea
M_l  = v_3
\left(
\begin {array}{ccc}
0 &0& B^3\\
D^1& D^2 & 0\\
C^1&C^2&0
\end {array}
\right) &\Rightarrow&
M_l\;M_l^\dagger
    = v_3^2
    \pmatrix {|{\bf B}|^2  & 0 & 0 \cr
     0 & |{\bf D}|^2 & {\bf D} \cdot {\bf C}\cr
    0 & {\bf C} \cdot {\bf D} & |{\bf C}|^2},
\label{ChargedMasstypeII3Higgslc2}
\eea
where ${\bf B}=(0,0,B^3)^T$, ${\bf D}=(D^1,D^2,0)^T$ and ${\bf C}=(C^1,C^2,0)^T$, and where the dot product is defined as ${\bf D} \cdot {\bf C} = \sum_{i=1}^{i=3}
D^iC^{i*}$.
Now, one can adjust the Yukawa couplings so that to require an infinitesimal rotation in order to diagonalize the squared charged lepton mass matrix and be in the flavor basis.
In fact, let us just assume
the magnitudes of the three vectors coming in ratios comparable to the lepton mass ratios:
\bea
\frac{|{\bf B}|}{|{\bf C}|} \equiv \la_e \sim \frac{m_e}{m_\tau} = 2.8 \times 10^{-4}&,&
\frac{|{\bf D}|}{|{\bf C}|} \equiv \la_\m \sim \frac{m_\m}{m_\tau} = 5.9 \times 10^{-2}, \label{ratiochargedmass}
\eea
 Then it is easy to see that the matrix:
\bea
U(\t,\a,\b)  &=&
\left(
\begin {array}{ccc}
1 &0& 0\\
0& c_\t e^{-i\a} & s_\t e^{-i\b}\\
0&-s_\t e^{-i\a}& c_\t e^{-i\b}
\end {array}
 \right)
\label{diagonalizingmatrix}: \\
\a-\b=\mbox{arg}({\bf D} \cdot {\bf C}) &,& \tan{2\t} = \frac{2{\bf D} \cdot {\bf C} }{|{\bf D}|^2 -|{\bf C}|^2} \simeq 2 \frac{|{\bf D}|}{|{\bf C}|} \cos{\psi} \label{infintesimal}
 \eea
 where $\psi$ is the angle between the two complex vectors ${\bf D}$ and ${\bf C}$, defined by $\cos\psi = {\bf D} \cdot {\bf C} /(|{\bf D}| \cdot |{\bf C}|)$, does diagonalize $M_lM_l^\dagger$. Note that one can absorb the individual phases $\a,\b$, using the freedom of multiplying the unitary diagonalizing matrix by a diagonal phase matrix, which would leave us with only one `physical' phase $\a-\b$:
\bea
U(\t,\a,\b)  &=&
\left(
\begin {array}{ccc}
1 &0& 0\\
0& c_\t  & s_\t e^{-i(\b-\a)}\\
0&-s_\t e^{i(\b-\a)}& c_\t
\end {array}
 \right)
\label{diagonalizingmatrixU}
 \eea
 Thus, we are in the flavor basis, as required, up to an infinitesimal rotation of angle less than $10^{-2}$  (See Eqs. \ref{ratiochargedmass} and \ref{infintesimal}).
\end{itemize}

\item {\bf SM plus three Higgs singlets}

One might keep the SM Higgs doublet $\Phi$, with the same flavor transformations of Eq. (\ref{typeIISMPhi}) but add three Higgs singlets $\D_k$ so that to contribute to the charged lepton mass through dimension-5 operators.
 The Lagrangian responsible for the charged lepton mass is given by:
\bea
 \label{L4}
 {\cal{L}}_4 = {\cal{L}}_1 + {\cal{L}}_3 &=& Y_{ij} \overline{L}_i \Phi  l^c_j  +  \frac{g^j_{ik}}{\Lambda} \overline{L}_i \Phi \Delta_k  l^c_j\,
 \eea
 where $\Lambda$ is a mass high scale characterizing the Higgs singlets.
We assume the Higgs singlet fields $\D_k$, $k=1,2,3$ transform as $L_i$ under $S\times Z_2$:
\bea \D \stackrel{S}{\longrightarrow} S \D &,& \D \stackrel{Z_2}{\longrightarrow} \mbox{diag} (1,-1,-1) \D \label{typeII3SingletsDelta}\eea
As in the previous enumeration,  the RH charged leptons are supposed to be singlets under $S$ (Eq. \ref{typeII3HigssSlc}), whereas for $Z_2$ we have the following options:

\begin{itemize}

  \item {\bf $l^c_j$ transforms similarly as $L$ under $Z_2$}

We have thus Eq.(\ref{typeII3HiggsZ2lc1}).  The invariance of ${\cal{L}}_1$ implies
\bea S Y  = Y &,& \overline{L}_i  l^c_j
 \stackrel{Z_2}{\sim} \left(
\begin {array}{ccc}
1&-1 & -1\\
-1 & 1 & 1\\
-1 &1 &1
\end {array}
\right)
\label{typeII3Singletslc2invariance}
\eea
This leads, when $\Phi$ acquires a vev, to a contribution to the mass matrix (see Eqs. \ref{FI3}, \ref{typeIISMlc1invariance}):
\bea
M_1  &=&
\left(
\begin {array}{ccc}
a &0& 0\\
0& e& f\\
0&e&f
\end {array}
\right)
\label{ChargedMasstypeII3SingletsM1lc1}
\eea
Eq. (\ref{typeII3SingletsDelta}) would lead, exactly as the three Higgs doublets did in the previous enumeration, to a mass contribution
$M_2$ of the form of Eq. \ref{ChargedMasstypeIISMlc1} when the Higgs singlets
acquire vevs ($\d_k$), with the hierarchy $\d_3 \gg \d_1, \d_2$. Thus we get the charged lepton mass
matrix in the form:
\bea
M_l = M_1+M_2  &=&
\left(
\begin {array}{ccc}
a &B^2& B^3\\
D^1& e & f\\
C^1&e&f
\end {array}
\right) \label{ChargedMasstypeII3Singletslc1}
\eea
with the condition that $D^1 \neq C^1$ in order not to make the determinant of the matrix equal to zero implying a vanishing mass.

 \item {\bf $l^c_j$ transforms differently from $L$ under $Z_2$}

We have thus Eq.(\ref{typeII3HiggsZ2lc2}).  The invariance of ${\cal{L}}_1$ implies:
\bea S Y  = Y &,& \overline{L}_i  l^c_j
 \stackrel{Z_2}{\sim} \left(
\begin {array}{ccc}
1&1 & -1\\
-1 & -1 & 1\\
-1 &-1 &1
\end {array}
\right)
\label{typeII3Singletslc2invariance}
\eea
so when $\Phi$ acquires a vev we get a contribution to the mass matrix (see Eqs. \ref{FI3}, \ref{typeII3Singletslc2invariance}):
\bea
M_1  &=&
\left(
\begin {array}{ccc}
a &b& 0\\
0& 0& f\\
0&0&f
\end {array}
\right)
\label{ChargedMasstypeII3SingletsM1lc2}
\eea
Eq. (\ref{typeII3SingletsDelta}) would lead, exactly as the three Higgs doublets did in the previous case, to
a mass contribution $M_2$ of the form of Eq. \ref{ChargedMasstypeII3Higgslc2} when the Higgs singlets
acquire vevs ($\d_k$), with the hierarchy $\d_3 \gg \d_1, \d_2$. Thus we get the charged lepton mass
matrix in the form:
\bea
M_l = M_1 + M_2 &=&
\left(
\begin {array}{ccc}
a &b& B^3\\
D^1& D^2 & f\\
C^1&C^2&f
\end {array}
\right) \label{ChargedMasstypeII3Singletslc2}
\eea

\item In both previous items we get a charged lepton mass matrix of the form
\bea
M_l  &=&
\left(
\begin {array}{c}
{\bf A}^T\\
{\bf B}^T\\
{\bf C}^T
\end {array}
\right) \label{ChargedMasstypeIIthreeSinglets}
\eea
adjustable so that the three vectors are linearly independent making the mass matrix invertible. The discussion in \cite{malkawi} on the charged lepton mass matrix of the same form showed the
possibility to adjust Yukawa couplings in order to get the charged lepton mass hierarchy, and then automatically the working basis will become the flavor basis up to order $\lambda_\m$. We shall not repeat the same analysis here, but just note that in case the parameters $a,b,f$ (corresponding to ${\cal{L}}_1$) are negligible compared to $B,C,D$
(related to ${\cal{L}}_3$) then the last item (Eq. \ref{ChargedMasstypeII3Singletslc2}) is similar to the last item of the past enumeration (Eq. \ref{ChargedMasstypeII3Higgslc2}), where we showed explicitly the charged lepton mass diagonalizing matrix being an infinitesimal rotation, which allows to consider the matrices as being those in the flavor basis, with a good approximation.

\end{itemize}

\end{enumerate}

Before we finish this subsection, we note that there is an  advantage for using the type-II seesaw mechanism in that the flavor changing neutral current due to the triplet is highly suppressed because of the
heaviness of the triplet mass scale, or equivalently the smallness of the neutrino masses.

\subsection{Type--I seesaw}

We proceed now to find a realization of the perturbed texture of pattern C1 (Eq. \ref{C1pattern}) in type-I seesaw mechanism where the effective
neutrino mass matrix ($M_\n$) is expressed in terms of the Dirac neutrino mass matrix ($M_D$) and the RH Majorana neutrino mass matrix ($M_R$) through:
\bea \label{seesaw} M_\n &=& M_D M_R^{-1} M_D^T \eea
For the flavor symmetry, we start by adding a new $Z_2$ symmetry (called $ Z_2^\prime$) to the flavor symmetry of the type II case, but we shall see that
it is not enough to achieve the desired form, and needs to be expanded to a larger group (say to $S \times Z_8$) for this.

\subsubsection{ $S \times Z_2 \times Z_2^\prime$-flavor symmetry}

We consider here a minimal extension to the flavor group of the type II seesaw by adding a new $Z_2$-symmetry so that to get the group $(Z_2)^3$.

\begin{enumerate}
\item{\bf Matter content and symmetry transformations}

We have three SM-like Higgs doublets ($\phi_i$, $i=1,2,3$) which would give mass to the charged leptons and another three Higgs doublets ($\phi^\prime_i$, $i=1,2,3$) for the
Dirac neutrino mass matrix. The RH neutrinos are denoted by ($\n_{Ri}$, $i=1,2,3$). These fields transform as follows.
\bea  &\n_R \stackrel{Z_2^\prime}{\longrightarrow} - \n_R ,\;\; \phi^\prime \stackrel{Z^\prime_2}{\longrightarrow} -\phi^\prime \label{typeIZ2Pone} &\\
&L \stackrel{Z_2^\prime}{\longrightarrow} L,\;\; l^c \stackrel{Z_2^\prime}{\longrightarrow} l^c , \;\; \phi \stackrel{Z_2^\prime}{\longrightarrow} \phi, \label{typeIZ2Ptwo}
 \eea
\bea  &\n_R \stackrel{Z_2}{\longrightarrow} \mbox{diag}(1,-1,-1) \n_R ,\;\; \phi^\prime \stackrel{Z_2}{\longrightarrow} \mbox{diag}(1,-1,-1)  \phi^\prime \label{typeIZ2one} &\\
&L \stackrel{Z_2}{\longrightarrow} \mbox{diag}(1,-1,-1)  L,\;\; l^c \stackrel{Z_2}{\longrightarrow} \mbox{diag}(1,1,-1)  l^c , \;\; \phi \stackrel{Z_2}{\longrightarrow} \mbox{diag}(1,-1,-1) \phi, \label{typeIZ2two}
 \eea
\bea  &\n_R \stackrel{S}{\longrightarrow} S \n_R , \;\; \phi^\prime \stackrel{S}{\longrightarrow} \mbox{diag}(1,1,-1)  \phi^\prime \label{typeISone} &\\
&L \stackrel{S}{\longrightarrow} S L, \;\; l^c \stackrel{S}{\longrightarrow}  l^c , \;\; \phi \stackrel{S}{\longrightarrow} S\phi, \label{typeIStwo}
 \eea
\item{\bf Charged lepton mass matrix-flavor basis}

As was the case of type-II seesaw with three SM-like Higgs doublets and where the RH charged lepton singlets transform differently from $L$ under $Z_2$, the Lagrangian responsible for the charged lepton mass is given by Eq. (\ref{L2}). The $Z_2^\prime$ does not play a role here, since all the fields involved are singlets under it, except for
the fact that it does forbid the trilinear coupling between $\phi^\prime, L$ and $l^c$. Again, assuming a hierarchy in the Higgs $\phi$'s fields vevs ($v_3 \gg v_2,v_1$)
we end up with a charged lepton mass matrix of the form (Eq. \ref{ChargedMasstypeII3Higgslc2}) which can be adjusted to be in the flavor basis to a good approximation.

\item{\bf Dirac neutrino mass matrix}

The Lagrangian responsible for the neutrino mass matrix is
\bea
 \label{L_D}
 {\cal{L}}_D &=& g^k_{ij} \overline{L}_i \tilde{\phi^\prime}_k  \n_{Rj} \,\, ,\mbox{ where  } \tilde{\phi^\prime} = i \s_2  \phi^{\prime *}
 \eea
This lagrangian is clearly invariant under $Z_2^\prime$ (see Eq. \ref{typeIZ2Pone}) which forces the existence of $\phi^\prime$ rather than $\phi$ in ${\cal{L}}_D$.
For the $S \times Z_2$ factor, we get via Eqs. (\ref{typeIZ2one},\ref{typeIZ2two}, \ref{typeISone} and \ref{typeIStwo}) then:
\bea S^{\mbox{\textsc{t}}} g^{(k=1,2)} S = g^{(k=1,2)} &,& S^{\mbox{\textsc{t}}} g^{(k=3)} S = - g^{(k=3)}, \overline{L}_i  \n_{Rj}
 \stackrel{Z_2}{\sim} \left(
\begin {array}{ccc}
1&-1 & -1\\
-1 & 1 & 1\\
-1 &1 &1
\end {array}
\right)
\label{typeIinvariance}
\eea
where $g^{(k)}$ is the matrix whose $(i,j)^{th}$-entry is the Yukawa coupling $g^k_{ij}$. Then,  Eqs. (\ref{FI1g}, \ref{FI2g}, \ref{typeIZ2one} and \ref{typeIinvariance}) lead
to the following forms of the Yukawa coupling matrices:
\bea
g^{(1)}  =
\left(
\begin {array}{ccc}
A^1 &0& 0\\
0& C^1& D^1\\
0&D^1&C^1
\end {array}\right) , g^{(2)}  =
\left(
\begin {array}{ccc}
0 &B^2& B^2\\
E^2& 0& 0\\
E^2&0&0
\end {array}\right), g^{(3)}  = \left(
\begin {array}{ccc}
0 &B^3& -B^3\\
E^3& 0& 0\\
-E^3&0&0
\end {array}\right)\eea
Upon acquiring vevs ($v_i^\prime$, $i=1,2,3$) for the Higgs fields ($\phi^\prime_i$), we get the following Dirac neutrino mass matrix:
\bea
M_D  = \Sigma_{k=1}^{k=3} v^\prime_k g^{(k)} &=& \left(
\begin {array}{ccc}
A_D &B_D& B_D (1+\a)\\
E_D& C_D& D_D\\
E_D (1+\b)&D_D&C_D
\end {array}\right) \label{DiracMatrix} \eea
with \bea
\a = \frac{-2v^\prime_3 B^3}{v^\prime_2 B^2 + v^\prime_3 B^3} &,& \b = \frac{-2v^\prime_3 E^3}{v^\prime_2 E^2 + v^\prime_2 E^3}
\label{alfabeta}
\eea
If the vevs satisfy $v^\prime_3 \ll v^\prime_2$ and the Yukawa couplings are of the same order then we get perturbative  parameters $\a, \b \ll 1$.

\item{\bf Majorana neutrino mass matrix}

The mass term is directly present in the Lagrangian
\bea
 \label{L_R}
 {\cal{L}}_R &=& M_{Rij} \n_{Ri}  \n_{Rj} \,\, \label{majorana}
 \eea
It is invariant under $Z_2^\prime$. Then Eqs. (\ref{typeISone},\ref{typeIZ2one}) lead to:
\bea S^{\mbox{\textsc{t}}} M_R S = M_R , \n_{Ri}  \n_{Rj}
 \stackrel{Z_2}{\sim} \left(
\begin {array}{ccc}
1&-1 & -1\\
-1 & 1 & 1\\
-1 &1 &1
\end {array}
\right)
 &\stackrel{\mbox{Eq.}\ref{FI1}}{\Longrightarrow}& M_R = \left(
\begin {array}{ccc}
A_R &0& 0\\
0& C_R& D_R\\
0&D_R&C_R
\end {array} \right)
 \label{M_R}\eea

\item{\bf Effective neutrino mass matrix}

One can see by direct computation that plugging Eqs. (\ref{DiracMatrix},\ref{M_R}) in the seesaw formula (Eq. \ref{seesaw}) would result in an effective
neutrino mass matrix of the form:
\bea
M_\n &=& \left(
\begin {array}{ccc}
M_{\n11} &M_{\n12}& M_{\n12} (1+\chi)\\
M_{\n12}& M_{\n22}& M_{\n23}\\
M_{\n12} (1+\chi)&M_{\n23}&M_{\n22} (1+\xi)
\end {array} \right)
\label{M_eff}
\eea
where ($Y=A,B,C,D,E$)
\bea& \chi = \chi(\a,\b,Y_D,Y_R) ,\;\; \xi = \xi(\b,Y_D,Y_R): \b=0 \Rightarrow \xi=0 &\label{perturbations}\eea
Thus, in general, we do not get the desired $C1$-pattern form (Eq. \ref{C1pattern})  corresponding to $\xi =0$. However, for some choices of the Yukawa couplings
satisfying $E^3=0$ we get
this form (see Eq. \ref{alfabeta}), with $\chi$, as $\a$, is a small parameter for moderate values of Yukawa couplings.

\end{enumerate}

\subsubsection{ $S \times Z_8 $-flavor symmetry}

In order to get a realization of the $C1$ pattern form with no need to tune the Yukawa couplings, we extend the flavor symmetry to be $S \times Z_8$.

\begin{enumerate}
\item{\bf Matter content and symmetry transformations}

The matter spectrum consists of three SM-like Higgs doublets ($\phi_i$, $i=1,2,3$) responsible for the charged lepton masses, and of four Higgs doublets ($\phi^\prime_j$, $j=1,2,3,4$) giving rise when acquiring a vev to Dirac neutrino mass matrix,
and, as before, of left doublets ($L_i$, $i=1,2,3$), RH charged singlets ($l^c_j$, $j=1,2,3$) and RH neutrinos ($\n_{Rj}$, $j=1,2,3$). We introduce also two Higgs singlet scalars
($\Delta_k$, $k=1,2$) related to Majorana neutrino mass matrix. We denote the octic root of the unity by $w=e^{\frac{i\pi}{4}}$. The fields transform under the flavor symmetry as follows.
\bea  &L \stackrel{S}{\longrightarrow} S L ,\;\; l^c \stackrel{S}{\longrightarrow}  l^c ,\;\;  \phi \stackrel{S}{\longrightarrow} S\phi, & \label{csawSone} \\ &
\n_R \stackrel{S}{\longrightarrow} S \n_R ,\;\; \phi^\prime \stackrel{S}{\longrightarrow} \mbox{diag}(1,1,1,-1)  \phi^\prime, \;\; \Delta  \stackrel{S}{\longrightarrow} \Delta & \label{csawStwo}
 \eea
\bea
&L \stackrel{Z_8}{\longrightarrow} \mbox{diag}(1,-1,-1)  L, \;\;l^c \stackrel{Z_8}{\longrightarrow} \mbox{diag}(1,1,-1)  l^c , \;\; \phi \stackrel{Z_8}{\longrightarrow} \mbox{diag}(1,-1,-1) \phi,& \label{csawZ8one}\\
 &\n_R \stackrel{Z_8}{\longrightarrow} \mbox{diag}(w,w^3,w^3) \n_R ,\;\; \phi^\prime \stackrel{Z_8}{\longrightarrow} \mbox{diag}(w,w^3,w^7,w^3)  \phi^\prime,\;\;
 \Delta  \stackrel{Z_8}{\longrightarrow} \mbox{diag}(w^6,w^2)\Delta
 \label{csawZ8two} &
 \eea
Note here that we have the following transformation rule for $\tilde{\phi^\prime} \equiv i \s_2 \phi^{\prime *}$:
\bea \label{tilderule}
\tilde{\phi^\prime}  \stackrel{S}{\longrightarrow} \mbox{diag}(1,1,1,-1) \tilde{\phi^\prime} &,& \tilde{\phi^\prime}  \stackrel{Z_8}{\longrightarrow} \mbox{diag}(w^7,w^5,w,w^5) \tilde{\phi^\prime}
\eea
\item{\bf Charged lepton mass matrix-flavor basis}

As in the previous case of $S\times Z_2 \times Z_2^\prime$-flavor symmetry,  the charged lepton mass Lagrangian is given again by Eq. (\ref{L2}). Since the transformations of the involved fields
($L,l^c,\phi$) are identical under $S$ in both flavor symmetry groups and are equally the same under $Z_8$ (in $S \times Z_8$) compared to $Z_2$ (in $S\times Z_2 \times Z_2^\prime$), we end up, assuming again a hierarchy in the Higgs $\phi$'s fields vevs ($v_3 \gg v_2,v_1$), with a charged lepton mass matrix of the form (Eq. \ref{ChargedMasstypeII3Higgslc2}) adjustable to be approximately in the flavor basis. Note also here that no terms of the form
$f^{k\prime}_{ij}  \overline{L}_i \phi^\prime_k  l^c_j$ can exist since we have:
\bea \overline{L}_i  l^c_j
 \stackrel{Z_8}{\sim} \left(
\begin {array}{ccc}
1&1 & -1\\
-1 & -1 & 1\\
-1 &-1 &1
\end {array}
\right) &\stackrel{\mbox{Eq.}\ref{csawZ8two}}{\Longrightarrow}& \nexists i,j,k: \overline{L}_i \phi^\prime_k  l^c_j=Z_8(\overline{L}_i \phi^\prime_k  l^c_j)
\label{noterm}\eea

\item{\bf Dirac neutrino mass matrix}

The Lagrangian responsible for the neutrino mass matrix is again given by Eq. (\ref{L_D}). By means of Eqs. (\ref{csawSone},\ref{csawStwo}, \ref{csawZ8one} , \ref{csawZ8two} and \ref{tilderule}) we have:
\bea S^{\mbox{\textsc{t}}} g^{(k=1,2)} S = g^{(k=1,2,3)} &,& S^{\mbox{\textsc{t}}} g^{(k=4)} S = - g^{(k=4)}, \overline{L}_i  \n_{Rj}
 \stackrel{Z_8}{\sim} \left(
\begin {array}{ccc}
w&w^3 & w^3\\
w^5 & w^7 & w^7\\
w^5 &w^7 &w^7
\end {array}
\right)
\label{csawinvariance}
\eea
where, as before,  $g^{(k)}$ is the matrix whose $(i,j)^{th}$-entry is the Yukawa coupling $g^k_{ij}$. Then,  Eqs. (\ref{FI1g}, \ref{FI2g} and \ref{tilderule} and \ref{csawinvariance}) impose the following forms on the Yukawa coupling matrices:
\bea
g^{(1)}  =
\left(
\begin {array}{ccc}
A^1 &0& 0\\
0& 0& 0\\
0&0&0
\end {array}\right) , g^{(2)}  =
\left(
\begin {array}{ccc}
0 &B^2& B^2\\
0& 0& 0\\
0&0&0
\end {array}\right), g^{(3)}  = \left(
\begin {array}{ccc}
0 &0& 0\\
0& C^3& D^3\\
0&D^3&C^3
\end {array}\right),
g^{(4)}  =
\left(
\begin {array}{ccc}
0 &B^4& -B^4\\
0& 0& 0\\
0&0&0
\end {array}\right)
\eea
When the Higgs fields ($\phi^\prime_i$) get vevs ($v_i^\prime$, $i=1,2,3,4$), we obtain the following Dirac neutrino mass matrix:
\bea
M_D  = \Sigma_{k=1}^{k=4} v^\prime_k g^{(k)} &=& \left(
\begin {array}{ccc}
A_D &B_D & B_D (1+\a)\\
0& C_D& D_D\\
0&D_D&C_D
\end {array}\right) \label{csawDiracMatrix} \eea
with \bea
\a &=& \frac{-2v^\prime_4 B^4}{v^\prime_2 B^2 + v^\prime_4 B^4}
\label{alfa}
\eea
If the vevs satisfy $v^\prime_4 \ll v^\prime_2$ and the Yukawa couplings are of the same order then we get a perturbative  parameter $\a \ll 1$.

\item{\bf Majorana neutrino mass matrix}

The mass term is generated from the Lagrangian
\bea
 \label{L_R}
 {\cal{L}}_R &=& h^k_{ij}\, \Delta_k\, \n_{Ri}\,  \n_{Rj} \,\, \label{csawLR}
 \eea
 Under $Z_8$ we have the bilinear:
\bea &\n_{Ri}\;  \n_{Rj}
 \stackrel{Z_8}{\sim} \left(
\begin {array}{ccc}
w^2&w^4 & w^4\\
w^4 & w^6 & w^6\\
w^4 &w^6 &w^6
\end {array}
\right) \stackrel{\mbox{Eq.}\ref{csawZ8two}}{\Longrightarrow}&  \nonumber \\&{\cal{L}}_R=h^1_{11}\, \Delta_1\, \n_{R1}\,  \n_{R1}
+ h^2_{11}\, \Delta_2\, \n_{R2}\,  \n_{R2} + h^2_{23}\, \Delta_2\, \n_{R2}\,  \n_{R3} + h^2_{32}\, \Delta_2\, \n_{R3}\,  \n_{R2} + h^2_{33}\, \Delta_2\, \n_{R3}\,  \n_{R3}&
\label{csawZ8lag}
\eea
If we call $h^{(k)}$  the matrix whose $(i,j)^{th}$-entry is the coupling $h^k_{ij}$ then we have (the cross sign denote a non-vanishing entry):
\bea
h^{(1)}  =
\left(
\begin {array}{ccc}
\times &0& 0\\
0& 0& 0\\
0&0&0
\end {array}
\right) &,& h^{(2)}  =
\left(
\begin {array}{ccc}
0 &0& 0\\
0& \times& \times\\
0&\times&\times
\end {array}
\right)
\label{csawZ8mass}
\eea
  Then Eq. (\ref{csawStwo}) leads to:
\bea S^{\mbox{\textsc{t}}} h^{(k)} S = h^{(k)} ,
 &\stackrel{\mbox{Eqs.}\ref{FI1},\ref{csawZ8mass}}{\Longrightarrow}& h^{(1)}  =
\left(
\begin {array}{ccc}
a_R &0& 0\\
0& 0& 0\\
0&0&0
\end {array}
\right) , h^{(2)}  =
\left(
\begin {array}{ccc}
0 &0& 0\\
0& c_R& d_R\\
0&d_R&c_R
\end {array}
\right)
 \eea
  Thus when the Higgs singlets $\Delta$ acquires vevs ($\d^0_1,\d^0_2$) we get the Majorana neutrino mass matrix:
\bea
M_R  = \sum_{k=1}^{2} \d^0_k h^{(k)} &=& \left(
\begin {array}{ccc}
A_R &0& 0\\
0& C_R& D_R\\
0&D_R&C_R
\end {array}\right) \label{csawMajoranaMatrix} \eea

\item{\bf Effective neutrino mass matrix}

By direct computation, plugging Eqs. (\ref{csawDiracMatrix},\ref{csawMajoranaMatrix}) into the seesaw formula (Eq. \ref{seesaw}) results in an effective
neutrino mass matrix of the desired $C1$-pattern form:
\bea
M_\n &=& \left(
\begin {array}{ccc}
M_{\n11} &M_{\n12}& M_{\n12} (1+\chi)\\
M_{\n12}& M_{\n22}& M_{\n23}\\
M_{\n12} (1+\chi)&M_{\n23}&M_{\n22}
\end {array} \right)
\label{csawM_eff}
\eea
where the perturbation parameter $\chi$ is given by
\bea \chi &=&  \frac{\a(C_D-D_D)(C_R+D_R)}{(1+\a)(C_R\,D_D-D_R\,C_D)+C_R\,C_D -D_R\,D_D} \label{perturbation}\eea

\end{enumerate}

Before ending this section, we would mention that introducing multiple Higgs doublets as we did in our constructions might display flavor-changing neutral currents. However, the effects are calculable in the models and in principle one can adjust the Yukawa couplings so that processes like $\m \rightarrow e \g$ are suppressed \cite{hagedorn}. Moreover, and as was discussed in the introduction, the RG running effects are expected to be small when multiple Higgs doublets are present, so that not to spoil the predictions of the symmetry at low scale.

%%%%%%%%%%%%%%%%%%%%%%%%%%%%%%%%%%%
\section{Summary and Discussion}
We have carried out a thorough phenomenological analysis for the patterns of the neutrino mass matrix meeting the $\mu$ -- $\tau$  symmetry. We found that exact
  symmetry leads to a totally degenerate spectrum and so is excluded on phenomenological grounds.
  %Thus, we did not study theoretical realizations
%of the $\mu$ -- $\tau$ exact symmetry, even though it is conceivable to impose discrete symmetries at the Lagrangian level leading to neutrino mass matrices displaying this
 % symmetry. One should however consider this kind of setups as a starting point to be perturbed at a later stage in order to
% reach viable patterns.

We thus introduced and in a minimal way perturbations such that the neutrino mass matrix satisfies an approximate $\mu$ -- $\tau$ symmetry. We got four such patterns and carried out
a complete phenomenological analysis of them. We found that all these `deformed' patterns can accommodate the current data without need to adjust the input parameters. However, no
singular such patterns could meet the experimental constraints.

All the four patterns can produce all types of hierarchy and all have complex entries able to show CP-violation effects. The mixing angle $\t_x$ can cover all its admissible
range in all four patterns. As to the angle $\t_y$, it is unconstrained in the patterns C3 except that it should not equal the value $45^0$, whereas it is
restricted to be around $45^0$, without taking this value, in the C1 pattern for the normal and inverted hierarchies, and around $36^0$ or $52^0$ in the C4 pattern of normal hierarchy type.
Again, $\t_y$ can not take the value $45^0$ in the C2 pattern of normal or inverted hierarchy types, where it is just mildly constrained in the normal type to be around $45^0$.
However, for this latter pattern C2, the mixing angle $\t_z$ can not be larger than $10^0$. Actually, there is a narrow interval $]4^0,4.7^0[$ for $\t_z$ in the C4 pattern of normal type, whereas this mixing angle is bounded by $8^0$ in the inverted type.

The phases are not constrained in the C3 or C4 patterns, except that in the C4 pattern of normal type the Dirac phase  $\d$  can not be in the interval $~]160^0,185^0[$ and
 the Majorana phase $\r \mbox{( mod } \pi)$ can not belong to $]-20^0,20^0[$. As to the C1 pattern of normal type, the phases $\s,\r \mbox{( mod } \pi)$ can not take values in the
 interval $]-4^0,4^0[$ around the origin, whereas the Dirac phase $\d$ in all hierarchy types is excluded from a narrow band $]177^0,180.5^0[$ around $\pi$. For the C2 pattern, the
 phase $\r$ is excluded from the interval $]94^0,99^0[$ in the degenerate case, and from broader intervals in the normal ($]90^0,111^0[$) and inverted ($]48^0,137^0[$) types.
 The phase $\s \mbox{( mod } \pi)$ is bound not to be around zero in the normal and inverted types, whereas the Dirac phase $\d$ in all hierarchy types is excluded from narrow bands
 around zero ($]-3^0,1^0[$) and around $\pi$ ($]178^0,185^0[$).

 There exist linear correlations between $\d,\r,\s$ for the patterns C1 and C3 in all types of hierarchy, and a linear correlation between $<m_{ee}>$ and the {\bf LNM} in the degenerate
 type for these two patterns.

 The strength of the hierarchies is characterized by the ratio $m_{23}$, and the normal type hierarchy is usually mild taking values of order 1 in all patterns. However,
 the inverted hierarchy type in the patterns C1 and C3 can be very acute taking values of order $O(10^2)$.

 All these features might help in distinguishing between the independent patterns. For example, if by measuring the mass ratios we find a very pronounced hierarchy, then we know that
 we have either C1 or C3 pattern, of an inverted hierarchy type. Consequently, if by measuring the angle $\t_y$ we find a value far from $45^0$ then we know we have a C3 pattern.
 Also if $\d$ gives a value around $\pi$ then again we have a C3 pattern. On the other hand, if by measuring the masses we get a mild hierarchy then we do not actually
 have enough signatures to determine the pattern. Rather, we have exclusion rules which help to drop as much patterns as possible. For example, if $\r \mbox{( mod } \pi) \in ]-20^0,20^0[$
 or $\t_z > 5^0$ or $\t_y \neq 36^0, 52^0$ then we can drop the C4 pattern of normal type, whereas if $\t_z > 8^0$ we exclude the C4 of inverted type possibility.
 If $|\r \mbox{( mod } \pi)| < 4^0 $ then no C1 pattern of normal type, while if $\r \in ]94^0,99^0[$ then we drop the possibility of a C2 pattern. Also if $\t_z \geq10^0$ then we conclude that we
  do not have a C2 pattern of normal type. Moreover, the knowledge of all the phase angles and other mass parameters jointly and referring to the `narrow' bands of the
  correlation plots can help in deciding which texture does fit the data.

  We note finally that the deformation parameter $|\chi|$ can cover all its `perturbative' range ($\leq 20\%$), except for the pattern C4 where it is bound to be a `tangible'
  deformation ($|\chi| \geq 16\%$) in order to fit the experimental data.

All the perturbed patterns can be realized assuming exact $\m-\tau$ symmetry augmented by new matter fields and abelian symmetries at the Lagrangian level, and we have presented
some concrete examples using both types I and II of seesaw mechanism.

Our analysis follows a bottom-up approach and, in view of the full parameter space we adopted for the observables, can be considered as new. In particular, it shows in a very transparent
way the correlation between the perturbation $\chi$ and the non-vanishing $\t_z$. We can summarize the mainly new results in our work as follows. First, we presented the
complete analytical expressions (full or expanded) for all the observables and in all patterns. Second, we raised the question of convergence of
the expansion series (Eq. \ref{m13exp}) and analyzed it. Third, we presented an exhaustive analysis plotting all the possible correlations. Fourth, we disentangled the
effects of the two perturbation parameters and presented detailed theoretical realizations of the resulting perturbed patterns. Fifth, we treated also the case of singular
neutrino mass matrix. Sixth, we reached different conclusions compared to some other works with far more restricted parameter space.

%%%%%%%%%%%%%%%%%%%%%%%%%%%%%%%%%%%%%

\section*{{\large \bf Acknowledgements}}
Part of the work was done
within the associate scheme and short visits program of ICTP.\\  N.C.
acknowledges funding provided by the Alexander von Humboldt Foundation.

\end{document}